%% file: main.tex
\documentclass[12pt]{article}
\usepackage{hyperref}
\usepackage{cite}
\usepackage[dvips]{graphicx}
\usepackage{epsfig}
\usepackage{amssymb}
\usepackage{rotating}
\usepackage{mathtools}
\usepackage{color,soul}
\usepackage{hhline}
\usepackage{longtable}
\usepackage{multirow}

\newcommand\ol[1]{\overline{#1}}  		  
\newcommand\nn[0]{\nonumber}			  
\newcommand\sqb[1]{\left[#1\right]}      
\newcommand\nba[1]{\!\left(#1\right)}      
\newcommand\ovbb[0]{0\nu\beta\beta}      

\newenvironment{rcasesdef}
  {\left.\begin{aligned}}
  {\end{aligned}\right\rbrace}

\begin{document}
\vspace*{-1in}
\renewcommand{\thefootnote}{\fnsymbol{footnote}}

\begin{flushright}
\scriptsize{\texttt{DO-TH 17/19\\
CP3-Origins-2017-055 DNRF90\\
}}
\end{flushright}
\vskip 5pt
\begin{center}
{\Large{\bf 
Neutrinoless Double Beta Decay and\\[0.2cm] the Baryon Asymmetry of the Universe}}
\vskip 20pt
{
Frank F. Deppisch$^1$\footnote{E-mail: f.deppisch@ucl.ac.uk},
Lukas Graf$^1$\footnote{E-mail: lukas.graf.14@ucl.ac.uk},
Julia Harz$^2$\footnote{E-mail: jharz@lpthe.jussieu.fr},
Wei-Chih Huang$^{3,4}$\footnote{E-mail: huang@cp3.sdu.dk}
}

\vskip 15pt

{\it \small $^1$Department of Physics and Astronomy, UCL, Gower Street, London WC1E 6BT, UK}\\
{\it \small $^2$Sorbonne Universit\'es, Institut Lagrange de Paris, 98bis Blvd. Arago, F-75014 Paris; Sorbonne Universit\'es, UPMC Univ Paris 06 \& CNRS, UMR 7589, LPTHE, F-75005 Paris; France}\\
{\it \small $^3$  Fakult\"at f\"ur Physik, Technische Universit\"at Dortmund,
44221 Dortmund, Germany}\\
{\it \small $^4$CP$^3$-Origins, University of Southern Denmark, Campusvej 55, DK-5230 Odense M, Denmark}\\

\medskip

\begin{abstract}
\noindent
We discuss the impact of the observation of neutrinoless double beta decay on the washout of lepton number in the early universe. Neutrinoless double beta decay can be triggered by a large number of mechanisms that can be encoded in terms of Standard Model effective operators which violate lepton number by two units. We calculate the contribution of such operators to the rate of neutrinoless double beta decay and correlate it with the washout of lepton number induced by the same operators in the early universe. We find that the observation of a non-standard contribution to neutrinoless double beta decay, i.e. not induced by the standard mass mechanism of light neutrino exchange, would correspond to an efficient washout of lepton number above the electroweak scale for many operators up to mass dimension 11. Combined with Standard Model sphaleron transitions, this would render many baryogenesis mechanisms at higher scales ineffective.
\end{abstract}

\end{center}
\medskip


\renewcommand{\thefootnote}{\arabic{footnote}}
\setcounter{footnote}{0}
\newpage
\input{intro}
\input{ovbb}
\input{sm-operators}
\input{ovbb-contributions}
\input{washout}
\input{results}
\input{conclusions}
\section*{Acknowledgements}
\appendix
The authors are grateful to Martin Hirsch and Heinrich P\"as for collaboration during the early stages of this project. The authors would like to thank Brian Henning and Carlos Tamarit for useful discussions on renormalization in effective field theories and  Marcela Paz Gonz\'alez for useful discussions on the QCD running of neutrinoless double beta decay operators. The work of FFD and LG was supported by an STFC Consolidated Grant and a Royal Society Exchanges Grant. JH was supported by the Labex ILP (reference ANR-10-LABX-63) part of the Idex SUPER, and received financial state aid managed by the Agence Nationale de la Recherche, as part of the programme Investissements d'avenir under the reference ANR-11-IDEX-0004-02. WCH was supported by DFG Grant No. PA 803/10-1 and by Danish Council for Independent Research Grant DFF-6108-00623.
\bibliography{literature}
\bibliographystyle{h-physrev}
\end{document}

%% file: intro.tex
\section{Introduction \label{sec:introduction}}
The dynamics of the Standard Model (SM) is determined by its gauge symmetry and chiral structure. Not only does the group $SU(3)_C \times SU(2)_L \times U(1)_Y$ explain the interactions we observe in nature, its breaking also provides masses to the charged fermions via the Higgs mechanism. The discovery of the Higgs boson at the Large Hadron Collider (LHC)~\cite{Aad:2012tfa, Chatrchyan:2012ufa} has put us into the position to probe and verify this mass mechanism in the SM. 

Yet, neutrinos continue to evade our understanding. Being only left-handed, neutrinos cannot acquire a so-called Dirac mass like the other SM fermions. Neutrino oscillation experiments \cite{Agashe:2014kda} have clearly shown though that at least two of the three known neutrinos have masses. While oscillations cannot probe the absolute masses of neutrinos, they point to mass scales of order $10^{-2}$~eV to $5\times 10^{-2}$~eV corresponding to the solar and atmospheric oscillation lengths, respectively. On the other hand, cosmological observations set an upper limit on the sum of all active neutrino masses $\Sigma m_\nu \lesssim 0.15$~eV \cite{Ade:2015xua}, assuming the standard cosmological model and depending on the observational data considered. Nevertheless, one can say that the masses of two of the neutrinos have been determined to be in the range $\approx 0.01$ -- $1$~eV, but a precise measurement of the absolute neutrino mass scale, e.g. represented by the lightest neutrino mass, is still outstanding.

Neutrino masses could be of Dirac type, but this requires the existence of a new `right-handed' neutrino field $\nu_R$ and tiny Yukawa couplings $\lesssim 10^{-12}$. Such small couplings look rather unnatural (although the next-smallest Yukawa coupling, that of the electron, is also rather small $\approx 10^{-6}$), but this scenario is perfectly acceptable from a theoretical point of view. On the other hand, because the right-handed neutrinos would be completely sterile in the SM, it is theoretically possible for them to acquire a so-called Majorana mass $M$ of the form $M \bar \nu_L C \bar\nu_L^T$; in fact, such a mass would be expected to exist because it is not forbidden by the SM gauge symmetry; instead, it violates lepton number $L$ (an accidental symmetry in the SM) by two units, $\Delta L = 2$. As a bare Majorana mass, $M$ is unrelated to SM physics and especially to the electroweak (EW) breaking scale $m_\text{EW} \approx 100$~GeV. It is thus generically expected to be of the order of a large new physics scale $\Lambda_L \approx M$ associated with the breaking of the lepton number $L$ symmetry. Together with the Yukawa couplings between left and right-handed neutrinos, this will induce an effective dimension-5 operator, $\Lambda^{-1}_L(L L H H)$ \cite{Weinberg:1979sa}, where $L$ and $H$ represent the $SU(2)_L$ doublets of the left-handed lepton and the Higgs fields, respectively. After EW symmetry breaking, this generates a small effective Majorana mass $m_\nu \sim m_\text{EW}^2 / \Lambda_L$ for the dominantly active neutrinos. This is of course the famous seesaw mechanism \cite{Minkowski:1977sc, mohapatra:1979ia, Yanagida:1979as, seesaw:1979, Schechter:1980gr}, with an $L$ breaking scale naturally of the order $\Lambda_L \approx 10^{14}$~GeV to achieve the light neutrino masses $m_\nu \approx 0.1$~eV. 

While certainly the most prominent case, the above scenario is just one example of how the effective $L$-violating Weinberg operator and thus a Majorana mass for the active neutrinos can be generated; at the tree level there are two further generic ways, via triplet scalars and fermions, respectively, and there are numerous other ways at higher loop order and when allowing higher-dimensional effective interactions beyond the Weinberg operator. What these models have in common is that in order to generate light Majorana masses for the active neutrinos, $L$ needs to be broken. This symmetry, along with baryon number $B$ symmetry, is accidentally conserved in the SM at the perturbative level. Weak non-perturbative instanton and sphaleron effects through the chiral Adler-Bell-Jackiw anomaly \cite{Adler:1969gk, Bell:1969ts} do in fact violate baryon and lepton number but only in the combination $(B+L)$~\cite{tHooft:1976up}. The `orthogonal' combination $(B-L)$ remains conserved and thus lepton number violation (LNV), or more generally $(B-L)$ violation, along with the generation of Majorana neutrino masses requires the presence of New Physics beyond the SM (BSM).

In this context, a clear hint for physics beyond the SM is the observation of a baryon asymmetry of our Universe, quantified in terms of the baryon-to-photon number density \cite{Ade:2013zuv}
\begin{align}
\label{eq:etaBobs}
  \eta_B^\text{obs} = \left(6.20 \pm 0.15\right) \times 10^{-10}.
\end{align}
In order to generate a baryon asymmetry the three Sakharov conditions have to be fulfilled, namely (1) $B$ violation, (2) $C$ and $CP$ violation and (3) out-of-equilibrium dynamics. Different mechanisms and models exist which exhibit these conditions. One popular scenario is baryogenesis via leptogenesis (LG) \cite{Fukugita:1986hr}. In the standard `vanilla' scenario, a right-handed heavy neutrino decays out of equilibrium via a lepton number and $CP$ violating decay. As long as this happens before the EW phase transition, the lepton asymmetry is translated into a baryon asymmetry via sphaleron processes.

While the violation of $B-L$ is a crucial ingredient, e.g. in the leptogenesis scenario, in order to satisfy the third Sakharov condition, any LNV interactions must not be too efficient. Otherwise they remove the lepton number asymmetry and, due to the presence of sphaleron transitions in the SM, also the baryon number asymmetry before it is frozen in at the EW breaking scale. The search for LNV processes, with neutrinoless double beta ($\ovbb$) as the most prominent example, therefore provides a potential pathway to probe or rather falsify certain baryogenesis scenarios, if the lepton number washout in the early universe can be correlated with the LNV process rate. In this paper, we take such an approach in a model-independent fashion and study SM invariant operators of mass dimension 5, 7, 9 and 11 that violate lepton number by two units. We correlate their contribution to $\ovbb$, either at tree level or induced by radiative effects, with the lepton number washout in the early universe. Assuming the observation of $\ovbb$ decay, where we take the expected sensitivity of $T_{1/2}^{\ovbb} \approx 10^{27}$~y of future $\ovbb$ experiments, we determine the temperature range where the corresponding lepton number washout is effective.

After the discovery of the sphaleron transitions, the constraint on LNV operators from the requirement to protect the observed baryon asymmetry was soon realized \cite{Fukugita:1990gb}, with the Weinberg operator as the most prominent example \cite{Fukugita:1990gb, Harvey:1990qw, Campbell:1992jd, Campbell:1990fa, Campbell:1991at, Fischler:1990gn, Nelson:1990ir}. More generic non-renormalizable operators were discussed in \cite{Campbell:1990fa, Campbell:1991at} while the argument can also be extended to baryon number violating $\Delta B = 2$ operators inducing neutron-antineutron oscillations \cite{Kuzmin:1970nx, Mohapatra:1980qe}. More recently, we have shown in \cite{Deppisch:2013jxa} that searches for resonant LNV processes at the LHC can be used to infer strong lepton number washout and in \cite{Deppisch:2015yqa, Harz:2015fwa} we have demonstrated the principle to correlate the washout rate with non-standard $\ovbb$ contributions. In this paper we discuss the latter approach in more detail and extend the analysis to more than the four example operators previously analyzed. Clearly, more stringent constraints on the scale of baryogenesis can be derived in specific models. For example, in left-right symmetric scenarios, the gauge interaction felt by the right-handed neutrinos inducing leptogenesis will lead to a more rapid equilibration and thus strong constraints on the left-right breaking scale \cite{Frere:2008ct, Dhuria:2015cfa}. A similar effect occurs in $U(1)_{B-L}$ models where additional Yukawa couplings can washout the lepton asymmetry \cite{Dev:2017xry}.

The paper is organized as follows. In Section~\ref{sec:0vbb}, we discuss basic properties of $\ovbb$. Section~\ref{sec:sm-operators} provides a list of effective SM invariant operators up to dimension 11 that violate lepton number by two units. These form the basis of the subsequent discussion, which first deals with contributions to $\ovbb$ from such effective operators in Section~\ref{sec:0vbbcontribs}. It describes an algorithm we employ to estimate the radiative generation of operators that trigger $\ovbb$ decay. Section~\ref{sec:washout} then describes the washout of lepton number in the early Universe by an effective operator. In Section~\ref{sec:results} we correlate the washout with $\ovbb$ decay, and determine the temperature range in the thermal history where a given operator effectively washes out lepton number under the assumption that it triggers $\ovbb$ decay at an observable rate. We discuss the results and comment on potential caveats in Section~\ref{sec:conclusions}.

%% file: ovbb.tex
\section{Neutrinoless Double Beta Decay}\label{sec:0vbb}
The search for $\ovbb$ decay, i.e. the decay of an even-even nucleus emitting two electrons, is the most sensitive tool for probing Majorana neutrino masses. For example, the currently most stringent lower limit on the decay half life $T_{1/2}$ is derived using the Xenon isotope ${}^{136}_{\phantom{1}54}$Xe,
\begin{align}
	T_{1/2}^\text{Xe} \equiv T_{1/2}\left({}^{136}_{\phantom{1}54}\text{Xe} \to {}^{136}_{\phantom{1}56}\text{Ba} + e^- e^-\right) \gtrsim 10^{26}~\text{y}.
\end{align}
However, while this so-called mass mechanism is certainly the best known that triggers the decay, Majorana neutrino masses are not the only element of BSM physics which can induce it. Other mechanisms of $\ovbb$ decay where the LNV does not directly originate from Majorana neutrino masses but rather due to LNV masses or couplings of new particles appearing in various possible extensions of the SM. While the same couplings will also induce Majorana neutrino masses, due to the Schechter-Valle black box argument \cite{Schechter:1981bd}, the $\ovbb$ decay half life will not yield direct information about the neutrino mass. We rather consider the $\ovbb$ decay rate by expressing the new physics contributions in terms of effective low-energy operators \cite{Pas:1999fc, Pas:2000vn}.

\begin{figure}
	\centering
	\includegraphics[clip,width=0.242\textwidth]{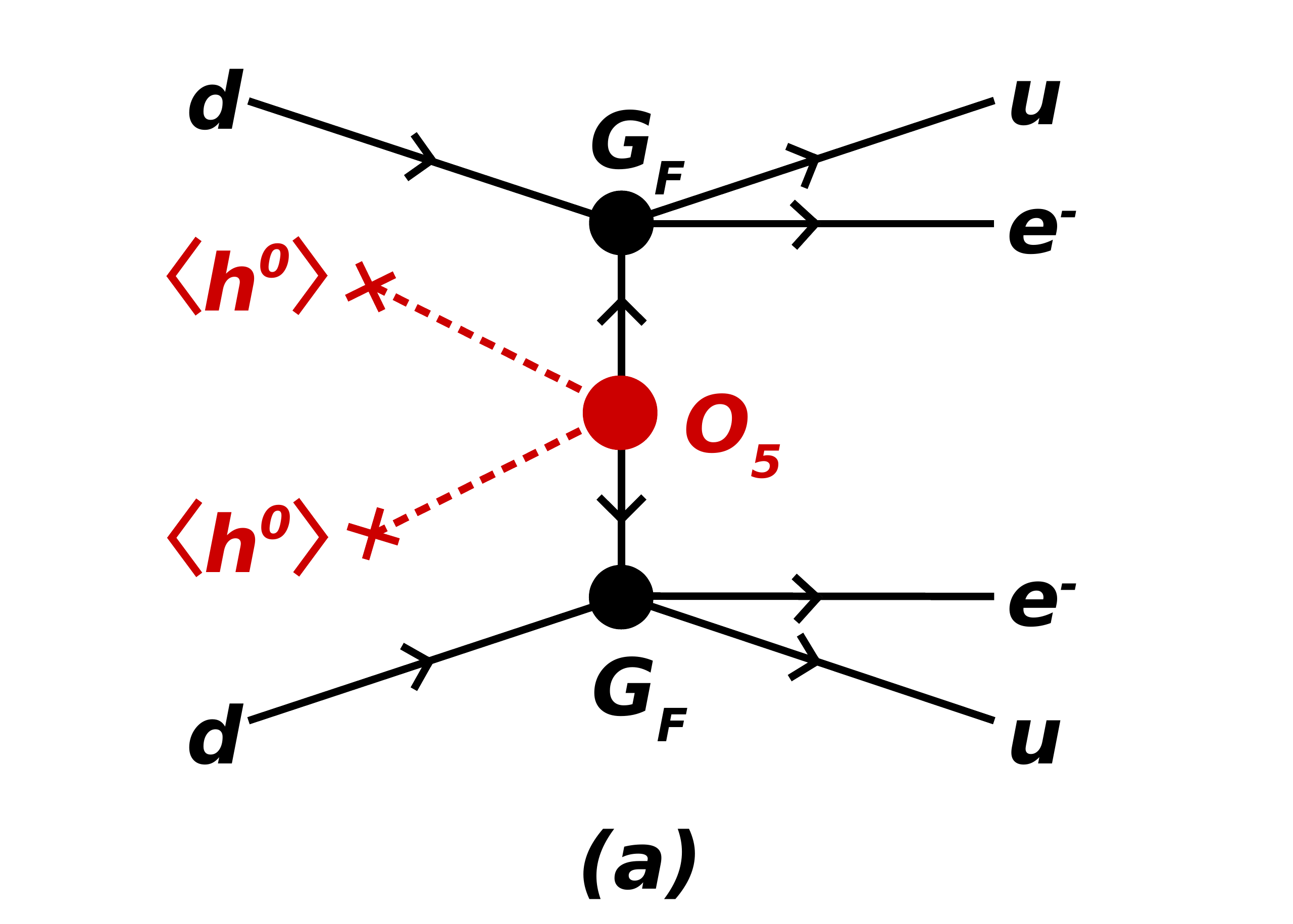}
	\includegraphics[clip,width=0.242\textwidth]{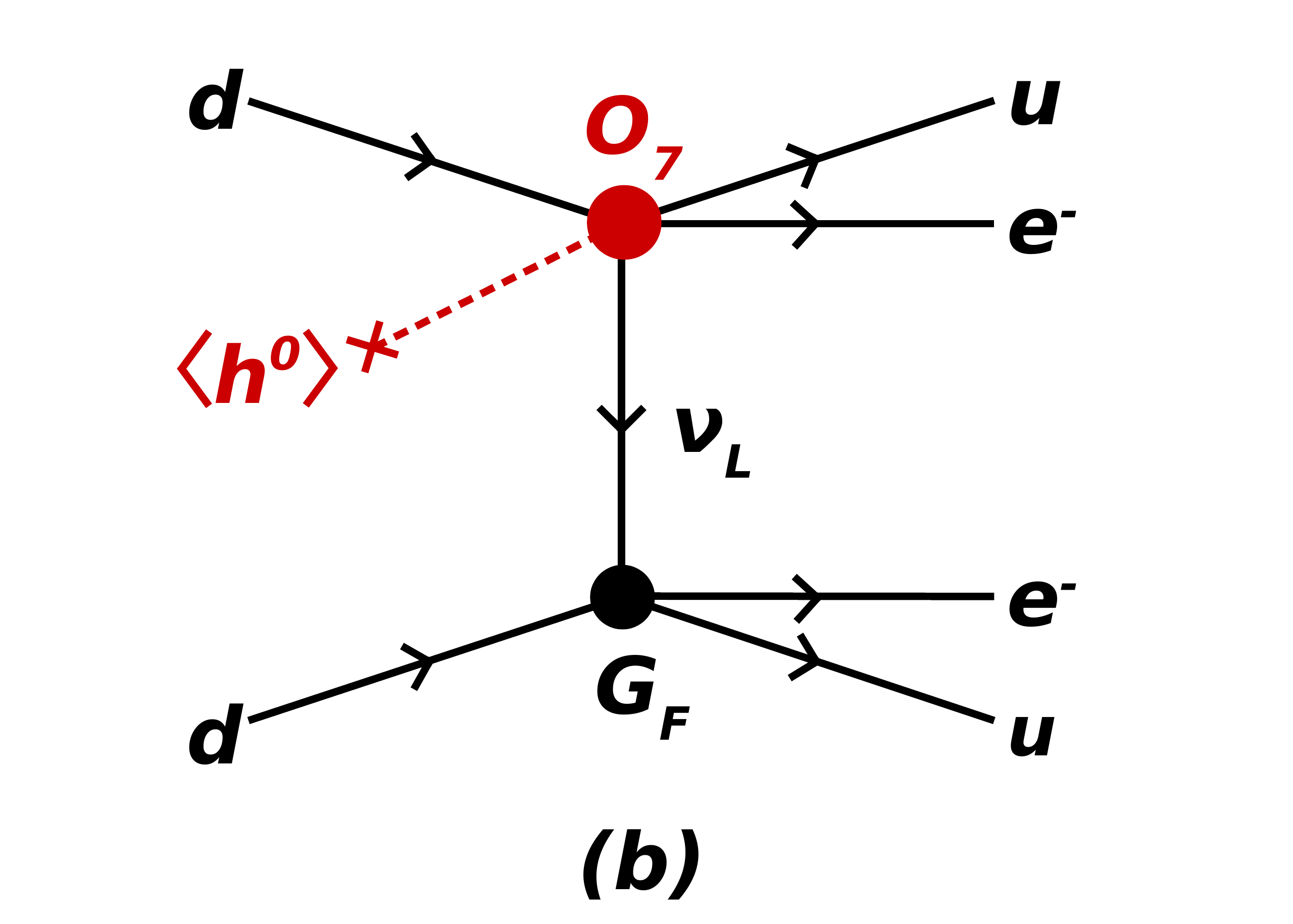}
	\includegraphics[clip,width=0.242\textwidth]{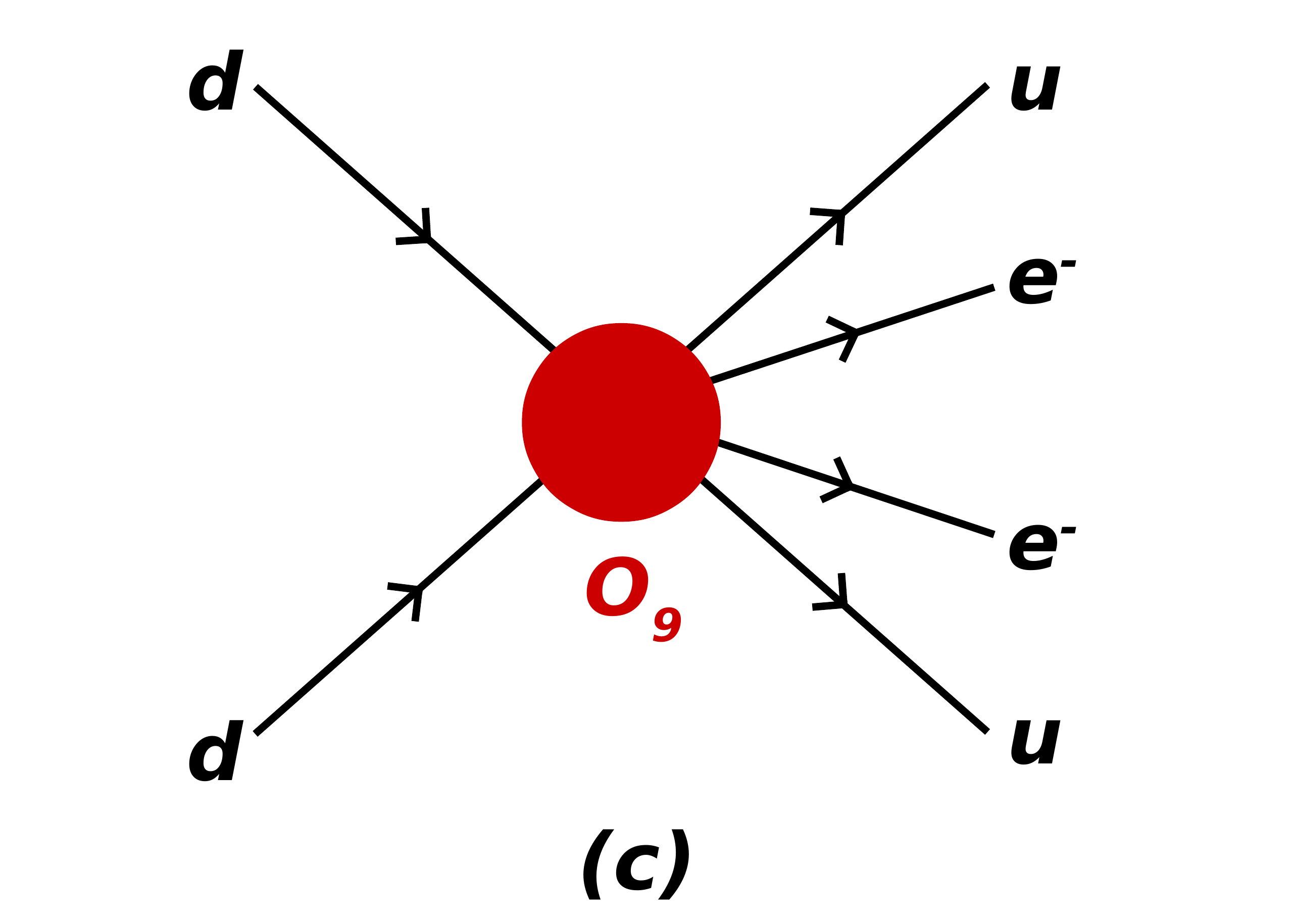}
	\includegraphics[clip,width=0.242\textwidth]{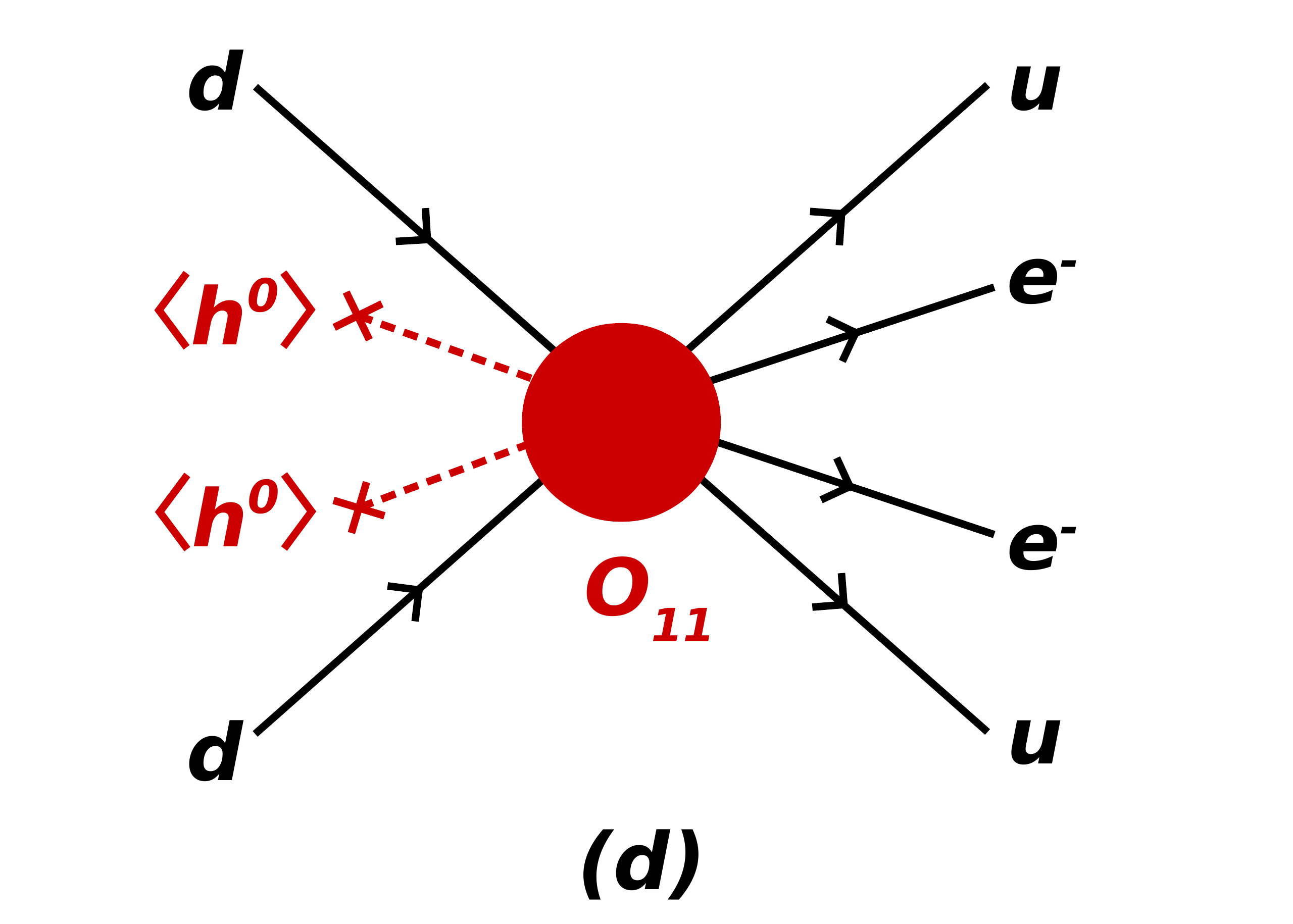}
	\caption{Contributions to $\ovbb$ from effective higher-dimensional LNV operators: (\emph{a}) 5-dim Weinberg operator (standard mass mechanism), (\emph{b}) 7-dim operator leading to long--range contribution, (\emph{c}, \emph{d}) 9-dim and 11-dim operators leading to short--range contributions.}
	\label{fig:graphs} 
\end{figure}
The nuclear matrix elements (NMEs) of the nuclear transition and their uncertainties are notoriously difficult to calculate and limits derived from $\ovbb$ decay are affected. Despite efforts devoted to the improvement of the nuclear modelling, the latest matrix elements obtained using various approaches differ in many cases by factors of ${\sim}(2-3)$. Experimentally, the most stringent bounds on $\ovbb$ decay are currently from $^{76}$Ge \cite{Agostini:2017iyd} and $^{136}$Xe \cite{KamLAND-Zen:2016pfg}. The results presented below are adapted from \cite{Deppisch:2012nb}, using the recent results in $^{76}$Ge of $T^\text{Ge}_{1/2} \ge 5.3\times 10^{25}$~y and in $^{136}$Xe of $T^\text{Xe}_{1/2} \ge 1.07\times 10^{26}$~y at $90\%$ confidence level (CL). Planned future experiments searching for $\ovbb$ decay are expected to reach sensitivities of the order of $T_{1/2} \approx 10^{27}$~y. For example, the recent comparative analysis \cite{Agostini:2017jim} quotes a discovery sensitivity at $3\sigma$ of $T^\text{Xe}_{1/2} = 4.1\times 10^{27}$~y for the planned nEXO experiment \cite{Mong:2016sza}. For definiteness, we will use a prospective sensitivity of
\begin{align}
	T^\text{Xe}_{1/2} = 10^{27}~\text{y},
\end{align} 
in $^{136}$Xe for our analysis. As we will outline below, our results are not very sensitive to the exact value of $T_{1/2}$, the isotope or the specific nuclear matrix elements.

As a basis of our subsequent discussion, we provide a brief overview of the possible effective contact interactions at the Fermi scale $m_F \approx 100$~MeV at which $\ovbb$ decay occurs. These are likewise triggered by effective SM invariant operators violating $\Delta L = 2$ of dimension 5, 7, 9, 11, etc., which we will discuss in the next section. Fig.~\ref{fig:graphs} shows the contribution of such operators schematically. For more details on the effective $\ovbb$ interaction, see for example the review \cite{Deppisch:2012nb} and references therein. General up-to-date reviews of $\ovbb$ decay and associated physics can be found in \cite{Vergados:2016hso}, while a more specific recent review on $0\nu\beta\beta$ NMEs is available in \cite{Engel:2016xgb}.

\paragraph{Standard Mass Mechanism}
Before discussing the exotic contributions of our interest, we remind the reader that the mass mechanism of $0\nu\beta\beta$ decay is sensitive to the effective Majorana neutrino mass
\begin{align}
	m_\nu = \sum_{j=1}^3 U_{ej}^2 m_{\nu_j} \equiv m_{ee},
\end{align}
where the sum runs over all active light neutrinos with masses $m_{\nu_j}$, weighted by the square of the charged-current leptonic mixing matrix $U$.\footnote{In this work we explicitly assume the particle content of the SM and we do not consider the presence of light ($\lesssim 100$~GeV), sterile neutrinos. Sterile neutrinos heavier than the electroweak scale are implicitly taken into account through their impact on the effective operators considered later.} This quantity is equal to the $(ee)$ entry of the Majorana neutrino mass matrix. The inverse $0\nu\beta\beta$ decay half life in a given isotope is then given by
\begin{align}
	T_{1/2}^{-1} = \left|\frac{m_\nu}{m_e}\right|^2 G_0 |M_\nu|^2,
\end{align}
where $G_0$ is the nuclear phase space factor and $|M_\nu|$ the corresponding NME. The effective Majorana neutrino mass is normalized with respect to the electron mass $m_e$ to yield a small dimensionless parameter $\epsilon_\nu = m_\nu/m_e$, comparable between different contributions. The current experimental results lead to a limit $m_\nu \lesssim 0.06 - 0.17$~eV \cite{KamLAND-Zen:2016pfg} with an uncertainty due to the nuclear matrix elements. Future experiments will aim to probe $m_\nu \approx 0.02$~eV.

\paragraph{Additional Long-Range Contributions} to $0\nu\beta\beta$ decay involve two vertices with the exchange of a light neutrino in between, cf.~Fig.~\ref{fig:graphs}~(b). The general Lagrangian can be written in terms of effective couplings $\epsilon^{\alpha}_{\beta}$ \cite{Pas:1999fc},
\begin{align}
\label{eq:L_longrange}
	{\cal L} = 
	\frac{G_F}{\sqrt{2}}
	\left( 
		j_{V-A}^{\mu} J^{\dagger}_{V-A,\mu} 
		+\sum_{\alpha,\beta} \epsilon_{\alpha}^{\beta} 
		j_{\beta} J^{\dagger}_{\alpha}
	\right),
\end{align}
where $j_{\beta} = \bar{e} {\cal O}_\beta \nu$ and $J_{\alpha}^{\dagger} = \bar{u} {\cal O}_\alpha d$ are the leptonic and hadronic currents, respectively. The sum runs over all possible Lorentz-invariant combinations with right-handed leptonic currents where the standard case $\alpha = \beta = (V-A)$ discussed above is shown separately. All currents are conventionally scaled relative to the strength of the ordinary ($V-A$) interaction, $G_F$, via the dimensionless $\epsilon$ factors. The operators ${\cal O}_{\alpha}$ in Eq.~\eqref{eq:L_longrange} are ${\cal O}_{V\pm A} = \gamma^{\mu}(1 \pm \gamma_5)$, ${\cal O}_{S\pm P} = (1 \pm \gamma_5)$ and ${\cal O}_{T_{R|L}} = \frac{i}{2}[\gamma_{\mu},\gamma_{\nu}](1 \pm \gamma_5)$.

\paragraph{Short--Range Contributions} to $0\nu\beta\beta$ decay involve a single effective contact interaction of dimension-9, cf.~Fig.\ref{fig:graphs}~(c,d). The possible operators are \cite{Pas:2000vn}
\begin{align}
\label{eq:L_shortrange}
	{\cal L} &= 
	\frac{G^2_F}{2m_p} \left(
		\epsilon_1 J          J           j
	  + \epsilon_2 J^{\mu\nu} J_{\mu\nu}  j
	  + \epsilon_3 J^{\mu}    J_{\mu}     j
	  + \epsilon_4 J^{\mu}    J_{\mu\nu}  j^{\nu}
	  + \epsilon_5 J^{\mu}    J           j_{\mu} 
\right),
\end{align}
with the hadronic currents $J = \overline{u}(1 \pm \gamma_5)d $, $J^\mu = \overline{u} \gamma^\mu(1 \pm \gamma_5)d$, $J^{\mu\nu} = \overline{u} \frac{i}{2}[\gamma^\mu,\gamma^\nu](1 \pm \gamma_5)d$ and the leptonic currents $j = \overline{e}(1 \pm \gamma_5) e^C$, $j^\mu = \overline{e}\gamma^\mu(1 \pm \gamma_5) e^C$. The operators are scaled with respect to a point-like, double beta decay-like interaction with the proton mass $m_p$.

\begin{table}[t!]
\centering
\begin{tabular}{c|c|cccc|cccccccc}
\hline
$|\epsilon|\times10^8$  & $\epsilon_\nu$ & $\epsilon^{V+A}_{V-A}$ &  $\epsilon^{V+A}_{V+A}$ & $\epsilon^{S+P}_{S\pm P}$ &  $\epsilon^{T_R}_{T_R}$ & $\epsilon_1$ & $\epsilon_2$ & 
$\epsilon_3^a$ & $\epsilon_3^b$ & 
$\epsilon_4$ & $\epsilon_5$ \\
\hline
$^{76}$Ge & 41 & 0.21 & 37 & 0.66 & 0.07 & 19 & 0.11 & 1.30 & 0.83 & 0.90 & 9.0 \\
$^{76}$Xe & 26 & 0.11 & 22 & 0.26 & 0.03 & 10 & 0.05 & 0.43 & 0.66 & 0.46 & 4.6 \\
\hline
\end{tabular}
\caption{Upper limits on effective $\ovbb$ interactions (in units of $10^{-8}$) from the current experimental bounds $T_{1/2}^\text{Ge} \gtrsim 5.3 \times 10^{25}$~y and $T_{1/2}^\text{Xe} \gtrsim 1.07 \times 10^{26}$~y. Only one $\epsilon$ is assumed to be present at a time. The coupling $\epsilon_\nu = m_{\nu}/m_e$ is due to the standard mechanism with the corresponding limits on the effective $\ovbb$ mass of 0.21~eV (Ge) and 0.13~eV (Xe). For the short-range interaction $\epsilon_3$, the limit depends on the chirality of the hadronic currents: $\epsilon_3^a$ (hadronic currents have same chirality), $\epsilon_3^b$ (hadronic currents have opposite chirality). Table adapted from \cite{Deppisch:2012nb}.}
\label{tab:limits}
\end{table}
In our subsequent analysis, the different non-standard operators, either long-range or short-range, originate from SM invariant LNV operators, themselves understood to be generated in a specific BSM scenario. Considering one operator with associated $\epsilon_{\alpha}^{\beta}$ at a time, the inverse $0\nu\beta\beta$ half life can be expressed as
\begin{align}
\label{eq:T12}
	T_{1/2}^{-1} = |\epsilon_{\alpha}^{\beta}|^2 G_{0k} |M_{\alpha\beta}|^2,
\end{align}
analogous to the standard case, where $G_{0k}$ denotes the corresponding nuclear phase space factors and $|M_{\alpha\beta}|$ the nuclear matrix elements, which depend both on the isotope as well the operator in question. Using the approach of \cite{Deppisch:2012nb}, the current limits on the effective $\ovbb$ interaction are shown in Tab.~\ref{tab:limits}\footnote{Compared to \cite{Deppisch:2012nb}, we omit the interaction $\epsilon^{T_R}_{T_L}$, which was shown to vanish \cite{Graesser:2016bpz}.}.

Having in mind the generation of the above long-range and short-range $\ovbb$ interactions from effective SM gauge-invariant LNV operators and ultimately from new physics where lepton number is broken at a high scale $\Lambda_\text{LNV}$, we neglect several effects. A complete scheme would include the QCD running and consequently generation of color-mismatched operators \cite{Mahajan:2013ixa, Gonzalez:2015ady, Peng:2015haa, Arbelaez:2016uto}. At the QCD scale, operators should be matched to a chiral effective field theory \cite{Cirigliano:2017ymo} and pion-induced transitions can in fact be dominant for short-range interactions, compared to the usually considered four-nucleon transitions. As already mentioned, the matrix elements calculated in nuclear structure theories continue to have large uncertainties. Despite the importance of such effects, we do not anticipate that our results depend on them strongly. This is because the underlying SM invariant operators that we consider are of dimension $D = 7,9,11$, and thus the uncertainty in the nuclear matrix element only enters as the third root for $D = 7$ when relating the $\ovbb$ half life with the operator scale, $\Lambda^{D=7}_\text{LNV} \propto 1/\sqrt[6]{T_{1/2}} \propto \sqrt[3]{|M|}$, and even weaker for the higher-dimensional operators.

%% file: sm-operators.tex
\section{Survey of LNV operators}
\label{sec:sm-operators}

So far the most exhaustive (but by no means complete) list of SM effective operators violating lepton number by two units has been published in \cite{deGouvea:2007xp}, based on an initial enumeration in \cite{Babu:2001ex}. In \cite{Deppisch:2015yqa} we studied four different operators contributing at tree-level to $\ovbb$ decay. In the following we will extend our discussion to all $\Delta L = 2$ operators up to dimension-9 and a representative fraction of dimension-11 operators. Most of the operators will not contribute directly at tree-level, but at various loop levels. Our goal is to study the contribution of all these operators to the effective $0\nu\beta\beta$ interactions, as depicted in Fig.~\ref{fig:graphs}.

The list of $\Delta L = 2$ SM effective operators we are using is largely based on \cite{deGouvea:2007xp}. In order to convince ourselves we are not missing important operators, we generate $\Delta L=2$ operator patterns using the Hilbert Series method \cite{Lehman:2014jma, Henning:2015alf}. This top-to-bottom approach to effective field theories is able to provide all possible independent operator patterns (i.e. solely the field content of particular operators, not the structure of contractions) that can be formed using a specified particle content respecting given symmetries. This ensures that we capture all possible types of $\Delta L = 2$ operators involving SM fermions and the Higgs. As in \cite{deGouvea:2007xp}, we omit operators involving gauge fields and derivatives and we are interested just in the possible $SU(2)$ contractions, which means we do not take into account all the possible Lorentz and $SU(3)$ structures. In general, we specify the effective operators considering just a single generation of fermions. The resulting sets of operators of dimensions 7, 9 and 11 are listed in Tabs.~\ref{tab:op7}, \ref{tab:op9} and \ref{tab:op11}, respectively, in addition to single Weinberg operator of dimension-5 in Eq.~\ref{eq:WeinbergOp}. Therefore, all the interactions we consider can be schematically summarized in a Lagrangian
\begin{align}
	\mathcal{L} = \mathcal{L}_{SM} + \frac{1}{\Lambda_5}\mathcal{O}_5 + 	
		\sum_{i}\frac{1}{\Lambda^3_{7_i}}\mathcal{O}_7^i + 
		\sum_{i}\frac{1}{\Lambda^5_{9_i}}\mathcal{O}_9^i + 
		\sum_{i}\frac{1}{\Lambda^7_{11_i}}\mathcal{O}_{11}^i,
\end{align}
where $\mathcal{L}_{SM}$ stands for the SM Lagrangian, $\mathcal{O}_5$ is the unique Weinberg operator of dimension-5 suppressed by the corresponding typical energy scale $\Lambda_5$ and $\mathcal{O}_D^i$ are the other SM effective operators of higher dimensions $D = 7,\ 9,\ 11$ suppressed by $\Lambda^{D-4}_{D_i}$. All the effective scales $\Lambda_{D_i}$ subsume any mass scales and couplings of an underlying UV-complete theory. In our calculations we will always consider just a single effective operator in addition to the SM Lagrangian at a time.

\begin{table}
	\centering
	\begin{tabular}{|c|c|c|c|c|c|}
		\hline
		\multicolumn{2}{|c|}{\bf Leptons} &
		\multicolumn{2}{|c|}{\bf Quarks} &
		\multicolumn{2}{|c|}{\bf Higgs Boson}\\
		\hline
		Label & Rep & Label & Rep & Label & Rep  \\
		\hline
		$L = \begin{pmatrix} \nu_L \\ e_L \end{pmatrix}$ & $\left(1,2,-\tfrac{1}{2}\right)$ & $Q = \begin{pmatrix} u_L \\ d_L \end{pmatrix}$ & $\left(3,2,\tfrac{1}{6}\right)$ & \multirow{3}{*}{$H = \begin{pmatrix} h^+ \\ h^0 \end{pmatrix}$} & \multirow{3}{*}{$\left(1,2,\tfrac{1}{2}\right)$} \\ \cline{1-4}
		\multirow{2}{*}{$e^c$} & \multirow{2}{*}{$\left(1,1,1\right)$} & $u^c$ & $\left(\bar{3},1,-\tfrac{2}{3}\right)$ & & \\ \cline{3-4}
		& & $d^c$ & $\left(\bar{3},1,\tfrac{1}{3}\right)$ & & \\ \hline
	\end{tabular}
	\caption{Notation used to label the corresponding representations of the fermions and the Higgs field of the SM given in the form $(SU(3)_c,SU(2)_L,U(1)_Y)$.}
	\label{tab:fields}
\end{table}
The notation used for the particle fields is summarized in Tab.~\ref{tab:fields}, where all the listed fermions are left-handed 2-component Weyl spinors. The right-handed hermitian conjugates are denoted by a bar, e.g. the conjugate partner for the electron singlet $e^c$ reads $\bar{e^c}$. The 2-component Weyl spinors used in this section are related to the 4-component spinors used e.g. in Eqs.~$\eqref{eq:L_longrange}$ and $\eqref{eq:L_shortrange}$ as
\begin{align}
	\nu = \begin{pmatrix} \nu_L \\ \bar{\nu_L} \end{pmatrix},\ \ 
	  e = \begin{pmatrix} e_L \\ \bar{e^c} \end{pmatrix},\ \ 
	  u = \begin{pmatrix} u_L \\ \bar{u^c} \end{pmatrix},\ \ 
	  d = \begin{pmatrix} d_L \\ \bar{d^c} \end{pmatrix},
\end{align}
where $e,u,d$ are Dirac spinors, while $\nu$ is a Majorana spinor.

\subsection{Dimension 5}
At dimension-5, there is a single $\Delta L = 2$ operator (modulo generations). The Hilbert series confirms this and reads
\begin{align}
\label{eq:WeinbergOp}
	\mathcal{H}_{5}^{\Delta L = 2} = L^2 H^2 + \text{h.c.}.
\end{align}
As noted, the Hilbert series method does not provide the actual Lorentz and gauge contractions involved in an operator. It is merely a polynomial in the given fields. The above of course corresponds to the unique Weinberg operator 
\begin{align}
	\mathcal{O}_1 = L^i L^j H^k H^l \epsilon_{ik} \epsilon_{jl},
\end{align} 
generating light Majorana neutrino masses after EW symmetry breaking and thus contributing to $\ovbb$ decay through the effective $\ovbb$ mass of the order $m_\nu = \frac{v^2}{\Lambda_5}$, where $v = 176$~GeV is the SM Higgs vacuum expectation value (VEV).

\subsection{Dimension 7}
At dimension-7, the Hilbert series of $\Delta L = 2$ operator terms reads
\begin{align}
	\mathcal{H}_{7}^{\Delta L = 2} 
	= d \bar{e^{c}} H L \bar{u^{c}} + 2 d^c H L^2 Q 
	+ e^c H L^3 + H^3 \bar H L^2 + H L^2 Q^{\dagger} \bar{u^{c}} + \text{h.c.}.
\end{align}
The integer coefficient in front of the term $d^c H L^2 Q$ indicates the multiplicity of the given pattern, which means there are two independent ways how to contract the fields so that the resulting operator is both Lorentz- and SM-invariant. In this particular case one can write down two independent operators with different $SU(2)_L$ structures - operators $\mathcal{O}_{3a}$ and $\mathcal{O}_{3b}$ in Tab.~\ref{tab:op7}.

\renewcommand*{\arraystretch}{1.2}
\begin{table}[t!]
	\centering
	{\small
		\begin{tabular}{|l|c|c|c|c|}
			\hline
			$\mathcal{O}$ & Operator & $m_\nu$ & LR & $\epsilon_\text{LR}$ \\
			\hline
			$\mathit{1^{H^2}}$ & $L^i L^j H^k H^l \overline{H}^{t}H_{t} \epsilon_{ik} \epsilon_{jl}$ & $\frac{v^2}{\Lambda}f\nba{\frac{v}{\Lambda}}$ & $-$ & $-$ \\
			$2$ & $ L^iL^jL^ke^cH^l\epsilon_{ij}\epsilon_{kl}$ & $\frac{y_e }{16\pi^2}\frac{v^2}{\Lambda}$ & $-$ & $-$ \\
			$3_a$ & $L^i L^j Q^k d^c H^l \epsilon_{ij} \epsilon_{kl}$ & $\frac{y_d g^2}{(16\pi^2)^2}\frac{v^2}{\Lambda}$ & $\frac{v}{\Lambda^3}$ & $\epsilon^{T_{R}}_{T_{R}}$ \\
			$3_b$ & $L^i L^j Q^k d^c H^l \epsilon_{ik} \epsilon_{jl}$ & $\frac{y_d}{16\pi^2}\frac{v^2}{\Lambda}$ & $\frac{v}{\Lambda^3}$ & $\epsilon^{\scriptscriptstyle S+P}_{\scriptscriptstyle S+P}$ \\
			$4_a$ & $L^i L^j \overline{Q}_i \bar{u^c} H^k \epsilon_{jk}$ & $\frac{y_u}{16\pi^2}\frac{v^2}{\Lambda}$ & $\frac{v}{\Lambda^3}$ & $\epsilon^{\scriptscriptstyle S+P}_{\scriptscriptstyle S-P}$ \\
			$4_b^{\dagger}$ & $\mathit{L^i L^j \overline{Q}_k\bar{u^c}H^k \epsilon_{ij}}$ & $\frac{y_u g^2}{(16\pi^2)^2}\frac{v^2}{\Lambda}$ & $\frac{v}{\Lambda^3}$ & $\epsilon^{\scriptscriptstyle S+P}_{\scriptscriptstyle S-P}$ \\
			$8$ & $L^i \bar{e^c} \bar{u^c} d^c H^j \epsilon_{ij}$ & $\frac{y_{e}^{\text{ex}} y_d y_u}{(16\pi^2)^2}\frac{v^2}{\Lambda}$ & $\frac{v}{\Lambda^3}$ & $2\epsilon^{\scriptscriptstyle V+A}_{\scriptscriptstyle V+A}$ \\
			\hline
		\end{tabular}
	}
	\caption{Effective $\Delta L = 2$ SM operators at dimension 7. Dominant contributions to $\ovbb$ decay via the effective neutrino mass ($m_\nu$) and long-range (LR) mechanisms are shown. The $\ovbb$ long-range interaction excited by a particular operator is denoted in column $\epsilon_\text{LR}$. The notation of the contributions is explained in Sec.~\ref{sec:identifydominant}.}
	\label{tab:op7}
\end{table}
The corresponding operators, including $SU(2)$ contractions are shown in Tab.~\ref{tab:op7}. Note that the operator $\mathcal{O}_{1^{H^2}}$ is highlighted in italic; it is simply the Weinberg operator with the singlet combination $H\bar H$ attached. We do not discuss these cases in detail but list them for completeness. Operator $\mathcal{O}_{4_b}$ is marked by a dagger because it is fierz-related to operator $\mathcal{O}_{4_a}$ and as such is not independent. Moreover, it vanishes for a single generation of fermions and, therefore on its own does not contribute to $\ovbb$ decay. On the other hand, it cannot be solely responsible for the observed neutrino oscillations for the same reason. Minimally, there should thus be a misalignment between the operator flavour structure and the charged-current mixing. Under this assertion, $\mathcal{O}_{4_a}$ may still contribute to $\ovbb$ with only a $\mathcal{O}(1)$ suppression which is why we retain it in our results.

\subsection{Dimension 9}
Similarly, at dimension-9, the Hilbert series of $\Delta L = 2$ operator terms reads
\begin{align}
	\mathcal{H}_{9}^{\Delta L=2} &= 
	\nba{d^c}^2 \bar{d^c} L^2 \bar{u^c} + \nba{d^c}^2 {\bar{e^c}}^2 {\bar{u^c}}^2
	+ 2\nba{d^c}^2 \bar{e^c} L Q \bar{u^c}+4 \nba{d^c}^2 L^2 Q^2+d^c e^c \bar{e^c} L^2 \bar{u^c} \nonumber\\
	&+ 2 d^c e^c L^3 Q + d^c \bar{e^c} H^2 \overline{H} L \bar{u^c}+2 d^c \bar{e^c} L \overline{Q} {\bar{u^c}}^2+3 d^c H^2 \overline{H} L^2 Q+d^c L^3 \overline{L} \bar{u^c}+4 d^c L^2 Q \overline{Q} \bar{u^c} \nonumber\\
	&+ d^c L^2 u^c{\bar{u^c}}^2 +\bar{d^c} H^3 L^2 \overline{Q}+\nba{e^c}^2 L^4+e^c H^2 \overline{H} L^3+e^c L^3 \overline{Q} \bar{u^c}+\bar{e^c} H^3 L^2 \overline{L}+\bar{e^c} H^3 L Q \overline{Q} \nonumber\\
	&+ H^4 {\overline{H}}^2 L^2 + H^3 L^2 Q u^c+2 H^2 \overline{H} L^2 \overline{Q} \bar{u^c}+2 L^2 {\overline{Q}}^2 {\bar{u^c}}^2  + \text{h.c.}.
\end{align}
The corresponding effective operators are listed in Tab.~\ref{tab:op9}. For completeness we again list operators `derived' from lower-dimensional ones, such as $\mathcal{O}_{1^{H^4}}$ which is the Weinberg operator with the singlet combination $H\bar H$ attached twice or $\mathcal{O}_{1^{y_e}}$ which is also the Weinberg operator with the singlet combination $\bar L H \bar{e^c}$ attached. Operator $\mathcal{O}_{12_b}$ is marked by an asterisk as it vanishes for a single generation of fermions and as such it does not contribute to $\ovbb$ decay. Like $\mathcal{O}_{4_a}$ discussed above, another source of lepton flavour violation is needed to produce the observed neutrino oscillations, in which case it may contribute which is why keep it in our results. It is also worth noting that the operator $\mathcal{O}_{76}$ does not appear in \cite{deGouvea:2007xp}.

\renewcommand*{\arraystretch}{1.45}
\begin{table}[t!]
\centering
{\scriptsize
\begin{tabular}{|l|c|c|c|c|c|c|}
\hline
$\mathcal{O}$ & Operator & $m_\nu$ & LR & $\epsilon_\text{LR}$ & SR & $\epsilon_\text{SR}$ \\
\hline
$\mathit{1^{H^4}}$ & $L^i L^j H^k H^l \overline{H}^{t}H_{t} \overline{H}^{u}H_{u} \epsilon_{ik} \epsilon_{jl}$  & $\frac{v^2}{\Lambda}f^2\nba{\frac{v}{\Lambda}}$ & $-$ & $-$ & $-$ & $-$ \\
$\mathit{1^{y_e}}$ & $L^i L^j H^k H^l (\overline{L}^tH_t\bar{e^c}) \epsilon_{ik} \epsilon_{jl}$ & $\frac{y_e}{(16\pi^2)^2} \frac{v^2}{\Lambda}$ & $-$ & $-$ & $-$ & $-$ \\
$\mathit{1^{y_d}}$ & $L^i L^j H^k H^l (\overline{Q}^tH_t\bar{d^c}) \epsilon_{ik} \epsilon_{jl}$ & $\frac{y_d}{(16\pi^2)^2} \frac{v^2}{\Lambda}$ & $\frac{y_{d}y_{d|u}^{\text{ex}}}{16\pi^2} \frac{v}{\Lambda^3} f\nba{\frac{v}{\Lambda}}$ & $\epsilon^{\scriptscriptstyle S+P}_{\scriptscriptstyle S\pm P}$ & $-$ & $-$ \\
$\mathit{2^{H^2}}$ & $ L^iL^jL^ke^cH^l \overline{H}^{t}H_{t} \varepsilon_{ij}\varepsilon_{kl}$ & $\frac{y_e }{16\pi^2}\frac{v^2}{\Lambda}f\nba{\frac{v}{\Lambda}}$ & $-$ & $-$ & $-$ & $-$ \\
$\mathit{3_a^{H^2}}$ & $L^i L^j Q^k d^c H^l \overline{H}^{t}H_{t} \epsilon_{ij} \epsilon_{kl}$ & $\frac{y_d g^2}{(16\pi^2)^2}\frac{v^2}{\Lambda}f\nba{\frac{v}{\Lambda}}$ & $\frac{v}{\Lambda^3}f\nba{\frac{v}{\Lambda}}$ & $\epsilon^{T_{R}}_{T_{R}}$  & $-$ & $-$ \\
$\mathit{3_b^{H^2}}$ & $L^i L^j Q^k d^c H^l \overline{H}^{t}H_{t} \epsilon_{ik} \epsilon_{jl}$ & $\frac{y_d}{16\pi^2}\frac{v^2}{\Lambda}f\nba{\frac{v}{\Lambda}}$ & $\frac{v}{\Lambda^3}f\nba{\frac{v}{\Lambda}}$ & $\epsilon^{\scriptscriptstyle S+P}_{\scriptscriptstyle S+P}$  & $-$ & $-$ \\
$\mathit{4_a^{H^2}}$ & $L^i L^j \overline{Q}_i \bar{u^c} H^k \overline{H}^{t}H_{t} \epsilon_{jk}$ & $\frac{y_u}{16\pi^2}\frac{v^2}{\Lambda}f\nba{\frac{v}{\Lambda}}$ & $\frac{v}{\Lambda^3}f\nba{\frac{v}{\Lambda}}$ & $\epsilon^{\scriptscriptstyle S+P}_{\scriptscriptstyle S-P}$ & $-$ & $-$ \\
$5$ & $L^i L^j Q^k d^c H^l H^m \overline{H}_i \epsilon_{jl}\epsilon_{km}$ & $\frac{y_d}{(16\pi^2)^2}\frac{v^2}{\Lambda}$ & $\frac{v}{\Lambda^3} f\nba{\frac{v}{\Lambda}}$ & $\epsilon^{\scriptscriptstyle S+P}_{\scriptscriptstyle S+P}$  & $-$ & $-$ \\
$6$ & $L^i L^j \overline{Q}_k\bar{u^c}H^l H^k \overline{H}_i\epsilon_{jl}$ & $\frac{y_u}{(16\pi^2)^2}\frac{v^2}{\Lambda}$ & $\frac{v}{\Lambda^3} f\nba{\frac{v}{\Lambda}}$ & $\epsilon^{\scriptscriptstyle S+P}_{\scriptscriptstyle S-P}$  & $-$ & $-$ \\
$7$ & $L^iQ^j \bar{e^c}\overline{Q}_kH^k H^l H^m\epsilon_{il} \epsilon_{jm}$ & $\frac{y_{e}^{\text{ex}}g^2}{(16\pi^2)^2}\frac{v^2}{\Lambda}f\nba{\frac{v}{\Lambda}}$ & $\frac{v^3}{\Lambda^5}$ & $2\epsilon^{\scriptscriptstyle V+A}_{\scriptscriptstyle V-A}$ & $-$ & $-$ \\
$\mathit{8^{H^2}}$ & $L^i \bar{e^c} \bar{u^c} d^c H^j \overline{H}^{t}H_{t} \epsilon_{ij}$ & $\frac{y_{e}^{\text{ex}} y_d y_u}{(16\pi^2)^2}\frac{v^2}{\Lambda}f\nba{\frac{v}{\Lambda}}$ & $\frac{v}{\Lambda^3}f\nba{\frac{v}{\Lambda}}$ & $2\epsilon^{\scriptscriptstyle V+A}_{\scriptscriptstyle V+A}$ & $-$ & $-$ \\
$9$ & $L^i L^j L^k e^c L^l e^c \epsilon_{ij}\epsilon_{kl}$ & $\frac{y_e ^2}{(16\pi^2)^2}\frac{v^2}{\Lambda}$ & $-$ & $-$  & $-$ & $-$ \\
$10$  & $\ L^i L^j L^k e^c Q^l d^c \epsilon_{ij}\epsilon_{kl}$ & $\frac{y_e y_d}{(16\pi^2)^2}\frac{v^2}{\Lambda}$ & $\frac{y_e}{16\pi^2} \frac{v}{\Lambda^3}$ & $\epsilon^{\scriptscriptstyle S+P}_{\scriptscriptstyle S+P}$ & $-$ & $-$ \\
$11_a$& $\ L^i L^j Q^k d^c Q^l d^c \epsilon_{ij} \epsilon_{kl}$ & $\frac{y_d^2 g^2}{(16\pi^2)^3}\frac{v^2}{\Lambda}$ & $\frac{y_d}{16\pi^2} \frac{v}{\Lambda^3}$ & $\epsilon^{T_{R}}_{T_{R}}$ & $\frac{g^2 }{16\pi^2}\frac{1}{\Lambda^5}$ & $\epsilon_{1}$ \\
$11_b$& $\ L^i L^j Q^k d^c Q^l d^c \epsilon_{ik}\epsilon_{jl}$ & $\frac{y_d^2}{(16\pi^2)^2}\frac{v^2}{\Lambda}$ & $\frac{y_d}{16\pi^2} \frac{v}{\Lambda^3}$ & $\epsilon^{\scriptscriptstyle S+P}_{\scriptscriptstyle S+P}$ & $\frac{1}{\Lambda^5}$ & $\epsilon_{1}$ \\
$12_a$& $L^iL^j\overline{Q}_i\bar{u^c}\overline{Q_j}\bar{u^c}$ & $\frac{y_u^2}{(16\pi^2)^2}\frac{v^2}{\Lambda}$ & $\frac{y_u}{16\pi^2} \frac{v}{\Lambda^3}$ & $\epsilon^{\scriptscriptstyle S+P}_{\scriptscriptstyle S-P}$  & $\frac{1}{\Lambda^5}$ & $\epsilon_{1}$ \\
$12_b^*$& $L^iL^j\overline{Q}_k\bar{u^c}\overline{Q}_l\bar{u^c}\epsilon_{ij}\epsilon^{kl}$ & $\frac{y_u^2 g^2}{(16\pi^2)^3}\frac{v^2}{\Lambda}$ & $\frac{y_u}{16\pi^2} \frac{v}{\Lambda^3}$ & $\epsilon^{\scriptscriptstyle S+P}_{\scriptscriptstyle S-P}$ & $\frac{g^2 y_d^{\text{ex}} y_u^{\text{ex}} }{(16\pi^2)^2}\frac{1}{\Lambda^5}$ & $\epsilon_{1}$ \\
$13$ & $L^i L^j \overline{Q}_i \bar{u^c}L^l e^c\epsilon_{jl}$ & $\frac{y_e y_u}{(16\pi^2)^2}\frac{v^2}{\Lambda}$ & $\frac{y_e}{16\pi^2} \frac{v}{\Lambda^3}$ & $\epsilon^{\scriptscriptstyle S+P}_{\scriptscriptstyle S-P}$ & $-$ & $-$ \\
$14_a$& $\ L^iL^j\overline{Q}_k\bar{u^c}Q^kd^c\epsilon_{ij}$ & $\frac{y_dy_ug^2}{(16\pi^2)^3}\frac{v^2}{\Lambda}$ & $\frac{y_{u|d}}{16\pi^2} \frac{v}{\Lambda^3}$ & $\epsilon^{T_{R}}_{T_{R}}$ & $\frac{g^2 }{(16\pi^2)^2}\frac{1}{\Lambda^5}$ & $\epsilon_{1}$ \\
$14_b$& $\ L^i L^j \overline{Q}_i \bar{u^c}Q^ld^c\epsilon_{jl}$ & $\frac{y_dy_u}{(16\pi^2)^2}\frac{v^2}{\Lambda}$ & $\frac{y_{u|d}}{16\pi^2} \frac{v}{\Lambda^3}$ & $\epsilon^{\scriptscriptstyle S+P}_{\scriptscriptstyle S\pm P}$ & $\frac{1}{\Lambda^5}$ & $\epsilon_{1}$ \\
$15$ & $L^i L^j L^k d^c \overline{L}_i \bar{u^c}\epsilon_{jk}$ & $\frac{y_dy_ug^2}{(16\pi^2)^3}\frac{v^2}{\Lambda}$ & $\frac{g^2 y_{u|d|e}^{\text{ex}}}{(16\pi^2)^2} \frac{v}{\Lambda^3}$ & $\epsilon^{\scriptscriptstyle S+P}_{\scriptscriptstyle S\pm P}|2\epsilon^{\scriptscriptstyle V+A}_{\scriptscriptstyle V+A}$ & $-$ & $-$ \\
$16$ & $L^i L^j e^c d^c \bar{e^c} \bar{u^c}\epsilon_{ij}$ & $\frac{y_dy_ug^4}{(16\pi^2)^4}\frac{v^2}{\Lambda}$ & $\frac{y_e}{16\pi^2} \frac{v}{\Lambda^3}$ & $2\epsilon^{\scriptscriptstyle V+A}_{\scriptscriptstyle V+A}$ & $-$ & $-$ \\
$17$ & $L^i L^j d^c d^c \bar{d^c} \bar{u^c}\epsilon_{ij}$ & $\frac{y_dy_ug^4}{(16\pi^2)^4}\frac{v^2}{\Lambda}$ & $\frac{g^2 y_{u|d|e}^{\text{ex}}}{(16\pi^2)^2} \frac{v}{\Lambda^3}$ & $\epsilon^{T_{R}}_{T_{R}}|2\epsilon^{\scriptscriptstyle V+A}_{\scriptscriptstyle V+A}$ & $\frac{y_d^{\text{ex}} y_e^{\text{ex}}}{16\pi^2} \frac{1}{\Lambda^5}$ & $2\epsilon_{5}$ \\
$18$ & $L^i L^j d^c u^c \bar{u^c} \bar{u^c}\epsilon_{ij}$ & $\frac{y_dy_ug^4}{(16\pi^2)^4}\frac{v^2}{\Lambda}$ & $\frac{g^2 y_{u|d|e}^{\text{ex}}}{(16\pi^2)^2}\frac{v}{\Lambda^3}$ & $\epsilon^{T_{R}}_{T_{R}}|2\epsilon^{\scriptscriptstyle V+A}_{\scriptscriptstyle V+A}$ & $\frac{y_e^{\text{ex}} y_u^{\text{ex}} }{16\pi^2} \frac{1}{\Lambda^5}$ & $2\epsilon_{5}$ \\
$19$ & $\ L^i Q^j d^c d^c \bar{e^c} \bar{u^c}\epsilon_{ij}$ & $\frac{y_{e}^{\text{ex}}y_d^2y_u}{(16\pi^2)^3}\frac{v^2}{\Lambda}$ & $\frac{y_d}{16\pi^2}\frac{v}{\Lambda^3}$ & $2\epsilon^{\scriptscriptstyle V+A}_{\scriptscriptstyle V+A}$ & $\frac{1}{\Lambda^5}$ & $2\epsilon_{5}$ \\
$20$ & $\ L^i d^c \overline{Q}_i \bar{u^c} \bar{e^c} \bar{u^c}$ &$\frac{y_{e}^{\text{ex}} y_dy_u^2}{(16\pi^2)^3}\frac{v^2}{\Lambda}$ & $\frac{y_u}{16\pi^2}\frac{v}{\Lambda^3}$ & $2\epsilon^{\scriptscriptstyle V+A}_{\scriptscriptstyle V+A}$ & $\frac{1}{\Lambda^5}$ & $2\epsilon_{5}$ \\
$61$ & $L^i L^j H^k H^l L^r e^c \overline{H}_r \epsilon_{ik} \epsilon_{jl}$ & $\frac{y_e }{16\pi^2}\frac{v^2}{\Lambda}f\nba{\frac{v}{\Lambda}}$ & $-$ & $-$  & $-$ & $-$ \\
$66$ & $L^i L^j H^k H^l \epsilon_{ik} Q^r d^c \overline{H}_r\epsilon_{jl}$ & $\frac{y_d}{16\pi^2}\frac{v^2}{\Lambda}f\nba{\frac{v}{\Lambda}}$ & $\frac{1}{16\pi^2} \frac{v}{\Lambda^3}$ & $\epsilon^{\scriptscriptstyle S+P}_{\scriptscriptstyle S+P}$  & $-$ & $-$ \\
$71$ & $L^i L^j H^k H^l Q^r u^c H^s \epsilon_{rs} \epsilon_{ik} \epsilon_{jl}$ & $\frac{y_u}{16\pi^2}\frac{v^2}{\Lambda}f\nba{\frac{v}{\Lambda}}$ & $\frac{y_u y^{\text{ex}}_{d|u}}{16\pi^2} \frac{v}{\Lambda^3} f\nba{\frac{v}{\Lambda}}$ & $\epsilon^{\scriptscriptstyle S+P}_{\scriptscriptstyle S\pm P}$  & $-$ & $-$ \\
$76$ & $\bar{e^c}\bar{e^c}d^cd^c\bar{u^c}\bar{u^c}$ & $\frac{y_{e}^{\text{ex}2}y_d^2y_u^2}{(16\pi^2)^4}\frac{v^2}{\Lambda}$ & $\frac{y_dy_uy_e^{\text{ex}}}{(16\pi^2)^2}\frac{v}{\Lambda^3}$ & $2\epsilon^{\scriptscriptstyle V+A}_{\scriptscriptstyle V+A}$ & $\frac{1}{\Lambda^5}$ & $2\epsilon_{3}^{a}$ \\
\hline
\end{tabular}
}
\caption{
Effective $\Delta L = 2$ SM operators at dimension 9. Dominant contributions to $\ovbb$ decay via the effective neutrino mass ($m_\nu$) as well as long-range (LR) and short-range (SR) mechanisms are shown. The $\ovbb$ long-range and short-range interactions excited by a particular operator are denoted in column $\epsilon_\text{LR}$ and $\epsilon_\text{SR}$. The notation of the contributions is explained in Sec.~\ref{sec:identifydominant}.
}
\label{tab:op9}
\end{table}

\subsection{Dimension 11}
As the number of operators of dimension 11 is quite large and many of them behave in a similar manner we restrict our calculations to the selection shown in Tab.~\ref{tab:op11}. It includes all 11-dimensional operators that trigger $\ovbb$ at tree level but we omit for example operators that do not appreciably contribute to $\ovbb$ through long-range or short-range interactions.

\renewcommand*{\arraystretch}{1.2}
\begin{table}[t!]
	\centering
	{\scriptsize
		\begin{tabular}{|l|c|c|c|c|c|c|}
			\hline
			$\mathcal{O}$ & Operator & $m_\nu$ & LR & $\epsilon_\text{LR}$ & SR & $\epsilon_\text{SR}$ \\
			\hline
			$21_a$& $L^iL^jL^ke^cQ^lu^cH^mH^n\epsilon_{ij}\epsilon_{km}\epsilon_{ln}$ & $\frac{y_e y_u}{(16\pi^2)^2}\frac{v^2}{\Lambda}f\nba{\frac{v}{\Lambda}}$ & $\frac{y_e y_{e}^{\text{ex}} y_u^{\text{ex}}}{(16\pi^2)^2} \frac{v^3}{\Lambda^5}$ & $2\epsilon^{\scriptscriptstyle V+A}_{\scriptscriptstyle V-A}$ & $-$ & $-$ \\
			$21_b$& $L^i L^j L^k e^c Q^l u^c H^m H^n \epsilon_{il} \epsilon_{jm}\epsilon_{kn}$ & $\frac{y_e y_u}{(16\pi^2)^2}\frac{v^2}{\Lambda}f\nba{\frac{v}{\Lambda}}$ & $\frac{y_e y_{e}^{\text{ex}} y_u^{\text{ex}}}{(16\pi^2)^2} \frac{v^3}{\Lambda^5}$ & $2\epsilon^{\scriptscriptstyle V+A}_{\scriptscriptstyle V-A}$ & $-$ & $-$ \\
			$23$ & $L^i L^jL^k e^c \overline{Q}_k\bar{d^c}H^lH^m \epsilon_{il} \epsilon_{jm}$ & $\frac{y_e y_d}{(16\pi^2)^2}\frac{v^2}{\Lambda}f\nba{\frac{v}{\Lambda}}$ & $\frac{y_{d|(e)}^{\text{ex}} y_d^{\text{ex}} y_e}{(16\pi^2)^2}\frac{v}{\Lambda^3} f\nba{\frac{v}{\Lambda}}$ & $\epsilon^{\scriptscriptstyle S+P}_{\scriptscriptstyle S-P}|\epsilon^{\scriptscriptstyle V+A}_{\scriptscriptstyle V+A}$ & $-$ & $-$ \\
			$24_a$& $L^i L^j Q^k d^c Q^l d^c H^m \overline{H}_i\epsilon_{jk} \epsilon_{lm}$ & $\frac{y_d^2}{(16\pi^2)^3}\frac{v^2}{\Lambda}$ & $\frac{y_d}{16\pi^2} \frac{v}{\Lambda^3} f\nba{\frac{v}{\Lambda}}$ & $\epsilon^{\scriptscriptstyle S+P}_{\scriptscriptstyle S+P}$ & $\frac{1}{\Lambda^5}f\nba{\frac{v}{\Lambda}}$ & $\epsilon_{1}$ \\
			$24_b$& $L^i L^j Q^k d^c Q^l d^cH^m \overline{H}_i \epsilon_{jm} \epsilon_{kl}$ & $\frac{y_d^2}{(16\pi^2)^3}\frac{v^2}{\Lambda}$ & $\frac{y_d}{16\pi^2} \frac{v}{\Lambda^3} f\nba{\frac{v}{\Lambda}}$ & $\epsilon^{\scriptscriptstyle S+P}_{\scriptscriptstyle S+P}$ & $\frac{g^2 }{(16 \pi^2)}\frac{1}{\Lambda^5}f\nba{\frac{v}{\Lambda}}$ & $\epsilon_{1}$ \\
			$25$ & $L^i L^j Q^k d^c Q^l u^c H^m H^n \epsilon_{im}\epsilon_{jn} \epsilon_{kl}$ & $\frac{y_dy_u}{(16\pi^2)^2}\frac{v^2}{\Lambda}f\nba{\frac{v}{\Lambda}}$ & $\frac{y_u}{(16\pi^2)^2} \frac{v}{\Lambda^3}$ & $\epsilon^{\scriptscriptstyle S+P}_{\scriptscriptstyle S+P}$ & $\frac{y_u^{\text{ex}2} }{(16 \pi^2)^2}\frac{1}{\Lambda^5}$ & $\epsilon_{1}$ \\
			$26_a$& $L^i L^j Q^k d^c \overline{L}_i \bar{e^c}H^l H^m \epsilon_{jl} \epsilon_{km}$ & $\frac{y_e y_d}{(16\pi^2)^3}\frac{v^2}{\Lambda}$ & $\frac{y_e}{16\pi^2} \frac{v}{\Lambda^3} f\nba{\frac{v}{\Lambda}}$ & $\epsilon^{\scriptscriptstyle S+P}_{\scriptscriptstyle S+P}$ & $-$ & $-$ \\
			$26_b$& $L^i L^j Q^k d^c \overline{L}_k \bar{e^c} H^lH^m\epsilon_{il}\epsilon_{jm}$ & $\frac{y_e y_d}{(16\pi^2)^2}\frac{v^2}{\Lambda}f\nba{\frac{v}{\Lambda}}$ & $\frac{y_e}{(16\pi^2)^2} \frac{v}{\Lambda^3}$ & $\epsilon^{\scriptscriptstyle S+P}_{\scriptscriptstyle S+P}$ & $-$ & $-$ \\
			$27_a$& $L^i L^j Q^k d^c \overline{Q}_i\bar{d^c} H^lH^m \epsilon_{jl} \epsilon_{km}$ & $\frac{g^2}{(16\pi^2)^3}\frac{v^2}{\Lambda}$ & $\frac{y_d}{16\pi^2} \frac{v}{\Lambda^3} f\nba{\frac{v}{\Lambda}}$ & $\epsilon^{\scriptscriptstyle S+P}_{\scriptscriptstyle S+P}$ & $\frac{y_d^{\text{ex}2} }{(16 \pi^2)^2}\frac{1}{\Lambda^5}$ & $\epsilon_{1}$ \\
			$27_b$& $L^i L^j Q^k d^c \overline{Q}_k\bar{d^c} H^l H^m \epsilon_{il} \epsilon_{jm}$ & $\frac{g^2}{(16\pi^2)^3}\frac{v^2}{\Lambda}$ & $\frac{y_d}{(16\pi^2)^2} \frac{v}{\Lambda^3}$ & $\epsilon^{\scriptscriptstyle S+P}_{\scriptscriptstyle S+P}$ & $\frac{y_d^{\text{ex}2} }{(16 \pi^2)^2}\frac{1}{\Lambda^5}$ & $\epsilon_{1}$ \\
			$28_a$& $L^i L^j Q^k d^c \overline{Q}_j \bar{u^c}H^l \overline{H}_i \epsilon_{kl}$ & $\frac{y_dy_u}{(16\pi^2)^3}\frac{v^2}{\Lambda}$ & $\frac{y_u}{16\pi^2} \frac{v}{\Lambda^3} f\nba{\frac{v}{\Lambda}}$ & $\epsilon^{\scriptscriptstyle S+P}_{\scriptscriptstyle S+P}$ & $\frac{v^2}{\Lambda^7}$ & $\epsilon_{1}$ \\
			$28_b$& $L^i L^j Q^k d^c \overline{Q}_k\bar{u^c} H^l \overline{H}_i \epsilon_{jl}$ & $\frac{y_dy_u}{(16\pi^2)^3}\frac{v^2}{\Lambda}$ & $\frac{y_{u|d}}{16\pi^2} \frac{v}{\Lambda^3} f\nba{\frac{v}{\Lambda}}$ & $\epsilon^{\scriptscriptstyle S+P}_{\scriptscriptstyle S\pm P}$ & $\frac{g^2}{(16 \pi^2)}\frac{1}{\Lambda^5}f\nba{\frac{v}{\Lambda}}$ & $\epsilon_{1}$ \\
			$28_c$& $L^i L^j Q^k d^c \overline{Q}_l \bar{u^c} H^l\overline{H}_i\epsilon_{jk}$ & $\frac{y_dy_u}{(16\pi^2)^3}\frac{v^2}{\Lambda}$ & $\frac{y_d}{16\pi^2} \frac{v}{\Lambda^3} f\nba{\frac{v}{\Lambda}}$ & $\epsilon^{\scriptscriptstyle S+P}_{\scriptscriptstyle S-P}$ & $\frac{1}{\Lambda^5}f\nba{\frac{v}{\Lambda}}$ & $\epsilon_{1}$  \\
			$29_a$& $L^i L^j Q^k u^c \overline{Q}_k \bar{u^c}H^l H^m \epsilon_{il} \epsilon_{jm}$ & $\frac{g^2}{(16\pi^2)^3}\frac{v^2}{\Lambda}$ & $\frac{y_u}{(16\pi^2)^2} \frac{v}{\Lambda^3}$ & $\epsilon^{\scriptscriptstyle S+P}_{\scriptscriptstyle S-P}$ & $\frac{y_d^{\text{ex}} y_u^{\text{ex}} }{(16 \pi^2)^2}\frac{1}{\Lambda^5}$ & $\epsilon_{1}$  \\
			$29_b$& $L^i L^j Q^k u^c \overline{Q}_l \bar{u^c} H^l H^m\epsilon_{ik} \epsilon_{jm}$ & $\frac{g^2}{(16\pi^2)^3}\frac{v^2}{\Lambda}$ & $\frac{y_u}{16\pi^2} \frac{v}{\Lambda^3} f\nba{\frac{v}{\Lambda}}$ & $\epsilon^{\scriptscriptstyle S+P}_{\scriptscriptstyle S-P}$ & $\frac{y_d^{\text{ex}} y_u^{\text{ex}} }{(16 \pi^2)^2}\frac{1}{\Lambda^5}$ & $\epsilon_{1}$  \\
			$30_a$& $L^i L^j \overline{L}_i \bar{e^c}\overline{Q}_k\bar{u^c} H^k H^l \epsilon_{jl}$ & $\frac{y_e y_u}{(16\pi^2)^3}\frac{v^2}{\Lambda}$ & $\frac{y_e}{16\pi^2} \frac{v}{\Lambda^3} f\nba{\frac{v}{\Lambda}}$ & $\epsilon^{\scriptscriptstyle S+P}_{\scriptscriptstyle S-P}$ & $-$ & $-$ \\
			$30_b$& $L^i L^j\overline{L}_m \bar{e^c} \overline{Q}_n \bar{u^c} H^k H^l\epsilon_{ik} \epsilon_{jl} \epsilon^{mn}$ & $\frac{y_e y_u}{(16\pi^2)^2}\frac{v^2}{\Lambda}f\nba{\frac{v}{\Lambda}}$ & $\frac{y_e}{(16\pi^2)^2} \frac{v}{\Lambda^3}$ & $\epsilon^{\scriptscriptstyle S+P}_{\scriptscriptstyle S-P}$ & $-$ & $-$ \\
			$31_a$& $L^i L^j \overline{Q}_i\bar{d^c}\overline{Q}_k\bar{u^c} H^k H^l \epsilon_{jl}$ & $\frac{y_dy_u}{(16\pi^2)^2}\frac{v^2}{\Lambda}f\nba{\frac{v}{\Lambda}}$ & $\frac{y_d}{16\pi^2} \frac{v}{\Lambda^3} f\nba{\frac{v}{\Lambda}}$ & $\epsilon^{\scriptscriptstyle S+P}_{\scriptscriptstyle S-P}$ & $\frac{y_d^{\text{ex}2} }{(16 \pi^2)^2}\frac{1}{\Lambda^5}$ & $\epsilon_{1}$ \\
			$31_b$& $L^i L^j \overline{Q}_m\bar{d^c} \overline{Q}_n\bar{u^c}H^k H^l\epsilon_{ik} \epsilon_{jl} \epsilon^{mn}$ & $\frac{y_dy_u}{(16\pi^2)^2}\frac{v^2}{\Lambda}f\nba{\frac{v}{\Lambda}}$ & $\frac{y_d}{(16\pi^2)^2} \frac{v}{\Lambda^3}$ & $\epsilon^{\scriptscriptstyle S+P}_{\scriptscriptstyle S-P}$ & $\frac{y_d^{\text{ex}2} }{(16 \pi^2)^2}\frac{1}{\Lambda^5}$ & $\epsilon_{1}$ \\
			$32_a$& $L^i L^j \overline{Q}_j \bar{u^c}\overline{Q}_k \bar{u^c} H^k \overline{H}_i$ & $\frac{y_u^2}{(16\pi^2)^3}\frac{v^2}{\Lambda}$ & $\frac{y_u}{16\pi^2} \frac{v}{\Lambda^3} f\nba{\frac{v}{\Lambda}}$ & $\epsilon^{\scriptscriptstyle S+P}_{\scriptscriptstyle S-P}$ & $\frac{1}{\Lambda^5}f\nba{\frac{v}{\Lambda}}$ & $\epsilon_{1}$ \\
			$32_b$& $L^i L^j\overline{Q}_m \bar{u^c} \overline{Q}_n \bar{u^c} H^k\overline{H}_i \epsilon_{jk} \epsilon^{mn}$ & $\frac{y_u^2}{(16\pi^2)^3}\frac{v^2}{\Lambda}$ & $\frac{y_u}{16\pi^2} \frac{v}{\Lambda^3} f\nba{\frac{v}{\Lambda}}$ & $\epsilon^{\scriptscriptstyle S+P}_{\scriptscriptstyle S-P}$ & $\frac{y_d^{\text{ex}2} }{(16 \pi^2)^2}\frac{1}{\Lambda^5}$ & $\epsilon_{1}$ \\
			$34$ & $\bar{e^c} \bar{e^c} L^i Q^j e^c d^c H^kH^l \epsilon_{ik} \epsilon_{jl}$ & $\frac{y_{e}^{\text{ex}}y_dg^2}{(16\pi^2)^4}\frac{v^2}{\Lambda}$ & $\frac{g^2 y_{e|u|(d)}^{\text{ex}}}{(16\pi^2)^2} \frac{v}{\Lambda^3} f\nba{\frac{v}{\Lambda}}$ & $\epsilon^{\scriptscriptstyle S+P}_{\scriptscriptstyle S+P}|2\epsilon^{\scriptscriptstyle V+A}_{\scriptscriptstyle V\pm A}$ & $-$ & $-$ \\
			$35$ & $\bar{e^c} \bar{e^c} L^i e^c\overline{Q}_j\bar{u^c} H^jH^k\epsilon_{ik}$ & $\frac{y_{e}^{\text{ex}}y_ug^2}{(16\pi^2)^4}\frac{v^2}{\Lambda}$ & $\frac{g^2 y_{e|d|(u)}^{\text{ex}}}{(16\pi^2)^2}\frac{v}{\Lambda^3} f\nba{\frac{v}{\Lambda}}$ & $\epsilon^{\scriptscriptstyle S+P}_{\scriptscriptstyle S-P}|2\epsilon^{\scriptscriptstyle V+A}_{\scriptscriptstyle V\pm A}$ & $-$ & $-$ \\
			$36$ & $\bar{e^c} \bar{e^c} Q^i d^c Q^j d^c H^k H^l \epsilon_{ik} \epsilon_{jl}$ & $\frac{y_{e}^{\text{ex}2}y_d^2g^2}{(16\pi^2)^5}\frac{v^2}{\Lambda}$ & $\frac{y_d y_e^{\text{ex}} y_{e|u|(d)}^{\text{ex}}}{(16\pi^2)^2} \frac{v}{\Lambda^3} f\nba{\frac{v}{\Lambda}}$ & $\epsilon^{\scriptscriptstyle S+P}_{\scriptscriptstyle S+P}|2\epsilon^{\scriptscriptstyle V+A}_{\scriptscriptstyle V\pm A}$ & $\frac{1}{\Lambda^5}f\nba{\frac{v}{\Lambda}}$ & $\epsilon_{1}$ \\
			$37$ & $\bar{e^c} \bar{e^c} Q^i d^c\overline{Q}_j\bar{u^c} H^j H^k \epsilon_{ik}$ & $\frac{y_{e}^{\text{ex}2}y_dy_ug^2}{(16\pi^2)^5}\frac{v^2}{\Lambda}$ & $\frac{g^2 y_e^{\text{ex}}}{(16\pi^2)^2}\frac{v^3}{\Lambda^5}$ & $2\epsilon^{\scriptscriptstyle V+A}_{\scriptscriptstyle V+A}$ & $\frac{1}{\Lambda^5}f\nba{\frac{v}{\Lambda}}$ & $\epsilon_{1}$ \\
			$38$ & $\bar{e^c} \bar{e^c} \overline{Q}_i\bar{u^c}\overline{Q}_j \bar{u^c} H^i H^j$ & $\frac{y_{e}^{\text{ex}2}y_u^2g^2}{(16\pi^2)^5}\frac{v^2}{\Lambda}$ & $\frac{y_{e|d|(u)}^{\text{ex}} y_e^{\text{ex}} y_u}{(16\pi^2)^2}\frac{v}{\Lambda^3} f\nba{\frac{v}{\Lambda}}$ & $\epsilon^{\scriptscriptstyle S+P}_{\scriptscriptstyle S-P}|2\epsilon^{\scriptscriptstyle V+A}_{\scriptscriptstyle V\pm A}$ & $\frac{1}{\Lambda^5}f\nba{\frac{v}{\Lambda}}$ & $\epsilon_{1}$ \\
			$40_a$& $L^i L^j L^k Q^l \overline{L}_i \overline{Q}_jH^m H^n \epsilon_{km} \epsilon_{ln}$ & $\frac{g^2}{(16\pi^2)^3}\frac{v^2}{\Lambda}$ & $\frac{g^2 y_{d|u|u|e}^{\text{ex}}}{(16\pi^2)^2}\frac{v}{\Lambda^3} f\nba{\frac{v}{\Lambda}}$ & $\epsilon^{\scriptscriptstyle S+P}_{\scriptscriptstyle S\pm P}|2\epsilon^{\scriptscriptstyle V+A}_{\scriptscriptstyle V\pm A}$ & $-$ & $-$ \\
			$43_a$ & $L^i L^j L^k d^c \overline{L}_l\bar{u^c} H^l\overline{H}_i \epsilon_{jk}$ & $\frac{y_dy_ug^2}{(16\pi^2)^4}\frac{v^2}{\Lambda}$ & $\frac{g^2y_{u|d|e}^{\text{ex}}}{(16\pi^2)^2}\frac{v}{\Lambda^3} f\nba{\frac{v}{\Lambda}}$ & $\epsilon^{\scriptscriptstyle S+P}_{\scriptscriptstyle S\pm P}|2\epsilon^{\scriptscriptstyle V+A}_{\scriptscriptstyle V+A}$ & $-$ & $-$ \\
			$44_c$& $L^i L^j Q^k e^c \overline{Q}_l \bar{e^c} H^l H^m\epsilon_{ij}\epsilon_{km}$ & $\frac{g^4}{(16\pi^2)^4}\frac{v^2}{\Lambda}$ & $\frac{y_e}{16\pi^2}\frac{v^3}{\Lambda^5}$ & $\epsilon^{V+A}_{V-A}$ & $-$ & $-$ \\
			$47_a$& $L^i L^j Q^k Q^l \overline{Q}_i\overline{Q}_j H^mH^n \epsilon_{km} \epsilon_{ln}$ & $\frac{g^2}{(16\pi^2)^3}\frac{v^2}{\Lambda}$ & $\frac{g^2 y^{\text{ex}}_{d|u|(e)}}{(16\pi^2)^2}\frac{v}{\Lambda^3} f\nba{\frac{v}{\Lambda}}$ & $\epsilon^{\scriptscriptstyle S+P}_{\scriptscriptstyle S\pm P}|2\epsilon^{\scriptscriptstyle V+A}_{\scriptscriptstyle V-A}$ & $\frac{v^2}{\Lambda^7}$ & $2\epsilon_3^{a}$ \\
			$47_d$& $L^i L^j Q^k Q^l \overline{Q}_i\overline{Q}_m H^m H^n\epsilon_{jk}\epsilon_{ln}$ & $\frac{g^2}{(16\pi^2)^3}\frac{v^2}{\Lambda}$ & $\frac{g^2 y^{\text{ex}}_{d|u|(e)}}{(16\pi^2)^2}\frac{v}{\Lambda^3} f\nba{\frac{v}{\Lambda}}$ & $\epsilon^{\scriptscriptstyle S+P}_{\scriptscriptstyle S\pm P}|2\epsilon^{\scriptscriptstyle V+A}_{\scriptscriptstyle V-A}$ & $\frac{v^2}{\Lambda^7}$ & $2\epsilon_3^{a}$ \\
			$53$ & $L^i L^j d^c d^c \bar{u^c} \bar{u^c}\overline{H}_i \overline{H}_j$ & $\frac{y_d^2y_u^2g^2}{(16\pi^2)^5}\frac{v^2}{\Lambda}$ & $\frac{y_d y_u y^{\text{ex}}_{u|d|e}}{(16\pi^2)^2}\frac{v}{\Lambda^3} f\nba{\frac{v}{\Lambda}}$ & $\epsilon^{\scriptscriptstyle S+P}_{\scriptscriptstyle S\pm P}|2\epsilon^{\scriptscriptstyle V+A}_{\scriptscriptstyle V+A}$ & $\frac{v^2}{\Lambda^7}$ & $2\epsilon_3^{a}$ \\
			$54_a$& $L^i Q^j Q^k d^c \overline{Q}_i \bar{e^c}H^l H^m \epsilon_{jl} \epsilon_{km}$ & $\frac{y_{e}^{\text{ex}}y_dg^2}{(16\pi^2)^4}\frac{v^2}{\Lambda}$ & $\frac{g^2 y^{\text{ex}}_{e|(d)|u}}{(16\pi^2)^2} \frac{v}{\Lambda^3} f\nba{\frac{v}{\Lambda}}$ & $\epsilon^{\scriptscriptstyle S+P}_{\scriptscriptstyle S+P}|2\epsilon^{\scriptscriptstyle V+A}_{\scriptscriptstyle V\mp A}$ & $-$ & $-$ \\
			$54_d$& $L^i Q^j Q^k d^c \overline{Q}_l \bar{e^c} H^lH^m \epsilon_{ij} \epsilon_{km}$ & $\frac{y_{e}^{\text{ex}}y_dg^2}{(16\pi^2)^4}\frac{v^2}{\Lambda}$ & $\frac{y_d}{16\pi^2}\frac{v^3}{\Lambda^5}$ & $\epsilon^{\scriptscriptstyle V+A}_{\scriptscriptstyle V-A}$ & $\frac{v^2}{\Lambda^7}$ & $2\epsilon_5$ \\
			$55_a$& $L^i Q^j \overline{Q}_i \overline{Q}_k\bar{e^c} \bar{u^c} H^k H^l \epsilon_{jl}$ & $\frac{y_{e}^{\text{ex}}y_ug^2}{(16\pi^2)^4}\frac{v^2}{\Lambda}$ & $\frac{y_u}{16\pi^2}\frac{v^3}{\Lambda^5}$ & $\epsilon^{\scriptscriptstyle V+A}_{\scriptscriptstyle V-A}$ & $\frac{v^2}{\Lambda^7}$ & $2\epsilon_5$ \\
			$59$ & $L^i Q^j d^c d^c \bar{e^c} \bar{u^c}H^k \overline{H}_i \epsilon_{jk}$ & $\frac{y_{e}^{\text{ex}}y_d^2y_u}{(16\pi^2)^4}\frac{v^2}{\Lambda}$ & $\frac{y_d}{(16\pi^2)^2}\frac{v}{\Lambda^3}$ & $\epsilon^{\scriptscriptstyle V+A}_{\scriptscriptstyle V+A}$ & $\frac{1}{\Lambda^5}f\nba{\frac{v}{\Lambda}}$ & $2\epsilon_5$ \\
			$60$ & $L^i d^c \overline{Q}_j \bar{u^c}\bar{e^c}\bar{u^c} H^j \overline{H}_i$ & $\frac{y_{e}^{\text{ex}}y_dy_u^2}{(16\pi^2)^4}\frac{v^2}{\Lambda}$ & $\frac{y_u}{(16\pi^2)^2}\frac{v}{\Lambda^3}$ & $\epsilon^{\scriptscriptstyle V+A}_{\scriptscriptstyle V+A}$ & $\frac{1}{\Lambda^5}f\nba{\frac{v}{\Lambda}}$ & $2\epsilon_5$ \\
			\hline
		\end{tabular}
	}
	\caption{As Tab.~\ref{tab:op9} but showing selected effective $\Delta L = 2$ SM operators at dimension 11.}
	\label{tab:op11}
\end{table}

%% file: ovbb-contributions.tex
\section{Contributions to Neutrinoless Double Beta Decay}
\label{sec:0vbbcontribs}

To determine the contributions of the operators listed in the previous section to $\ovbb$ decay, we start by identifying those which trigger this rare nuclear process at tree level. Clearly, the Weinberg operator $\mathcal{O}_1$ is such an operator through the effective neutrino mass, cf. Fig.~\ref{fig:graphs}~(a). At higher dimensions, the following operators trigger $\ovbb$ directly after EW symmetry breaking,
\begin{align}
\label{eq:treelevelops}
	\text{Dimension-7\phantom{0}: }
		&\mathcal{O}_{3a},\ \mathcal{O}_{3b},\ \mathcal{O}_{4a},\ \mathcal{O}_{8}; \\
	\text{Dimension-9\phantom{0}: }
		&\mathcal{O}_{5},\ \mathcal{O}_{6},\ \mathcal{O}_{7},\
		\mathcal{O}_{11b},\ \mathcal{O}_{12a},\ \mathcal{O}_{14b},\
		\mathcal{O}_{19},\ \mathcal{O}_{20},\ \mathcal{O}_{76}; \\
	\text{Dimension-11: }
		&\mathcal{O}_{24a},\ \mathcal{O}_{28a},\ \mathcal{O}_{28c},\ \mathcal{O}_{32a},\ 
		\mathcal{O}_{36},\ \mathcal{O}_{37},\ \mathcal{O}_{47a},\ 
		\mathcal{O}_{47d}, \nonumber\\
		&\mathcal{O}_{53},\ \mathcal{O}_{54a},\ \mathcal{O}_{54d},\ 
		\mathcal{O}_{55a},\ \mathcal{O}_{59},\ \mathcal{O}_{60}.
\end{align}
They contribute as in Fig.~\ref{fig:graphs}~(b), (c) and (d), respectively, except the dim-9 operators $\mathcal{O}_{5}$, $\mathcal{O}_{6}$ and $\mathcal{O}_{7}$ which trigger long-range interactions at tree level after all three Higgs fields acquire their vacuum expectation value. In order to estimate the contribution of a single $D$-dimensional operator to $\ovbb$ decay, we consider radiative corrections to all other LNV operators of the same and lower dimension. This implies a huge number of possibilities in reducing a single operator such that we utilize an algorithm as outlined below.

\subsection{$SU(2)$ Decomposition and Effective $\ovbb$ Couplings}
\label{sec:SU2decomposition}
To understand how the $\Delta L=2$ SM effective operators contribute at low energy to $\ovbb$ decay, we first decompose them into the $SU(2)$ components. As a result, for each operator we obtain $2^{d/2}$ components, where $d$ is the number of $SU(2)$ doublets present in the given operator. For example, the operator $\mathcal{O}_{3a}$ splits into 4 different $SU(2)$ components,
\begin{align} \label{eq:decomposition}
	\mathcal{O}_{3a} 
	= L^i L^j Q^k d^c H^l \epsilon_{ij} \epsilon_{kl} 
	= \nu_Le_L u_Lh^0 d^c 
	- e_L\nu_L u_Lh^0 d^c 
	- \nu_Le_L d_Lh^+ d^c 
	+ e_L\nu_L d_Lh^+ d^c.
\end{align}
The $\Delta L=2$ SM effective operators contributing to $\ovbb$ decay at tree level must correspond in the broken phase to one of the terms in the effective $\ovbb$ decay Lagrangians in Eqs.~\eqref{eq:L_longrange} and \eqref{eq:L_shortrange}, which e.g. means that the $SU(2)_L$-components in Eq.~\eqref{eq:decomposition} including $h^0$ can be (after EW symmetry breaking) mapped to one of the terms in Eq.~\eqref{eq:L_longrange}. Contributions to $\ovbb$ decay triggered by other operators contributing to $\ovbb$ decay at loop level are determined by finding their relation to the tree-level-contributing ones, which will be achieved by employing the algorithmic approach described later on. The effective couplings $\epsilon$ appearing in Eqs.~\eqref{eq:L_longrange} and \eqref{eq:L_shortrange} can be restricted by current limits on the $\ovbb$ decay half-life. For this we relate the broken-phase contributions to those triggered by the SM effective operators and obtain thus bounds on the new physics scales $\Lambda$ suppressing the SM effective operators. To identify these relations among unbroken-phase and broken-phase contributions we proceed as follows. 

\subsubsection{6D Long-Range Contributions}
Let us first focus on $\Delta L=2$ SM effective operators that trigger (after EW symmetry breaking) the long-range contributions to $\ovbb$ decay at tree level. Using the above list, applying the $SU(2)$ decomposition, breaking the phase and checking the non-vanishing components we conclude there are in total 7 such operators. Each of them corresponds in the broken phase to one of four different 6-dimensional low-energy $\ovbb$ decay operators (all formed by 4 relevant fermions - $u$, $d$, $e$ and $\nu$). This correspondence can be summarized as follows:
\begin{align}
	\mathcal{O}_{3a}, \mathcal{O}_{3b}, \mathcal{O}_{5} 
	&\to \bar{e_L} \bar{\nu_L} \bar{u_L} \bar{d^c}, \label{eq:condition1} \\
	\mathcal{O}_{4a}, \mathcal{O}_{6} 
	&\to \bar{e_L} \bar{\nu_L} u^c d_L, \label{eq:condition2} \\
	\mathcal{O}_{7} 
	&\to \bar{u_L} \bar{\nu_L} e^c d_L, \label{eq:condition3} \\
	\mathcal{O}_{8} 
	&\to \bar{d^c} \bar{\nu_L} e^c u^c. \label{eq:condition4}
\end{align}
Operators $3a$, $3b$, $4a$ and $8$ are of dimension 7, while operators $5$,  $6$ and $7$ are 9-dimensional. Therefore, the 7D operators will contribute to $\ovbb$ decay with a single power of the EW vev $v$, while the contributions triggered by 9D operators will be proportional to $v^3$. As such, the 9D operators are relevant just in cases for which their contribution is comparable with the leading-order contribution generated by a competing 7-dimensional operator\footnote{If there is such an operator; as is apparent, there is no 7-dimensional SM-invariant operator corresponding to Eq.~\eqref{eq:condition3}. This can be understood from considerations of reasonable UV completions of this vertex. Therefore, for this particular operator the leading contribution will always be proportional to $v^3$.}. Moreover, one can imagine contributions proportional to $v^3$ coming from the 7-dimensional operators multiplied by decoupled invariant $H\bar{H}$, forming a compound effective operator of dimension-9 as listed in Tab.~\ref{tab:op9}. In some cases, i.e. when the relevant 7D operator contributes at second or higher loop level, one might need to take into account even $v^5$-dependent contributions produced by a 9D operator times $H\bar{H}$, a contribution corresponding to a compound effective operator of dimension 11. Therefore, as an example, the total contribution to the 6D $\ovbb$ decay operator in Eq.~\eqref{eq:condition1} for the case when the leading contributions coming from operators $3a$ and $3b$ are suppressed by two loops can be generated after EW symmetry breaking from
\begin{align} \label{eq:contribsum}
	\mathcal{L}_{7+9+11} &= \frac{1}{(16\pi^2)^2}\frac{\mathcal{O}_{3a}}{\Lambda^3} 
	+ \frac{1}{(16\pi^2)^2}\frac{\mathcal{O}_{3b}}{\Lambda^3} 
	+ \frac{1}{16\pi^2}\frac{\mathcal{O}_{5}}{\Lambda^5} \nonumber\\
   &+ \frac{1}{16\pi^2}\frac{\mathcal{O}_{3a}(H\ol{H})}{\Lambda^5} 
    + \frac{1}{16\pi^2}\frac{\mathcal{O}_{3b}(H\ol{H})}{\Lambda^5} 
    + \frac{\mathcal{O}_{5}(H\ol{H})}{\Lambda^7}.
\end{align}

We now relate the 7 different SM effective operators that trigger the long-range $\ovbb$ decay at tree level (on the left-hand side of Eqs.~(\ref{eq:condition1} - \ref{eq:condition4})) to the low-energy $\ovbb$ decay Lagrangian \eqref{eq:L_longrange}. Taking the four 6D operators on the right-hand sides of Eqs.~(\ref{eq:condition1} - \ref{eq:condition4}) and rewriting them using the four-spinor notation we get
\begin{align}
	\bar{e_L} \bar{\nu_L} \bar{u_L} \bar{d^c} &\leftrightarrow \ol{e}\left(1+\gamma_5\right)\nu\ \ol{u}\left(1+\gamma_5\right)d, \\
	\bar{e_L} \bar{\nu_L} u^c d_L &\leftrightarrow \ol{e}\left(1+\gamma_5\right)\nu\ \ol{u}\left(1-\gamma_5\right)d, \\
	\bar{u_L} \bar{\nu_L} e^c d_L &\leftrightarrow \ol{u}\left(1+\gamma_5\right)\nu\ \ol{e}\left(1-\gamma_5\right)d, \label{eq:4Dcondition3} \\
	\bar{d^c} \bar{\nu_L} e^c u^c &\leftrightarrow \ol{d}\left(1+\gamma_5\right)\nu\ \ol{e}\left(1-\gamma_5\right)u, \label{eq:4Dcondition4}
\end{align}
where the right-hand sides of Eqs.~\eqref{eq:4Dcondition3} and \eqref{eq:4Dcondition4} can be further Fierz-transformed to the conventional field ordering prescribed by Eq.~\eqref{eq:L_longrange}. Thus, we obtain
\begin{align}
	\ol{e}\left(1-\gamma_5\right)d\ \ol{u}\left(1+\gamma_5\right)\nu &= 
	\frac{1}{2} \ol{e}\gamma^{\mu}\left(1+\gamma_5\right)\nu\ \ol{u}\gamma_{\mu}\left(1-\gamma_5\right)d, \\
	\ol{e}\left(1-\gamma_5\right)u\ \ol{d}\left(1+\gamma_5\right)\nu &= 
	\frac{1}{2} \ol{e}\gamma^{\mu}\left(1+\gamma_5\right)\nu\ \ol{u}\gamma_{\mu}\left(1+\gamma_5\right)d.
\end{align}
Taking these equalities into account, one can relate the scale of the SM invariant operators on the left-hand sides of Eqs.~(\ref{eq:condition1} - \ref{eq:condition4}) to the effective couplings $\epsilon_{\alpha}^{\beta}$ as follows:
\begin{align}
	\mathcal{O}_{3a}: \frac{\lambda_{BSM}^3v}{\Lambda_{3a}^3} = \frac{G_F\epsilon_{T_{R}}^{T_{R}}}{\sqrt{2}},
\end{align}
\vspace{-0.4cm}
\begin{align}
\begin{rcasesdef} 
	&\mathcal{O}_{3b}: \frac{\lambda_{BSM}^3 v}{\Lambda_{3b}^3} \\ 
	&\mathcal{O}_{5}: \frac{\lambda_{BSM}^5 v^3}{\Lambda_{5}^5}
\end{rcasesdef} &= \frac{G_F\epsilon_{S+P}^{S+P}}{\sqrt{2}}, &
\begin{rcasesdef} 
	&\mathcal{O}_{4a}: \frac{\lambda_{BSM}^3 v}{\Lambda_{4a}^3} \\ 
	&\mathcal{O}_{6}: \frac{\lambda_{BSM}^5 v^3}{\Lambda_{6}^5} 
\end{rcasesdef} &= \frac{G_F\epsilon_{S-P}^{S+P}}{\sqrt{2}}, \label{eq:corresp2} \\
	\mathcal{O}_{7}: \frac{\lambda_{BSM}^5 v^3}{\Lambda_{7}^5} &= 2\frac{G_F\epsilon_{V-A}^{V+A}}{\sqrt{2}}, & 
	\mathcal{O}_{8}: \frac{\lambda_{BSM}^3 v}{\Lambda_{8}^3} &= 2\frac{G_F\epsilon_{V+A}^{V+A}}{\sqrt{2}}. 
\label{eq:corresp4}
\end{align}
The contributions of the SM effective operators on the left-hand side of the above equations are simply given by a certain power of $v$ (determined by the number of Higgs fields present in the particular operator) divided by $\Lambda^{D-4}$, where $\Lambda$ is the typical energy scale of the effective operator of dimension $D$. The powers of a generic coupling $\lambda_{BSM}$ illustrates the scaling generated by a typical tree-level UV completion of the given operator and in the following calculations we simply set $\lambda_{BSM} = 1$.

\subsubsection{9D Short-Range Contributions}
Analogously, the correspondence between $\Delta L=2$ SM effective operators contributing to $\ovbb$ decay at tree level and the terms in the short-range part of the low-energy $\ovbb$ decay effective Lagrangian can be determined. While the term in Eq.~\eqref{eq:L_shortrange} proportional to $\epsilon_1$ is formed by scalar Lorentz bilinears by definition, the other terms proportional to the remaining four epsilons include $\gamma$-matrices and must be Fierz transformed. For the terms with couplings $\epsilon_3$ and $\epsilon_5$, vector currents are present; thus, the same type of Fierz transformation as the one used for the 6D operators can be employed, which results in an extra factor of $2$ in front of the $\epsilon$-coupling. For the $\epsilon_2$-terms of Eq.~\eqref{eq:L_shortrange} one could consider the following identity
\begin{align}
	&\sqb{\overline{u}_a \tfrac{i}{2}[\gamma^{\mu},\gamma^{\nu}](1 \pm \gamma_5)d^a} \sqb{\overline{u}_b\tfrac{i}{2}[\gamma_{\mu},\gamma_{\nu}](1 \pm \gamma_5)d^b} \nn \\
	&= -2\sqb{\overline{u}_a (1 \pm \gamma_5)d^b} [\overline{d}^a (1 \pm \gamma_5)u_b ] - \sqb{\overline{u}_a (1 \pm \gamma_5)d^a} \sqb{\overline{u}_b (1 \pm \gamma_5)d^b}.
\end{align}
The two terms on the right-hand side of the above equation cannot be combined into a single one, as they differ by their $SU(3)_c$ structures represented by indices $a,\,b$. Therefore, to excite the effective coupling $\epsilon_2$, one needs to combine these two different contractions. However, since we always assume just a single $\Delta L=2$ effective operator at a time, we will not discuss this kind of contribution. The situation is similar for the terms of Eq.~\eqref{eq:L_shortrange} proportional to $\epsilon_4$. If we for simplicity assume just one specific combination of chiral currents (i.e. one specific term proportional to $\epsilon_4$), we can employ the following Fierz transformation
\begin{align}
	&\overline{u}_a \gamma_{\mu}(1 + \gamma_5)d^a \sqb{\overline{u}_b \tfrac{i}{2}[\gamma^{\mu},\gamma^{\nu}](1 - \gamma_5)d^b} \overline{e} \gamma_{\nu}(1 + \gamma_5)e^c \nn \\
 	&= -2i \overline{u}_a (1 - \gamma_5)d^b \sqb{\overline{u}_b (1 - \gamma_5)e^c} \overline{e} (1 + \gamma_5)d^a \nn \\
 	&\hspace{0.45cm}- i \overline{u}_a (1 - \gamma_5)e^c \sqb{\overline{u}_b (1 - \gamma_5)d^b} \overline{e} (1 + \gamma_5)d^a
\end{align}
and the conclusion is the same as for the $\epsilon_2$ terms.

We will map all operators to the effective couplings $\epsilon_1$, $\epsilon_3$ or $\epsilon_5$. Omission of $\epsilon_2$ and $\epsilon_4$ does not cause any problems, as there are no $\Delta L = 2$ effective operators contributing uniquely to these terms. Every operator that can be mapped to a term of Eq.~\eqref{eq:L_shortrange} proportional to $\epsilon_2$ and $\epsilon_4$ excites also effective couplings $\epsilon_1$ and $\epsilon_5$, respectively. Consequently, effective scales of every operator listed in Eq.~\eqref{eq:treelevelops} can be related to one of the three effective couplings $\epsilon_1$, $\epsilon_3$ and $\epsilon_5$ as follows:
\begin{align}
\centering
\begin{rcasesdef} 
	\mathcal{O}_{11b}, \mathcal{O}_{12a}, \mathcal{O}_{14b}:\ & \frac{\lambda_{BSM}^4}{\Lambda_{i}^5} \\ 
	\mathcal{O}_{24a}, \mathcal{O}_{28a}, \mathcal{O}_{28c}, \mathcal{O}_{32a}, \mathcal{O}_{36}, \mathcal{O}_{37}, \mathcal{O}_{38}:\ & \frac{\lambda_{BSM}^6v^2}{\Lambda_{i}^7} \\
\end{rcasesdef} &= \frac{G_F^2\epsilon_{1}}{2m_p}, \\ 
\begin{rcasesdef}
	\mathcal{O}_{47a}, \mathcal{O}_{47d}, \mathcal{O}_{53}:\ & 
	\frac{\lambda_{BSM}^6v^2}{\Lambda_i^7} \\
	\mathcal{O}_{76}:\ &
	\frac{\lambda_{BSM}^4}{\Lambda_i^5}
\end{rcasesdef} &= 2\frac{G_F^2\epsilon_3^{LL,RR}}{2m_p}, \\
\begin{rcasesdef} 
	\mathcal{O}_{19}, \mathcal{O}_{20}:\ & \frac{\lambda_{BSM}^4}{\Lambda_{i}^5} \, \\ 
	\mathcal{O}_{54a}, \mathcal{O}_{54d}, \mathcal{O}_{55a}, \mathcal{O}_{59}, \mathcal{O}_{60}:\ & \frac{\lambda_{BSM}^6v^2}{\Lambda_{i}^7} \\
\end{rcasesdef} &= 2\frac{G_F^2\epsilon_{5}}{2m_p}.
\end{align}

\subsection{Estimation of Wilson Coefficients}
\label{sec:loopclosing}

Given a higher-dimensional operator, we want to estimate the value of same- and lower-dimensional Wilson coefficients induced by radiative effects. To this end, we consider for each operator all loop diagrams that could lead to the corresponding operators. As these contributions would be absorbed by the Wilson coefficient of the contribution during the matching procedure, we are able to estimate the size of contributions by next-to-leading order diagrams. Important to note at this stage is that this implies a certain assumption about the underlying UV theory. Estimating the Wilson coefficient by loops of heavy particles implies an underlying `natural' theory, meaning that the contributions are determined by the heavy mass of new physics. A prominent example where this approach does not work is e.g. in the Higgs sector leading to the famous `hierarchy problem'. This means that our estimation would for instance fail when having a UV model that features certain cancellations between loop contributions (as e.g. in supersymmetry). However, this guiding principle turned out to be successful many times in the history of particle physics, e.g. for hadronic resonances or the charm quark \cite{Giudice:2008bi}, such that our approach seems to be justified as long as one is aware of its limitations. Given these assumptions, we can estimate the Wilson coefficients as follows:
\begin{table}[t!]
	\centering
	\begin{tabular}{|cc|c||c|}
		\hline
		$f_L$ & $\bar{f}_L$ &  $Z$ & $g /(16 \pi^2)$\\
		$f_L$ & $\bar{f}^\prime_L$ &  $W^-$ & $g/(16 \pi^2)$\\
		$\bar{f}^c$ & $\bar{f}^\prime_L$ & $H^-$ & $y_f/(16 \pi^2)$\\
		$\bar{f}^c$ & $\bar{f}_L$ &  $h^0$ & $y_f/(16 \pi^2)$\\
		\hline
		$Z$ & $\bar{f}_L$ &  $\bar{f}_L$ & $g /(16 \pi^2)$\\
		$W^-$ & $\bar{f}_L$ &  $\bar{f}^\prime_L$ & $g/(16 \pi^2)$\\
		$H^-$ & ${f}^c$ &  $\bar{f}^\prime_L$ & $y_f/(16 \pi^2)$\\
		$h^0$ & $\bar{f}_L$ &  ${f}^c$ & $y_f/(16 \pi^2)$\\
		\hline
		$\langle h \rangle$ & $\bar{f}_L$ &  ${f}^c$ & $v y_f/(16 \pi^2)$\\
		\hline
		$ h^0 | W^- | H^-$ & $\bar{h}^0 | W^+ | H^+$ &  $-$ & $1/(16 \pi^2)$\\
		\hline
	\end{tabular}
	\caption{Effective Feynman rules contracting the fields in the first two columns via a loop and radiating the field listed in the third column. The coefficient in the last column indicates the corresponding contribution. Here, $f_L$ denotes a left-handed and $f^c$ a right-handed Weyl-spinor with $f$ indicating a fermion according to Tab.~\ref{tab:fields}.}
	\label{tab:rules1}
\end{table}
\begin{table}[t]
	\centering
	\begin{tabular}{|ccc||c|}
		\hline
		$f_L$ & $f^c$ &  $h^0$ & $y_f/(16 \pi^2)^2$ \\
		$f^c$ & $f^\prime_L$ &  $H^+$ & $y_f/(16 \pi^2)^2$ \\
		\hline
		$f_L$ & $\bar{f}^\prime_L$ &  $W^+$ & $g/(16 \pi^2)^2$ \\
		$f_L$ & $\bar{f}_L$ &  $Z$ & $g/(16 \pi^2)^2$ \\
		\hline
		$h$ & $Z\hphantom{^+} | W^+$ &$Z\hphantom{^+} | W^-$ & $v g^2/(16 \pi^2)^2$ \\
		$Z$ & $H^+ | W^+$ &  $W^- | H^-$ & $2v g^2/(16 \pi^2)^2$ \\
		\hline
	\end{tabular}
	\caption{Effective Feynman rules contracting the fields in the first three columns leading to a double loop. The coefficient in the last column indicates the corresponding contribution. Here, $f_L$ denotes a left-handed and $f^c$ a right-handed Weyl-spinor with $f$ indicating a fermion according to Tab.~\ref{tab:fields}.}
	\label{tab:rules2}
\end{table}

\begin{enumerate}
	\item First, we specify the SU(2) component of the SM invariant operator (A) that we want to study, as well as the SU(2) component of the operator (B) we want it to reduce to. This step will be performed for each SU(2) component of each operator (A) to all lower or same dimensional operators (B).
	
	\item To match lower-dimensional operators (B), we apply all possible Feynman rules that reduce the dimension and contain SM fermions, gauge fields or the Higgs boson. Tab.~\ref{tab:rules1} lists the Feynman rules leading to one-loop contractions, whereas Tab.~\ref{tab:rules2} shows all considered two-loop contractions that are necessary in order to obtain all possible contributions to $0\nu\beta\beta$ (see Fig.~\ref{fig:graphs}). In the following we want to discuss the included Feynman rules in detail.
	
	\item We introduce for every closed loop a factor $1/(16\pi^2)$. We regulate further the loops via a momentum cut-off $\Lambda$. Each loop integral contributes with the power of $\Lambda$ equal to the mass dimension of the integral. We want to clearly state that this treatment would be wrong for loop corrections within a pure EFT approach. If we would be interested in the estimation of loop corrections within the EFT, we should use dimensional regularisation involving only SM masses. As mentioned before, we do not consider a pure EFT approach for the estimation of the lower dimensional Wilson coefficients and approximate the size by the assumption of a `natural' UV theory. Thus we are allowed to introduce heavy masses (or a cut-off scale $\Lambda$) that would be integrated out in a pure EFT following the Appelquist-Carazzone decoupling theorem \cite{Appelquist:1974tg}.

	\begin{figure}[h!]
		\centering
		\includegraphics[clip,width=0.28\textwidth]{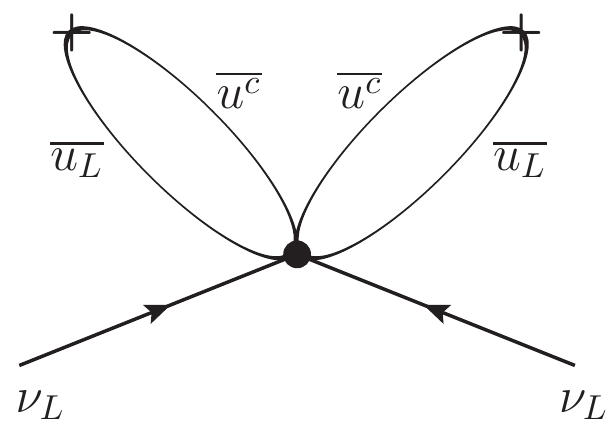}
		\includegraphics[clip,width=0.28\textwidth]{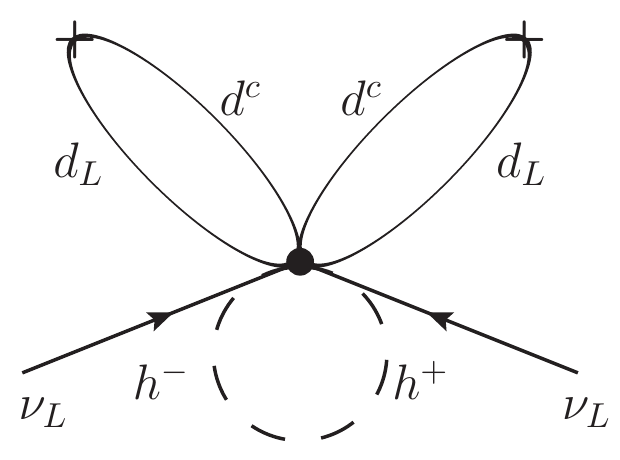}
		\includegraphics[clip,width=0.28\textwidth]{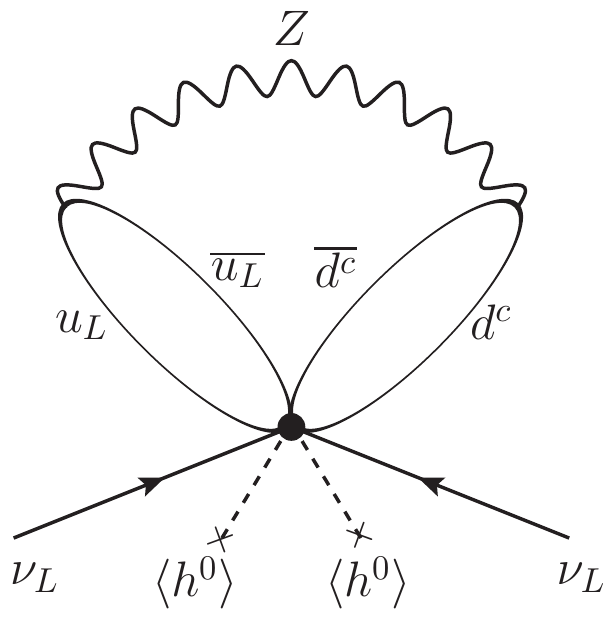}
		\caption{Diagrams showing the reduction of $\mathcal{O}_{12a}$ (left), $\mathcal{O}_{24a}$ (centre) and $\mathcal{O}_{27a}$ (right) to the neutrino mass.}
		\label{fig:rules1a} 
	\end{figure}
	\item Fig.~\ref{fig:rules1a} shows a few examples of the rules listed in Tab.~\ref{tab:rules1}. The left diagram shows the dominant contribution to the Weinberg operator resulting from the 9-dimensional operator $\mathcal{O}_{12a} = L^i L^j \bar{Q}_i \bar{u^c} \bar{Q}_j\bar{u^c}$ that we considered in \cite{Deppisch:2013jxa}. The centre diagram shows the contribution from the 11-dimensional operator $\mathcal{O}_{24a} = L^i L^j \bar{Q}_i \bar{u^c} \bar{Q}_j \bar{u^c}$ discussed in \cite{Deppisch:2013jxa} as well. The right diagram shows the 11-dimensional operator $\mathcal{O}_{27a} = L^i L^j Q^k d^c \bar{Q}_i \bar{d^c} H^l H^m \epsilon_{jl} \epsilon_{km}$. The algorithm merges the fermions into a neutral Higgs boson in the left and centre diagram, which will acquire a VEV in a second iteration. In the right diagram, however, two fermions are merged into a vector boson. In a further iteration the algorithm will merge the vector bosons, forming a one-loop contribution. A similar process happens to the charged Higgs bosons that are merged in a loop in the centre diagram. The resulting contributions to the neutrino mass are
	\begin{align}
		m_\nu^{12a} = \frac{y_u^2 v^2}{(16\pi^2)^2 \Lambda}, \quad 
		m_\nu^{24a} = \frac{y_d^2 v^2}{(16\pi^2)^3 \Lambda}, \quad 
		m_\nu^{27a} = \frac{g^2 v^2}{(16\pi^2)^3 \Lambda}.
	\end{align}
	In \emph{each} iteration, \emph{all} $n$-rules (where $n$ indicates the number of legs on that the rule has effect) are tested if applicable to \emph{all} combinations of $n$ legs of the diagram with in total $m$ legs.
	
	\item As mentioned in the previous subsection, a number of operators can be reduced to an effective operator contributing at tree-level multiplied by an $\bar{H}H$ pair, which can be either contracted to the vacuum giving a factor of $\frac{v^2}{\Lambda^2}$, or closed into a Higgs loop producing an extra factor of $\frac{1}{16\pi^2}$. As a result, we get a contribution to $\ovbb$ decay proportional to $f\nba{\frac{v}{\Lambda}} \equiv \nba{\frac{1}{16\pi^2}+\frac{v^2}{\Lambda^2}}$. Moreover, some of the operators can be reduced to the desired low-energy $\ovbb$ operator in several different ways through different tree-level contributing $\Delta L=2$ SM effective operators, which results in a sum of contributions as illustrated in Eq.~\eqref{eq:contribsum}. However, in the present approach only qualitatively different contributions to $\ovbb$ decay are studied and compared; therefore, this multiplicity is neglected in the obtained results.

	\begin{figure}[h!]
		\centering
		\includegraphics[clip,width=0.3\textwidth]{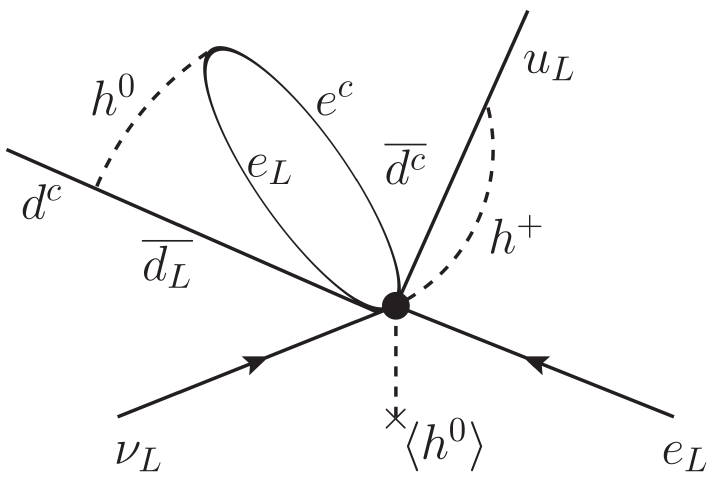}
		\includegraphics[clip,width=0.3\textwidth]{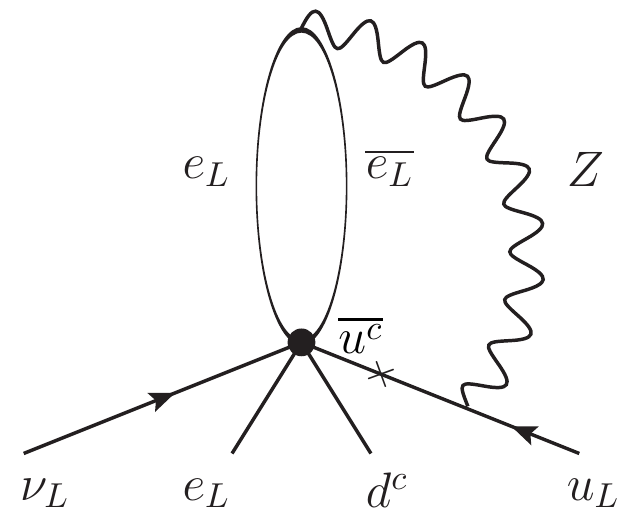}
		\caption{Diagrams showing the reduction of $\mathcal{O}_{23}$ (left) and $\mathcal{O}_{15}$ (right) to a long-range $\ovbb$ decay contribution.}
		\label{fig:rules1b} 
	\end{figure}
	\item Taking the example of reducing a 9-dimensional and an 11-dimensional operator to a long-range contribution, Fig.~\ref{fig:rules1b} demonstrates another set of necessary Feynman rules, listed in Tab.~\ref{tab:rules1}. The left diagram shows a long-range contribution induced by $\mathcal{O}_{23} = L^i L^j L^k e^c \bar{Q}_k \bar{d^c} H^l H^m \epsilon_{il} \epsilon_{jm}$ and the right diagram for $\mathcal{O}_{15} = L^i L^j L^k d^c \bar{L}_i \bar{u^c} \epsilon_{jk}$. In order to recover the long-range contribution for operator $\mathcal{O}_{23}$, we have to merge one fermion with a Higgs to create the fermion needed. As shown in Fig.~\ref{fig:rules1b}~(right), a similar Feynman rule is needed for $\mathcal{O}_{15}$. We have to merge a vector boson with a fermion in order to generate the correct quark. In this case, however, we have to additionally insert a Higgs VEV beforehand. In order to provide a converging algorithm, the latter rule is added explicitly to the algorithm as indicated in Tab.~\ref{tab:rules1}. We allow for up to three additional Higgs VEV insertions per diagram. This leads to the following contributions
	\begin{align}
		\frac{G_F \epsilon_{7}^{15}}{\sqrt{2}} 
		= \frac{y_u^\text{ex} g^2 v}{(16\pi^2)^2 \Lambda^3}, \quad 
		\frac{G_F \epsilon_{7}^{23}}{\sqrt{2}} 
		= \frac{y_e (y_d^\text{ex})^2 v^2}{(16\pi^2)^3 \Lambda^3}\,,
	\end{align}
	with $\epsilon_{7}^{15} = \epsilon_{S+P}^{S+P}$ and $\epsilon_{7}^{23} = \epsilon_{S+P}^{S-P}$.
	
	\item Fig.~\ref{fig:rules1b} and the above contributions demonstrate another important feature. We need to distinguish between external and internal Yukawa couplings. While the flavour of Yukawa couplings associated with external fields is fixed to the first generation in order to generate $\ovbb$ decay, internal Yukawa couplings can be summed over all flavours (we assume by default internal Yukawa couplings and indicate external Yukawa couplings, fixed to the first fermion generation, as $y^\text{ex}$). This can make a significant impact in the discussion of the results, as we will see later in more detail.

	\begin{figure}[t!]
		\centering
		\includegraphics[clip,width=0.3\textwidth]{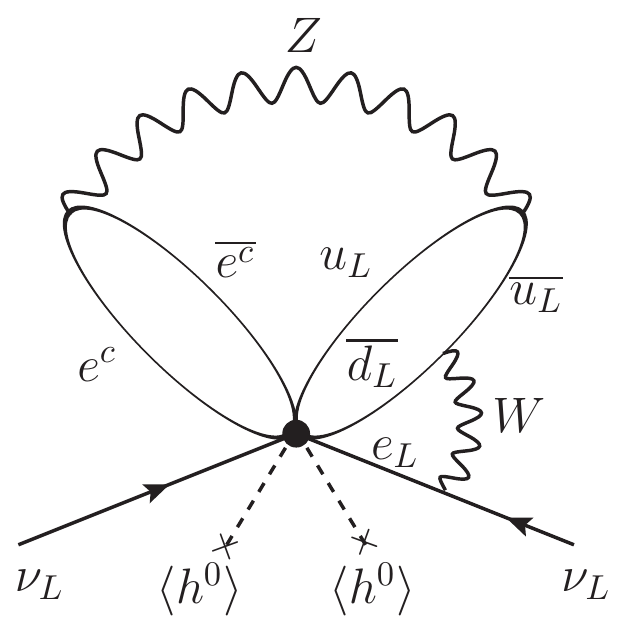}
		\caption{Diagram showing the reduction of $\mathcal{O}_{44c}$ to the neutrino mass.}
		\label{fig:rules2} 
	\end{figure}
	\item Fig.~\ref{fig:rules2} shows three features that require a special treatment. Due to the given $SU(2)$ structure $L^i L^j \epsilon_{ij}$ in $\mathcal{O}_{44c} = L^i L^j Q^k e^c \bar{Q}_l \bar{e^c} H^l H^m \epsilon_{ij} \epsilon_{km}$, the charged electron has to be converted into a neutrino. This requires the introduction of `$t$-channel' rules. In this example, an additional $W$ boson loop is used. As the exchange does not reduce the dimension of the operator and would naively applied lead to an infinite number of iterations, it is treated separately. The corresponding cases to be considered are depicted in Fig.~\ref{fig:prerun}. Very rare rules are treated manually. 
	\begin{figure}[t!]
		\centering
		\includegraphics[clip,width=0.24\textwidth]{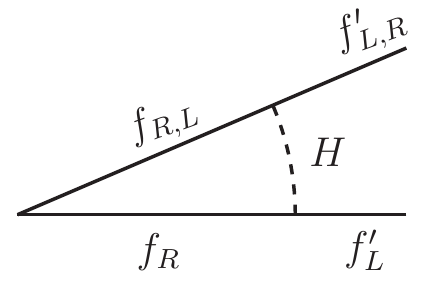}
		\includegraphics[clip,width=0.24\textwidth]{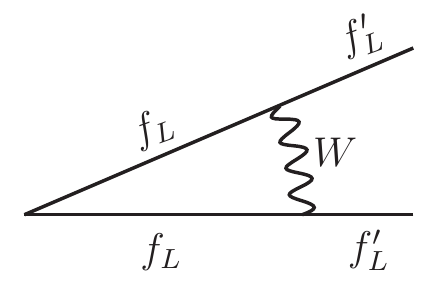}
		\includegraphics[clip,width=0.24\textwidth]{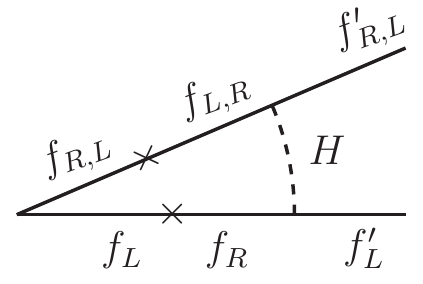}
		\includegraphics[clip,width=0.24\textwidth]{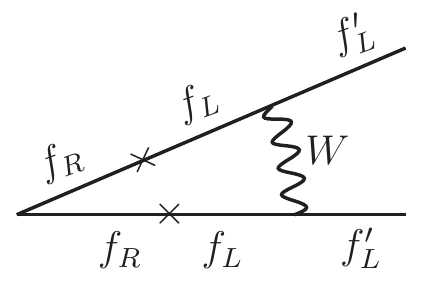}\\
		\includegraphics[clip,width=0.24\textwidth]{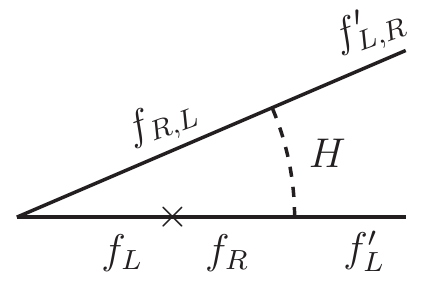}
		\includegraphics[clip,width=0.24\textwidth]{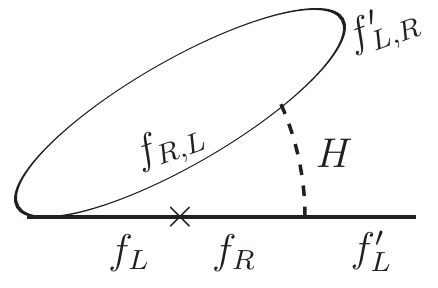}
		\includegraphics[clip,width=0.24\textwidth]{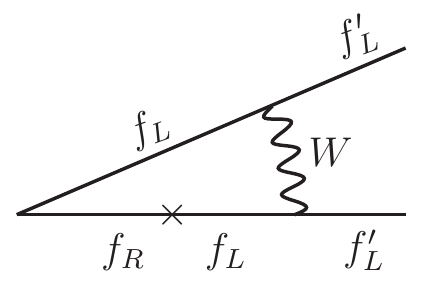}
		\includegraphics[clip,width=0.24\textwidth]{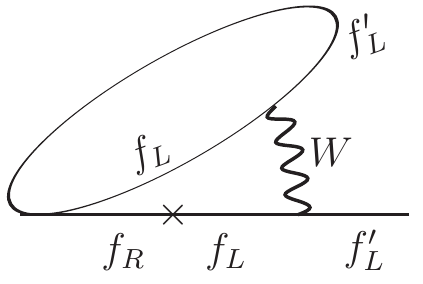}
		\caption{Additional `$t$-channel' cases that have to be separately considered in order to obtain a converging algorithm.}
		\label{fig:prerun} 
	\end{figure}

	\item Another feature needs to be carefully considered when applying Feynman rules within an automatized algorithm.  In order to obtain a valid contribution, diagrams should contain only loops with an even number of fermions (those without mass insertion). Otherwise diagrams will be proportional to the external neutrino momentum and/or will vanish. Thus the algorithm tracks during all iterations the number of fermions within an isolated loop contribution. In case a final diagram would consist out of loops with odd numbers of fermions, a neutral gauge boson exchange has to be added. This has consequences on the previously described step. While in Fig.~\ref{fig:prerun} (upper row) an even number of fermions is generated in the $t$-channel, the lower row features an odd number. Thus the latter allows for closing a loop directly with another fitting leg of the initial operator without the radiation of any other particle. As closing a loop without radiating a fermion would generate a vanishing contribution, it is not included in the default algorithm.
	
	\item A final feature, demonstrated in Fig.~\ref{fig:rules2}, is the necessity of Feynman rules in the algorithm which close a multi-loop that results from merging three legs. After applying the $t$-channel rules of step (8) and closing two fermion legs by radiating a $Z$ boson, the remaining two fermions of the operator have to be connected with the $Z$ boson. The corresponding rules are listed in Tab.~\ref{tab:rules2}. Taking the example of $\mathcal{O}_{44c}$, we obtain finally the following contribution: 
	\begin{align}
		m_\nu^{44c} = \frac{g^4 v^2}{(16\pi^2)^4 \Lambda}.
	\end{align}

	\item Given the described algorithm, each SU(2) decomposed operator is reduced to every possible lower-dimensional or equal-dimensional operator via all possible loop diagrams that can contribute to $\ovbb$ decay. In a final step the most dominant contribution is identified as described in the following.
\end{enumerate}
	
\begin{figure}[t!]
\centering
\includegraphics[clip,width=0.28\textwidth]{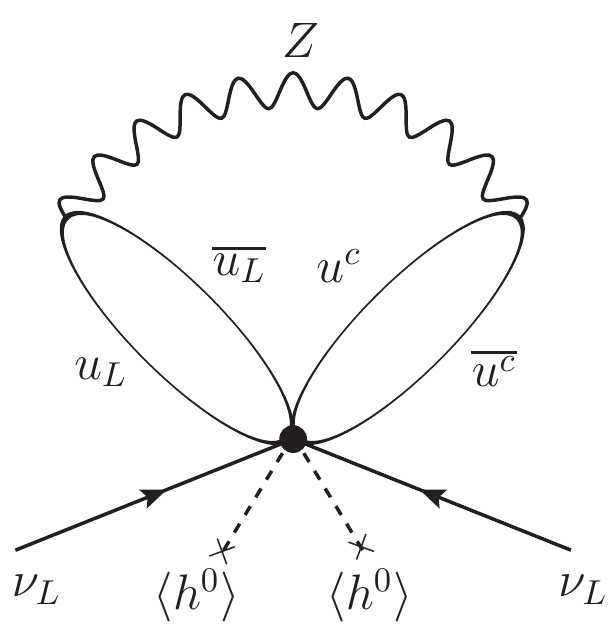}
\includegraphics[clip,width=0.28\textwidth]{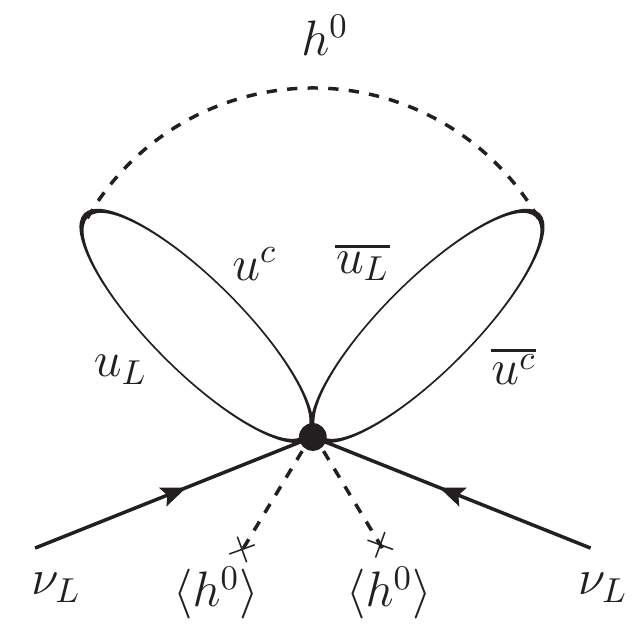}
\includegraphics[clip,width=0.28\textwidth]{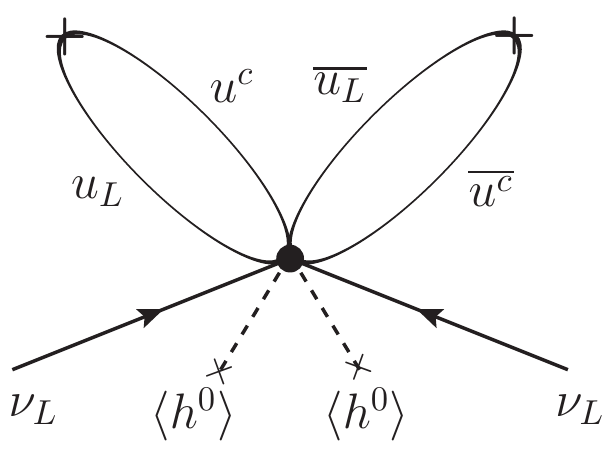}
\caption{Dominant mass contribution of $\mathcal{O}_{29}$ for first generation internal Yukawa couplings includes the gauge coupling (left). For third generation internal Yukawa couplings the corresponding contributions with Yukawa couplings are dominant (centre and right).}
\label{fig:op29}
\end{figure}

We can compare our results for the mass mechanism with the result found in \cite{deGouvea:2007xp}. However, for certain operators discrepancies will occur (and are expected). Comparing $\mathcal{O}_{29a}$ with our results in Tab.~\ref{tab:op11}, we obtain
\begin{align}
	m^{29a,1st}_{\nu} &=\frac{g^2}{(16\pi^2)^3}\frac{v^2}{\Lambda},
\label{eq:29a1} 
\end{align}
while in \cite{deGouvea:2007xp} the contribution 
\begin{align}
	m^{29a,3rd}_{\nu} 
	&=\frac{y_u^2}{(16\pi^2)^2}\frac{v^2}{\Lambda}f\nba{\frac{v}{\Lambda}},
\label{eq:29a3}
\end{align}
is found. This results from the determination of the dominant contribution. While Eq.~\ref{eq:29a3} is dominant for third generation internal Yukawa couplings, Eq.~\eqref{eq:29a1} is the dominant one for only first generation Yukawa couplings, cf. Fig.~\eqref{fig:op29}. As we store all contributions that result from our algorithm, we are able to vary the generation of Yukawa couplings while using the correct contribution of Eq.~\eqref{eq:29a3} and Eq.~\eqref{eq:29a1}. The same behaviour occurs for contributions of operators $\mathcal{O}_{74a}$, $\mathcal{O}_{74b}$ and $\mathcal{O}_{75}$.

\subsection{Determination of the Operator Scale}
\label{sec:identifydominant}
\begin{table}[b!] 
	\centering
	{
	\begin{tabular}{c|c|c|c|c|c|c}
		\hline
		$y_e$ & $y_{\tau}$ & $y_d$ & $y_b$ & $y_u$ & $y_t$ & $g$ \\ \hline
		$2.1\times 10^{-6}$ & $0.01$ & $2.0\times 10^{-5}$ & 
		$2.4\times 10^{-2}$ & $9.4\times 10^{-6}$ & $0.99$ & $0.46$ \\
		\hline
	\end{tabular}}
	\caption{Values of couplings in the numerical evaluation of the contributions to $\ovbb$ decay.}
	\label{tab:numvals}
\end{table}
Having all the possible contributions of each operator, we need to identify the most dominant ones. To do so we just simply compare their numerical values considering the values of involved quantities summarized in Tab.~\ref{tab:numvals}. For the purpose of comparison we consider just first generation of fermions. As for the operator scale $\Lambda$, we assume its value to be $\Lambda = 2186\ \mathrm{GeV}$, when doing the comparison. At this value, $\frac{1}{16\pi^2} = \frac{v^2}{\Lambda^2}$ is satisfied. In other words, we assume such a value of $\Lambda$ for which the suppression at one-loop corresponds to the suppression by a factor $\frac{v^2}{\Lambda^2}$. The reason is that for a number of operators one gets a contribution involving an $H\bar{H}$ pair, which in the broken phase gives such a factor. However, it can be at the same time closed into a trivial Higgs loop, thus contributing a factor of $\frac{1}{16\pi^2}$. As we want to keep both these contributions, we assume for convenience the above stated value of $\Lambda$ ensuring their equality.

We determine the dominant contribution to the light neutrino mass, the dominant long-range contribution and the dominant short-range contribution separately. It is important to note that just these are then used for further calculations, which is, of course, just an approximation. We do not sum over all possible contributions, and thus we do not take into account the multiplicity (given e.g. by several different radiative reductions of a specific operator to the same tree-level-contributing $\ovbb$ decay operator) of any of the contributions either. For each operator, the dominant contributions are listed in Tabs.~\ref{tab:op7}, \ref{tab:op9} and \ref{tab:op11}. The corresponding $\epsilon$ couplings excited by a particular $\ovbb$ decay contribution are also shown therein. The relation to each coupling $\epsilon$ is made assuming a specific (convenient) scalar Lorentz contraction (i.e. a simple Lorentz contraction not involving any $\gamma$ matrices) of the given operator. Considering other Lorentz structures of the initial operator different $\epsilon$ couplings could be also triggered.

In the $\ovbb$ decay contributions listed in Tabs.~\ref{tab:op7}, \ref{tab:op9} and \ref{tab:op11} the short-hand notation $f\nba{\frac{v}{\Lambda}}\equiv \nba{\frac{1}{16\pi^2}+\frac{v^2}{\Lambda^2}}$ is used and the numbers in brackets in front of several operators (in the `Operator' column) denotes the number of possible Lorentz or $SU(3)_c$ contractions allowed for the given operator, appearing just when more than one such possibility exists. Certain operators give several qualitatively different (but numerically similar) long-range contributions that differ just by the involved Yukawa couplings. The possible flavours are shown in a single subscript separated by vertical lines. The respective couplings $\epsilon$ excited by given operators are also presented. For multiple long-range contributions we show multiple epsilons ordered accordingly with Yukawa coupling labels. Moreover, in some cases one of the Yukawa flavour indices is shown in brackets, which means that the contribution proportional to that particular type of Yukawa coupling includes only the factor $\frac{v^2}{\Lambda^2}$ and not the loop factor $\frac{1}{16\pi^2}$ in $f\nba{\frac{v}{\Lambda}}$. For example, the long-range contribution of operator $34$ reading $\frac{y_{e|u|(d)}^{\text{ex}}}{(16\pi^2)^2 g^2} \frac{v}{\Lambda^3} f\nba{\frac{v}{\Lambda}}$ and the corresponding excited $\epsilon$ couplings $\epsilon^{\scriptscriptstyle S+P}_{\scriptscriptstyle S+P}|2\epsilon^{\scriptscriptstyle V+A}_{\scriptscriptstyle V\pm A}$ in fact stand for 3 individual contributions $\frac{y_{e}^{\text{ex}}}{(16\pi^2)^2 g^2} \frac{v}{\Lambda^3} f\nba{\frac{v}{\Lambda}}$, $\frac{y_{u}^{\text{ex}}}{(16\pi^2)^2 g^2} \frac{v}{\Lambda^3} f\nba{\frac{v}{\Lambda}}$ and $\frac{y_{d}^{\text{ex}}}{(16\pi^2)^2 g^2} \frac{v^3}{\Lambda^5}$ exciting $\epsilon$-couplings $\epsilon^{\scriptscriptstyle S+P}_{\scriptscriptstyle S+P}$, $\epsilon^{\scriptscriptstyle V+A}_{\scriptscriptstyle V+A}$ and $\epsilon^{\scriptscriptstyle V+A}_{\scriptscriptstyle V-A}$, respectively. Whenever a dash instead of a contribution is shown, such a reduction of a particular operator is not possible in our approach and to be able to obtain such a contribution one would have to consider what we call an `s-channel' exchange rule, which is described in Section~\ref{sec8:schannel} in more detail.

Although all these contributions are presented, just the dominant one is used as input in further calculations. For a given value of the $\ovbb$ decay half-life, e.g. assuming a hypothetical observation at a value of $T_{1/2}^\text{Xe} = 10^{27}$~y, it is then easy to determine the corresponding operator scale $\Lambda$ that will be the basis to calculate the washout of lepton number in the early Universe. In practice, we collect all contributions for a given operator, fix the SM gauge and Yukawa couplings according to Tab.~\ref{tab:numvals}, possibly selecting between first and third generation values for internal Yukawa couplings. We then express the inverse $\ovbb$ half-life as a function of the operator scale $\Lambda$, defined as the maximum among all contributions. We thus neglect any enhancement from two or more contributions of similar size but also potential interference effects. The former will have little impact on the derived operator scale; for the latter we would like to stress that we always assume (a currently hypothetical) observation of $\ovbb$ decay. If two or more contributions partially cancel each other, it would in fact strengthen our argument as a given $\ovbb$ measurement will correspond to a stronger washout. Finally, assuming such a measurement, specifically $T_{1/2}^\text{Xe} = 10^{27}$~y, we determine the corresponding operator scale from the dominant contribution. As the inverse half-life is $\propto \Lambda^{4-D}$, the dominant contribution corresponds to the highest scale of the operator for a given half-life (if the scale were lower, the dominant contribution would induce a more rapid $\ovbb$ decay).

%% file: washout.tex
\section{Lepton Number Washout}
\label{sec:washout}

In this section, we study the washout effect of a pre-existing net lepton asymmetry from the aforementioned operators. For simplicity, we assume only one LNV operator being active at a time. We will use the classic Boltzmann equation formalism to calculate the net lepton number density in the early expanding universe. Generically, the Boltzmann equation for a particle species $N$ reads\footnote{See, for instance, Refs.~\cite{Griest:1990kh, Edsjo:1997bg, Giudice:2003jh} for more detailed discussions on the Boltzmann equation formalism.} 
\begin{align}
\label{eq:BoltzmannY}
	z H n_\gamma \frac{d\eta_N}{dz} 
	= -\sum_{a,i,j,\cdots} [Na \cdots\leftrightarrow ij\cdots],
\end{align}
where $\eta_N$ is the number density of $N$ normalized to the photon density, $\eta_N \equiv n_N/n_\gamma$, and
\begin{align}
	[Na\cdots\leftrightarrow ij\cdots] 
	&= \frac{n_N n_a \cdots}{n_N^{\rm eq}n_a^{\rm eq}\cdots} 
	\gamma^\text{eq}(Na \cdots \to ij\cdots)                   \nonumber\\
	&- \frac{n_i n_j}{n_i^{\rm eq}n_j^\text{eq}\cdots}
    \gamma^\text{eq}\left(ij \cdots \to Na \cdots \right).
\end{align}
The space-time scattering density in thermal equilibrium, $\gamma^{\mathrm{eq}}$, with $n$ initial and $m$ final particles is defined as
\begin{align}
\label{eq:gammageneral}
	\gamma^\text{eq}(N a\cdots \to ij\cdots) 
	&= \int \frac{\mathrm{d}^3 p_N}{2 E_N (2\pi)^3} e^{-\frac{E_N}{T}} 
	\times \prod\limits_{a=1}^{n-1} 
	\left[\int \frac{\mathrm{d}^3 p_a}{2 E_a (2\pi)^3} e^{-\frac{E_a}{T}} \right]  
	\nonumber\\
	&\times \prod\limits_{i=1}^m 
	\left[\int \frac{\mathrm{d}^3 p_i}{2 E_i (2\pi)^3}\right] 
    \times (2\pi)^4\delta^4 \left(p_N + \sum_{a=1}^{n-1} p_a - \sum_{i=1}^m p_i\right) |M|^2, 
\end{align}
with $|M|^2$ being the squared amplitude of the process summed over initial and final spins. As shown in \cite{Giudice:2003jh}, by inserting unity into $\gamma^\text{eq}$,
\begin{align}
	1 &= \int \mathrm{d}^4 P \delta^4\left(P - p_N - \sum_{a=1}^{n-1} p_a\right)
	     \nonumber\\
  	  &= \int \frac{1}{2} \sqrt{P_0^2 - s}\ 
  	  \delta^4\left(P - p_N - \sum_{a=1}^{n-1} p_a\right)\ 
  	  \mathrm{d}s\ \mathrm{d}P_0\ \mathrm{d}\Omega,
\end{align}
where $s = P_0^2 - |\vec{P}|^2$ and $\Omega$ is the two-dimensional solid angle of the three-momentum $\vec{P}$, the scattering density can be expressed as
\begin{align}
	\gamma^\text{eq}(N a\cdots \to ij\cdots) 
	&= \frac{1}{2}\frac{1}{(2 \pi)^4}\int \mathrm{d}s \int \mathrm{d}\Omega 
	\int \mathrm{d}P_0 \sqrt{\frac{P_0^2}{s} - 1}\ \sqrt{s}\ e^{-P_0/T} \nonumber\\
	&\times \int \frac{\mathrm{d}^3 p_N}{2 E_N (2\pi)^3}
	\times \prod\limits_{a=1}^{n-1} 
	\left[\int \frac{\mathrm{d}^3 p_a}{2 E_a (2\pi)^3} \right] 
	(2 \pi)^4 \delta^4\left(P - p_N - \sum_{a=1}^{n-1} p_a\right) \nonumber\\
	&\times \prod\limits_{i=1}^m 
	\left[\int \frac{\mathrm{d}^3 p_i}{2 E_i (2\pi)^3} \right]
	(2 \pi)^4 \delta^4 \left(P - \sum_{i=1}^m p_i\right) \times |M|^2 \nonumber\\can 
	&= \frac{1}{2 (2 \pi)^4} \int \mathrm{d}s \sqrt{s}  \int \mathrm{d}P_0 \sqrt{P_0^2/s - 1} e^{-P_0/T}
	\int \mathrm{d}\Omega \mathrm{d} PS^n \mathrm{d} PS^m \times |M|^2,
\end{align}
in which $\int \mathrm{d} PS^n~(\int\mathrm{d} PS^m)$ is the  initial~(final) state phase space integral.

Assuming that $|M|^2$ does not depend on the relative motion of particles with respect to the thermal plasma, one obtains after integrating over $P_0$ and $\Omega$, 
\begin{align}
\label{eq:thermal_rate}
	\gamma^\text{eq}(N a\cdots \to ij\cdots) 
	= \frac{1}{(2 \pi)^3} \int \mathrm{d}s\ \sqrt{s} 
	K_1\left(\frac{\sqrt{s}}{T}\right)
	\mathrm{d} PS^n \mathrm{d} PS^m \times |M|^2,
\end{align}
with $K_n$ being the modified Bessel function of the second kind and $\int \mathrm{d} \Omega = 4 \pi$. 

We now consider the process to be mediated by an effective contact interaction involving all $N = n+m$ particles. In the case that all particles involved are scalars, $|M|^2$ is then simply proportional to $1/\Lambda^{2(N-4)}$, where $\Lambda$ is the cut-off scale of the corresponding effective operator. Since $|M|^2$ has no dependence on the phase space integral variables, $\gamma^\text{eq}$ can be easily computed in this case,
\begin{align}
\label{eq:gammascalar}
	\gamma^\text{eq} = \frac{1}{2^2 (2\pi)^{2N-3}} 
	\times \frac{\Gamma(N-2) \Gamma(N-3)}{\Gamma(n) \Gamma(n-1)
	\Gamma(N-n) \Gamma(N-n-1)} \times \frac{T^{2N-4}}{\Lambda^{2N-8}},
\end{align}
where we have used the phase space integration
\begin{align}
\label{eq:ph-sp}
	PS^n = \int \mathrm{d} PS^n 
	= \frac{1}{2(4\pi)^{2n-3}} \frac{s^{n-2}}{\Gamma(n)\Gamma(n-1)},
\end{align}
applicable in the limit where all particles are massless, $\sqrt{s} \gg m_i~(i = 1,\cdots,N)$.

In the case where fermions are involved, the matrix element of the process will depend in general on their energies. For each fermion, the squared amplitude receives an additional factor of $E/\Lambda$ compared to the scalar-only case simply based on naive dimensional analysis, where $E$ has a dimension of energy and is determined by the details of interaction kinematics. For interactions with a large number of particles involved, integration of $|M|^2$ over the phase space becomes complex in the presence of fermions. To obtain a reasonable approximation for integration, we apply two simple schemes where we replace each fermions energy $E$ i) by the centre-of-mass energy $\sqrt{s}$ and ii) by the average energy $\sqrt{s}/n$~($\sqrt{s}/m$) for an initial~(final) state fermion, respectively. In both cases, integration over the phase space proceeds as in the scalar case. Clearly, the above are only approximations. By comparing the results with exact calculations for a few select operators, we have found that the geometric mean of the above schemes approximates well the scattering rate.

To be more concrete, assuming there exist $n_f$ fermions within the $n$-particle
initial state and $m_f$ fermions within the $m$-particle final state\footnote{The dimension of the effective operator, $D$, is correlated with the number of particles it contains -- $N + N_f/2 = D$.}, the first method leads to $|M_1|^2 = \frac{\sqrt{s}^{N_f/2}}{\Lambda^{N-4+N_f/2}}$ ($N_f \equiv n_f + m_f$) while the second one results in $|M_2|^2 = \frac{(\sqrt{s}/n)^{n_f/2}(\sqrt{s}/m)^{m_f/2}}{\Lambda^{N - 4 + N_f/2}}$. As a result, one obtains
\begin{align}
\label{eq:gammaapprox}
	\gamma^\text{eq}_{1(2)}(N a\cdots \to ij\cdots) 
	&= \frac{2^{N_f-2}}{(2\pi)^{2N-3}} \times \overline{c}_{1(2)} \nonumber\\ 
	&\times \frac{\Gamma(N+N_f/2-3)\Gamma(N+N_f/2-2)}{\Gamma(n)\Gamma(n-1) \Gamma(N-n)\Gamma(N-n-1)} 
	\times \frac{T^{2N + N_f - 4}}{\Lambda^{2N + N_f -8}},
\end{align}
with
\begin{align}
	\overline{c}_1 = 1 \qquad \mathrm{and} \qquad 
	\overline{c}_2 = \frac{1}{n^{n_f} (N-n)^{N_f - n_f}},
\end{align}
using the two schemes.

Furthermore, it is necessary to include a symmetry factor to account for identical particles in the initial and final state due to the phase space integral, and also take into account the number of different ways for creation and annihilation, given any identical particles. In addition, given an operator, there exist physically distinctive lepton number washout processes by interchanging particles in the initial and final states. One thus needs to sum up all contributions from each of the permutations. The final result for the thermal rate $\gamma^\text{eq}$ is estimated as
\begin{align}
	\gamma^\text{eq} 
	= \sqrt{ \left( \Sigma \gamma^\text{eq}_{1} \right) 
	  \times \left( \Sigma \gamma^\text{eq}_{2} \right)},
\label{eq:ga_ave}
\end{align}
where the summations indicate the inclusion of permutations as well as the symmetry factors. We have checked that the approximation used here is in agreement with the true results up to $10\%$ discrepancy for some of the dimension-7 operators. 

Equipped with the approximate formulae for $\gamma^\text{eq}$, we now compute the $L$ washout rate from the operator $\mathcal{O}_8$, $L^i \bar{e^c} \bar{u^c} d^c H^j \epsilon_{ij}$ chosen as an illustrative example. The operator induces, for instance, the process $L \bar{e^c} \to u^c \bar{d^c} \bar{H}$~(symbols denote particles) while its complex conjugate yields the inverse process $L \bar{e^c} \leftarrow u^c \bar{d^c} \bar{H}$. On the other hand, by a permutation of the field operators, a physically different process $\bar{u^c} d^c H \to \bar{L} e^c$~($\bar{u^c} d^c H \leftarrow \bar{L} e^c$) is also created by $\mathcal{O}_8~(\mathcal{O}_8^\dag)$. The operator $\mathcal{O}_8$ can induce $3\leftrightarrow 2$ and $1\leftrightarrow 4$ processes, but the $1\leftrightarrow 4$ processes are suppressed in the phase space integral compared to those of $3\leftrightarrow 2$, as can be seen from Eq.~\eqref{eq:ph-sp}. Again, to compute the total $L$ washout from $\mathcal{O}_8$, one should sum over all the distinguishable permutations, thirty of them in total: twenty come from $(n,m) = (2,3)$ and $(3,2)$, and ten arise from $(n,m) = (1,4)$ and $(4,1)$ where $n$~($m$) denotes the number of the initial~(final) state particles. Note that $(2,3)$ and $(3,2)$ correspond to physically different processes; e.g. $L \bar{e^c} \leftrightarrow u^c \bar{d^c} \bar{H}$ is not equivalent to $\bar{u^c} d^c H \leftrightarrow \bar{L} e^c$.

Assuming that the SM Yukawa interactions and the sphalerons are in thermal equilibrium, all relevant chemical potentials can be expressed in terms of the chemical potential of the lepton doublet $L_\ell$~($\ell = e, \mu, \tau$)~\cite{Harvey:1990qw},\footnote{For experimentally testable $0\nu\beta\beta$ decay rates, all operators we consider should have cut-off scales around or below $10^5$ GeV (except for the 5-dim Weinberg  operator). All SM Yukawa couplings and the EW sphalerons are in thermal equilibrium
in this temperature range. As a result, it is a well-justified assumption.}
\begin{gather}
\label{eq:mu_rel_1}
	\mu_H = \frac{4}{21} \sum_{\ell} \mu_{L_\ell}, \,\,\,
	\mu_{\bar{u^c}} = \frac{5}{63} \sum_{\ell} \mu_{L_\ell}, \,\,\,
   	\mu_{\bar{e_\ell^c}} = \mu_{L_\ell} - \frac{4}{21} \sum_{\ell} \mu_{L_\ell}, \,\,\,
   	\mu_{\bar{d^c}} = - \frac{19}{63} \sum_{\ell} \mu_{L_\ell}.
\end{gather}
The chemical potential is related to the normalized density $\eta$ in the limit of a small asymmetry $\vert n-\bar{n} \vert \ll {n^\text{eq}}$ as
\begin{align}
\label{eq:mu_eta_1}
	\frac{n}{n^\text{eq}} = \frac{\eta}{\eta^\text{eq}} 
	\approx e^{\mu/T} \approx 1 + \frac{\mu}{T} \,\,\,\text{and}\,\,\,
	\frac{\bar{n}}{n^\text{eq}} = \frac{\bar{\eta}}{\eta^\text{eq}} 
	\approx 1 - \frac{\mu}{T}
	\,\,\, \Rightarrow  \,\,\, 
	\frac{\eta_\Delta}{\eta^\text{eq}} \equiv 
	\frac{\eta - \bar{\eta}}{\eta^\text{eq}}  = 2 \frac{\mu}{T},
\end{align} 
where $\eta^\text{eq} \equiv n^\text{eq}/n^\text{eq}_\gamma = 1/2$~($3/2$ due to the color factor) for $e_\ell^c$~($u^c$, $d^c$) while $\eta^\text{eq}=1$ for the doublets $L_\ell$ and $H$. In light of the chemical potential dependence, one only needs to compute the time evolution of the lepton doublet density since the densities of the other particles can be inferred from $\eta_L$ based on Eqs.~\eqref{eq:mu_rel_1} and \eqref{eq:mu_eta_1}. The Boltzmann equation of $L_e$ then reads\footnote{If the operator being considered contains identical doublets ($LL$ or $HH$), one should express the doublet in terms of its components in order to obtain correctly the symmetry factor. In this case, $\eta^{eq}=1/2$~(3/2) for the (colored) $SU(2)_L$ doublet components.}
\begin{align}
	z H n_\gamma \frac{d\, \eta_{L_e} }{d\, z} 
	&= - \left[ L_e \bar{e^c} \leftrightarrow u^c \bar{d^c} \bar{H} \right] 
	   + \left(\text{other permutations}\right) \nonumber\\
	&= - \left(\frac{n_{L_e}n_{\bar{e^c}}}{n^\text{eq}_{L_e}
		 n^\text{eq}_{\bar{e^c}}} 
	   - \frac{n_{u^c} n_{\bar{d^c}} n_{\bar{H}}}{n^\text{eq}_{u^c}  
	   	 n^\text{eq}_{\bar{d^c}} n^\text{eq}_{\bar{H}}} \right) 
	     \gamma^\text{eq} (L_e \bar{e^c} \to  u^c \bar{d^c} \bar{H}) 
	     + \cdots \nonumber\\
	&= -  \frac{22 \, \mu_{L_e}}{7\, T}  \gamma^\text{eq} 
	(L_e \bar{e^c} \to u^c \bar{d^c} \bar{H})  + \cdots \nonumber\\
&= - \frac{11}{7}
\eta_{\Delta L_e}  \gamma^\text{eq} 
	(L_e \bar{e^c} \to u^c \bar{d^c} \bar{H})  + \cdots,
\end{align}
where we assumed first generation fermions and a universal chemical potential among three lepton flavours. All possible permutations of $2\leftrightarrow 3$ and 
$1\leftrightarrow 4$ should be included. The last two equalities come from Eqs.~\eqref{eq:mu_rel_1} and \eqref{eq:mu_eta_1}. One can obtain the Boltzmann equations in a similar way for the antiparticle, $\bar{L}_e$.
Finally, the thermal rate $\gamma^{\text{eq}}$ can be computed  based on Eq.~\eqref{eq:ga_ave} and the total washout effect from the operator $\mathcal{O}_8$ is
\begin{align}
	z H n_\gamma \frac{d\eta_{\Delta L_e} }{dz} 
         & = - \frac{ 11 \sqrt{195} \, T^{10} }{ 7 \pi^7 \Lambda^6}  
         \eta_{\Delta L_e}.
\end{align}
Generalizing, the washout effect from a dimension-$D$ operator can be expressed as
\begin{align}
\label{eq:washout_gen}
	z H n_\gamma \frac{d\eta_{\Delta L_e} }{dz} 
         &= - c_D \frac{T^{2D-4}}{\Lambda^{2D-8}_D} \eta_{\Delta L_e},
\end{align}
where the equilibrium photon density is $n_\gamma \approx 2T^3/\pi^2$ and the Hubble parameter is $H \approx 1.66 \sqrt{g_*} \, T^2 / \Lambda_\text{Pl}$ with the effective number of relativistic degrees of freedom $g_* \approx 107$ in the SM and the Planck scale $\Lambda_\text{Pl} = 1.2\times 10^{19}$~GeV. The washout processes with an interaction rate $\Gamma_W$ can be regarded to be in equilibrium if 
\begin{align}
\label{eq:washout}
  \frac{\Gamma_W}{H} &\equiv \frac{c_D}{n_\gamma H}\frac{T^{2D-4}}{\Lambda_D^{2D-8}} 
	= c_D' \frac{\Lambda_\text{Pl}}{\Lambda_D}\left(\frac{T}{\Lambda_D}\right)^{2D-9} \gtrsim 1,
\end{align}
with $c_D' = \pi^2 c_D/(3.3 \sqrt{g_*}) \approx 0.3 \, c_D$. This approximately implies that the washout is effective within the temperature interval
\begin{align}
\label{eq:temp_limit}
  \Lambda_D \left( \frac{\Lambda_D}{c_D' \Lambda_\text{Pl}} \right)^{\frac{1}{2D-9}} 
	\equiv \lambda_D \lesssim T \lesssim \Lambda_D.
\end{align}
The upper limit $T \lesssim \Lambda_D$ is imposed to ensure the validity of the effective operator approach, but washout may continue above $\Lambda_D$ in an underlying UV theory as discussed in Section~\ref{sec:conclusions}. Furthermore, the lower bound on the scale of baryogenesis can be more precisely determined by solving the Boltzmann Eq.~\eqref{eq:washout_gen} from the baryogenesis scale down to the EW scale to see if the observed baryon asymmetry can be reproduced. This leads to the more accurate lower limit
\begin{align}
\label{eq:lambda_hat}
	\hat\lambda_D \approx 
	\left[(2D-9) \ln\left(\frac{10^{-2}}{\eta_B^\text{obs}}\right) \lambda_D^{2D-9} + v^{2D-9}\right]^{\frac{1}{2D-9}},
\end{align}
that is larger than $\lambda_D$ obtained simply based on $\Gamma_W \gtrsim H$. We here conservatively assume a primordial asymmetry of order one, perhaps generated in a non-thermal fashion.

One obvious question concerns the range of efficient washout of the dim-5 Weinberg operator. It has been shown~\cite{Nelson:1990ir} that if neutrinos have Majorana masses, there exists an upper bound on the scale of baryogenesis $T \lesssim 10^{12}~\text{GeV} (1\text{ eV}/m_\nu)^2$. That is because the underlying LNV mechanism which induces the Majorana masses will erase both the lepton and baryon asymmetry with the help of the sphalerons. The same approach was adopted in our previous work~\cite{Deppisch:2015yqa}, where we instead use the current $0\nu\beta\beta$ constraints leading to $T\lesssim 2\times 10^{12}$~GeV. As we shall see later, the analysis of the washout effect based on the effective approach will not be valid above the cut-off scales at which new particles can be produced on-shell. In the following, we in any case focus on LNV operators other than the Weinberg operator, and correlate their washout effect with the induced $0\nu\beta\beta$ rate. 

%% file: results.tex
\section{Results}\label{sec:results}
In our previous paper \cite{Deppisch:2013jxa}, we discussed the contribution of the four exemplary operators $\mathcal{O}_1,\mathcal{O}_8, \mathcal{O}_{12a}, \mathcal{O}_{24a}$ contributing at tree level to $0 \nu \beta \beta$ decay as depicted in Fig.~\ref{fig:graphs}. In the following we study the contributions that can be obtained by each of the $\Delta L =2$ operators listed in Section~\ref{sec:sm-operators} leading to $0 \nu \beta \beta$ decay via short- or long-range contributions triggered either via tree-level or higher-order contributions.

Our main results are presented in Figs.~\ref{fig:washout-all-1stgen}, \ref{fig:washout-lr-sr} and \ref{fig:washout-all-3rdgen}. We show the temperature range of highly effective washout for each given operator. We indicate 7-dimensional operators in \emph{purple}, 9-dimensional ones in \emph{green} and 11-dimensional ones in \emph{magenta}. The upper limit of each bar indicates the scale $\Lambda$ of the given operator assuming an observation of $\ovbb$ at the future sensitivity $T_{1/2}^{\mathrm{Xe}} = 10^{27}y$. Depending on the figure, we either take into account all contributions as outlined before or we take only the long- or short-range contributions leading to an effective scale $\Lambda_\text{long}$ or $\Lambda_\text{short}$, respectively. As we only consider effective operators at this point do not know the washout above the operator scale and the upper limit is imposed to ensure the validity of the effective operator approach, see Eq.~\eqref{eq:temp_limit}. 

The \emph{dark} bar segments then depict the interval [$\hat{\lambda},\Lambda$] of strong washout, whereas the lower limit of the \emph{light} segment is given by $\lambda$. Here, $\lambda$ gives the temperature where $\Gamma_W/H=1$, see Eq.~\eqref{eq:temp_limit}, and $\hat{\lambda}$ denotes the temperature at which an asymmetry of order one can be injected to yield the observed baryon asymmetry, cf. Eq.~\eqref{eq:lambda_hat}.

\subsection{Long-Range Contribution}\label{sec8:longrange}
We commence our discussion with operators that clearly trigger dominantly a long-range contribution. We discuss different aspects that have to be considered in order to correctly estimate the scale of the operators constraint from $\ovbb$ and describe their impact on the identified washout interval.

\begin{figure}[t!]
\centering
\includegraphics[clip,width=0.99\linewidth]{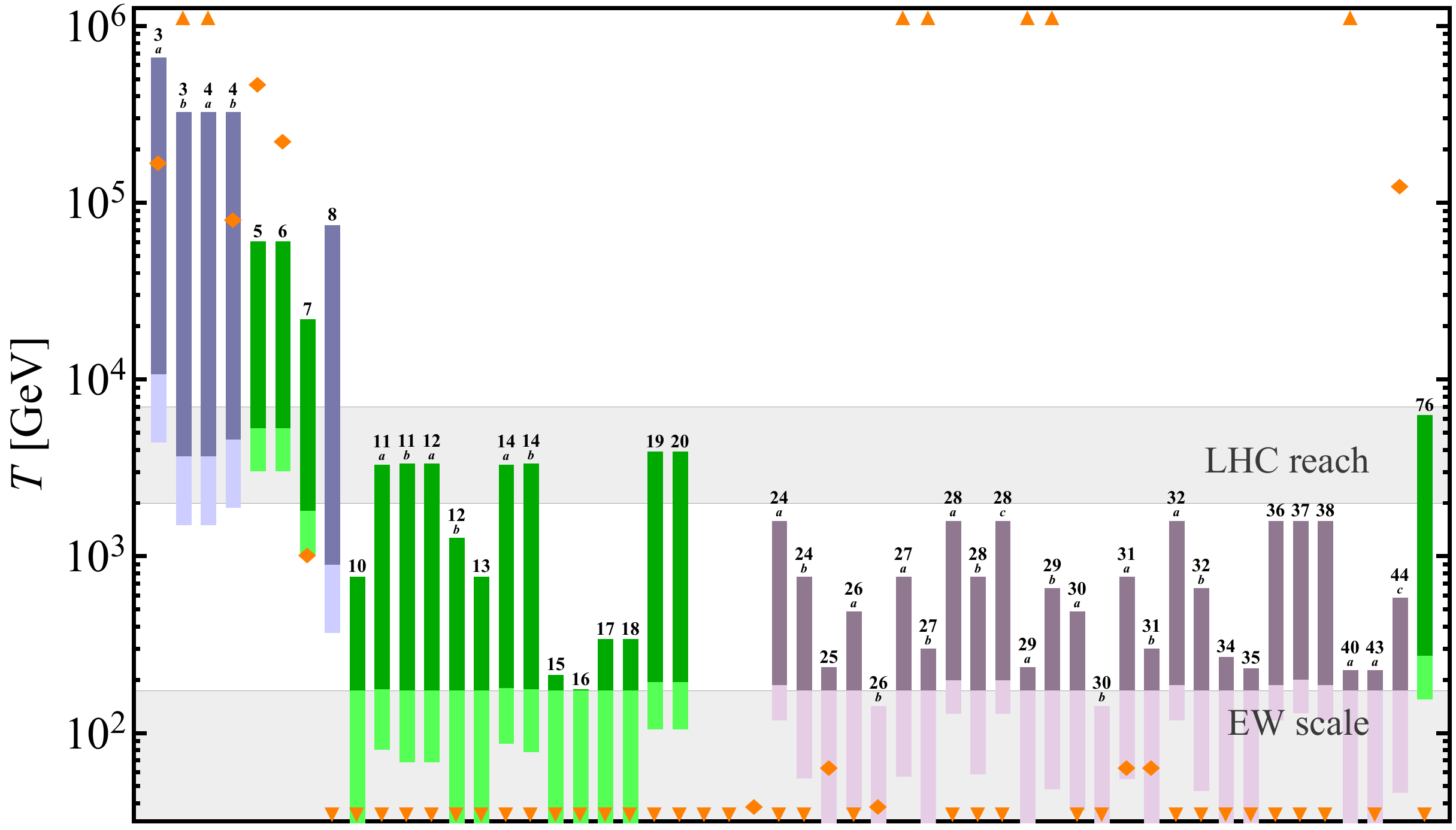}
\caption{Temperature range of the effective washout for each given operator assuming observation of $\ovbb$ at $T_{1/2}^\text{Xe} = 10^{27}$~y. We indicate 7-dimensional operators in \emph{purple}, 9-dimensional ones in \emph{green} and 11-dimensional ones in \emph{magenta}. The \emph{darker} bar segment depicts the strong washout range between $\Lambda$ and $\hat{\lambda}$. The lower limit of the \emph{lighter} segment is given by $\lambda$, see Eqs.~\eqref{eq:temp_limit} and \eqref{eq:lambda_hat}. Both long-and short-range contributions to $0\nu\beta\beta$ decay induced by the given operator are taken into account. The \emph{orange diamond} shows the corresponding scale of the Weinberg operator whereas the \emph{orange arrows} pointing up or down indicate a scale larger or smaller than the range of the plot. All SM Yukawa couplings are chosen at their 1st generation values. The two grey bars indicate the temperature range below the EW scale where sphaleron transitions become inefficient and the rough mass reach of LHC searches for resonances, respectively.}
\label{fig:washout-all-1stgen}  
\end{figure}

\paragraph{Impact of Sensitivity on $\boldsymbol{\ovbb}$ Couplings}
While the operators $\mathcal{O}_{3a,3b,4a}$ lead to the same scaling as $\mathcal{O}_8$ (discussed in Ref.~\cite{Deppisch:2013jxa}),
\begin{align} 
	\frac{G_F \epsilon_7^{3a,3b,4a,8}}{\sqrt{2}}=\frac{v}{\Lambda^3},
\end{align}
their specific hadronic and leptonic current structure that was derived in Section~\ref{sec:SU2decomposition} leads to different effective couplings
\begin{align}
        \epsilon_7^{3a} =  \epsilon_{T_R}^{T_R}, \quad 
	\epsilon_7^{3b} =  \epsilon_{S+P}^{S+P}, \quad 
	\epsilon_7^{4a}     =  \epsilon_{S-P}^{S+P}, \quad 
	\epsilon_7^{8}      = 2\epsilon_{V+A}^{V+A},
\end{align}
cf. Eq.~\ref{eq:corresp2} and Eq.~\ref{eq:corresp4}. Given the different sensitivities listed in Tab.~\ref{tab:limits} due to the impact of the $0\nu\beta\beta$ nuclear matrix elements, the corresponding operator scales differ significantly:  $\Lambda_{3a} = 6.6\times 10^5$~GeV vs. $\Lambda_{3b,4a} = 3.3\times 10^5$~GeV vs. $\Lambda_8 = 7.5\times 10^4$~GeV, see Fig.~\ref{fig:washout-all-1stgen}.
The more stringent limit on $\epsilon_{T_R}^{T_R}$ for $\mathcal{O}_{3a}$ and on $\epsilon_{S\pm P}^{S+P}$ for $\mathcal{O}_{3b,4a}$ leads to a higher operator scale and thus to a suppressed washout rate in comparison to $\mathcal{O}_8$. Specifically, under the assumption of observing LNV at the future sensitivity of $T_{1/2}^{\mathrm{Xe}} = 10^{27}y$, $\mathcal{O}_{8}$ would exclude baryogenesis models above $\hat{\lambda}_8 \approx 900$~GeV, while $\mathcal{O}_{3b,4a}$ can only exclude models above $\hat{\lambda}_{3b,4a} \approx 4$~TeV and $\mathcal{O}_{3a}$ above $\hat{\lambda}_{3a} \approx 10$~TeV. This difference is significant, e.g. when considering searches for corresponding models at the LHC.

\begin{figure}[t!]
\centering
\includegraphics[clip,width=0.32\linewidth]{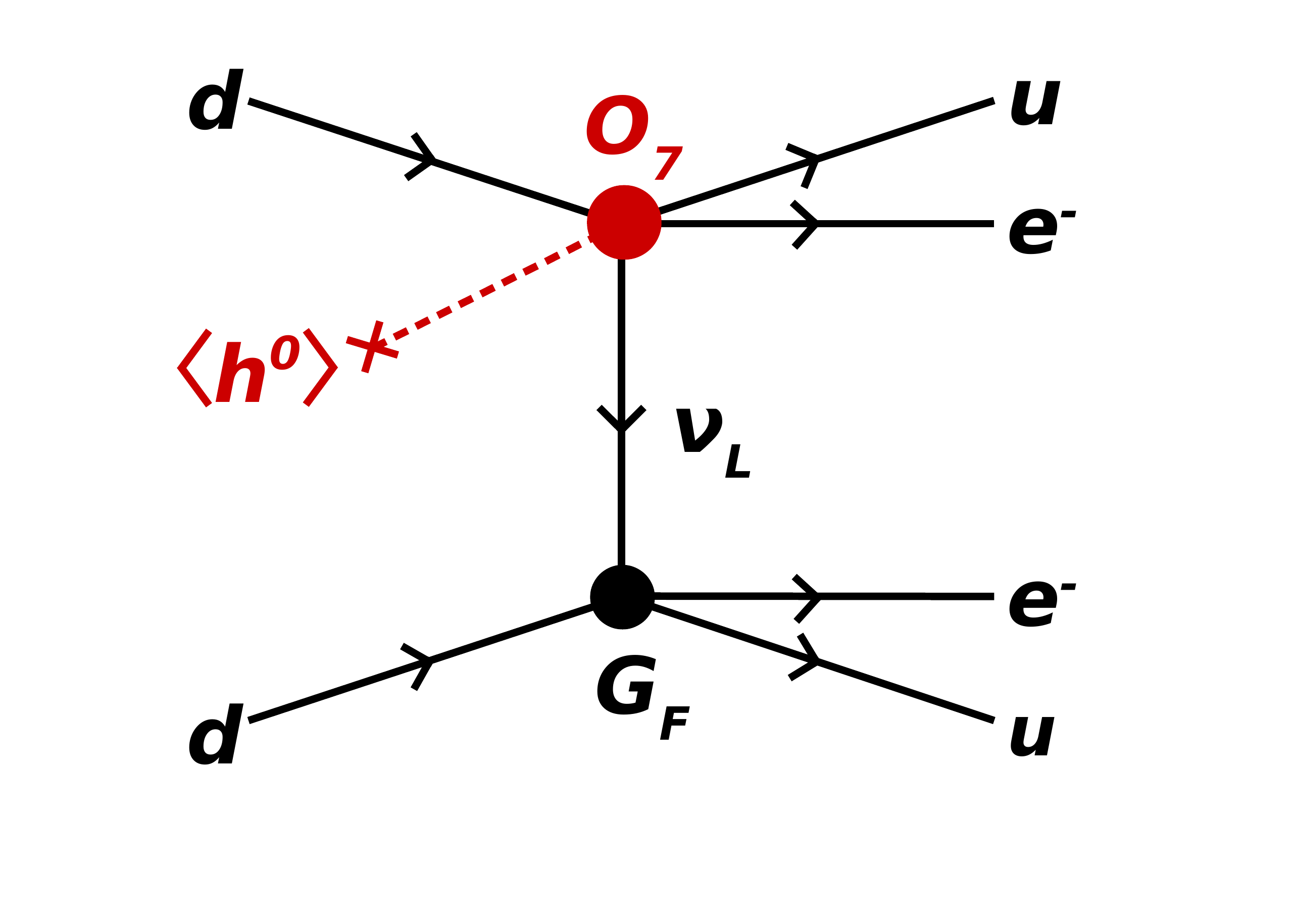}
\includegraphics[clip,width=0.32\linewidth]{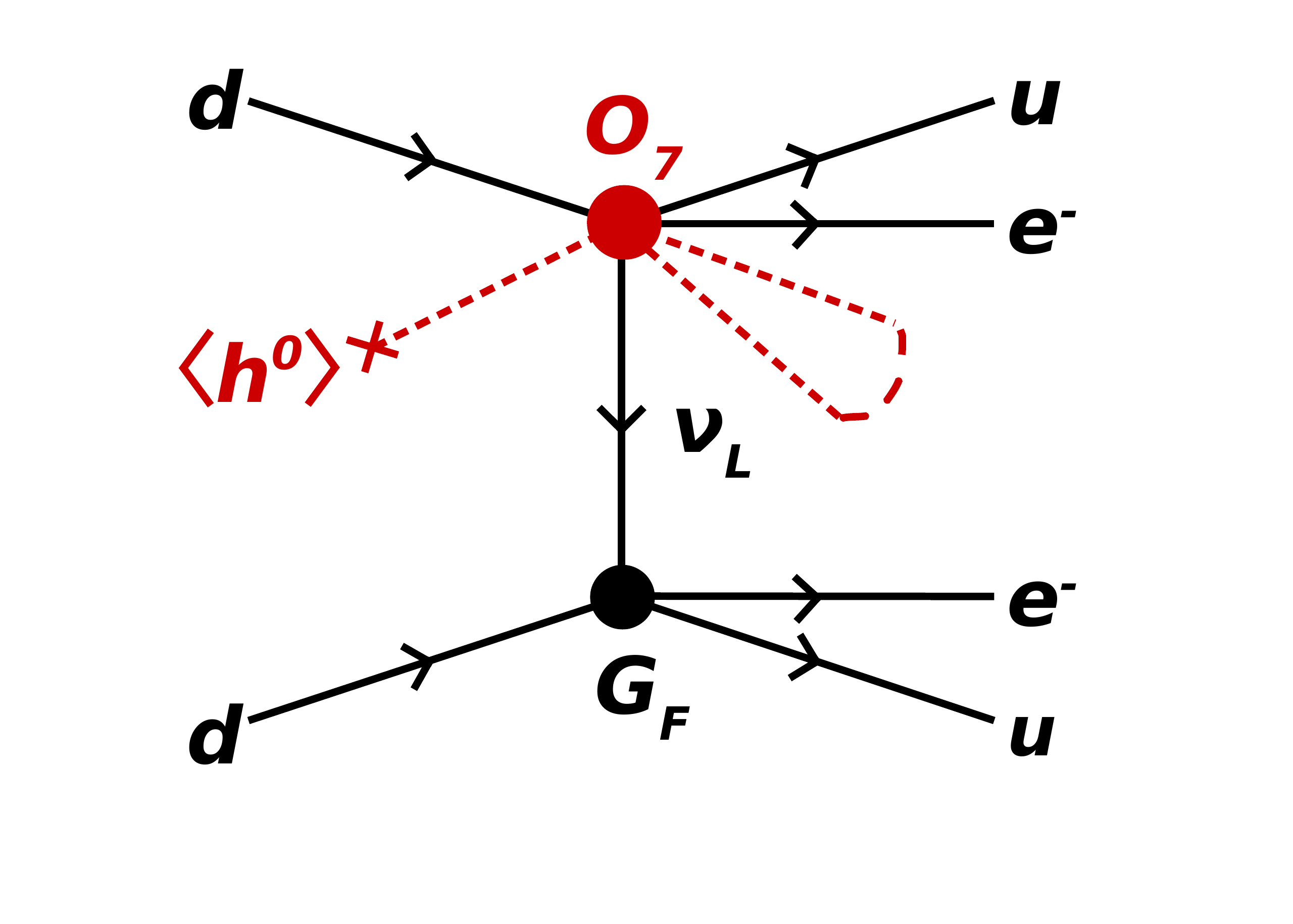}
\includegraphics[clip,width=0.32\linewidth]{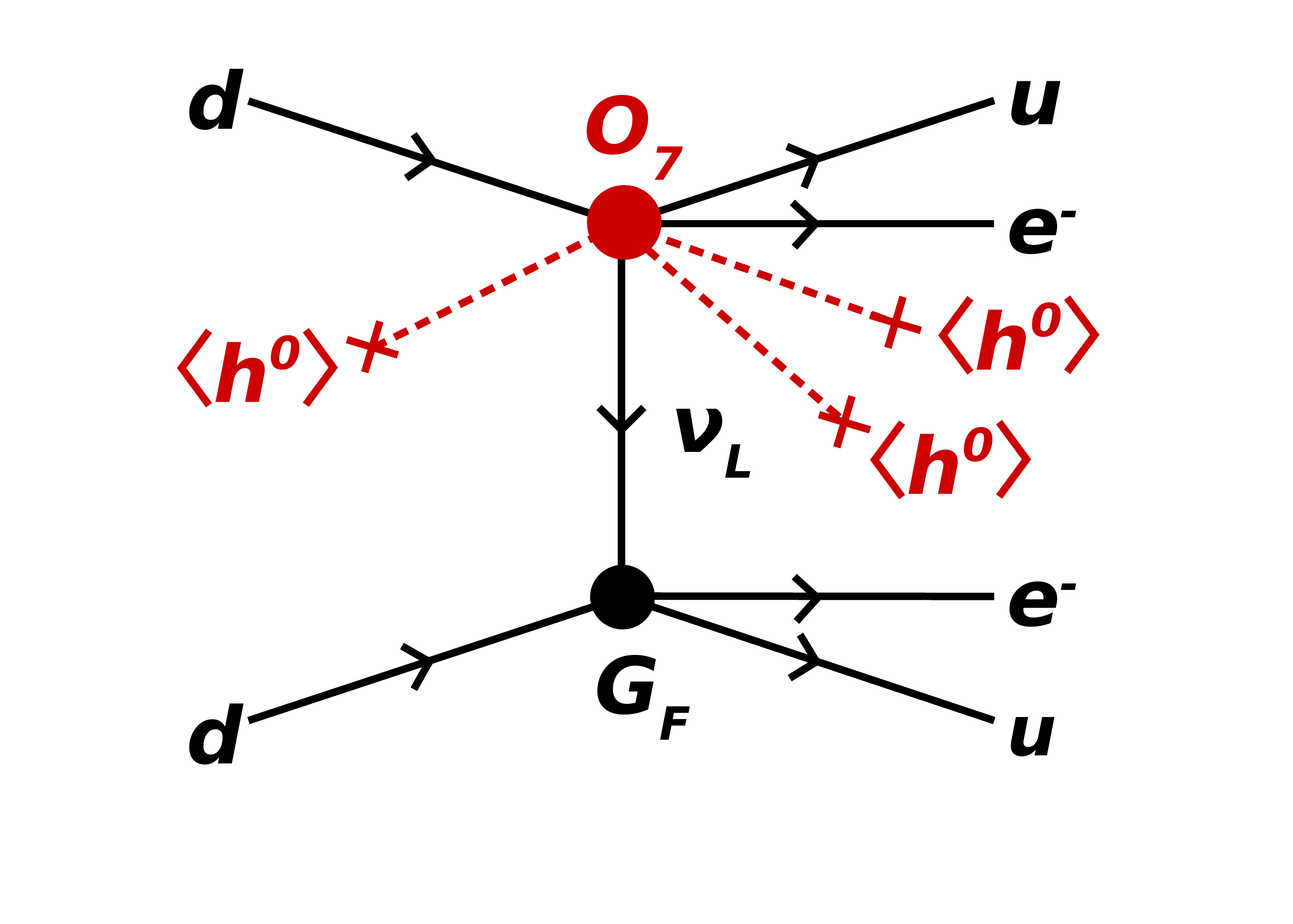}
\caption{The 7-dimensional operators $\mathcal{O}_{3a}$, $\mathcal{O}_{3b}$, $\mathcal{O}_{4a}$ and $\mathcal{O}_8$ generate a long-range contribution with one Higgs VEV (left). The 9-dimensional operators $\mathcal{O}_{5,6}$ allow for both, i.e. closing a Higgs loop (centre) and the insertion of Higgs VEVs (right), while the 9-dimensional operator $\mathcal{O}_7$ allows only for Higgs mass insertions (right).}
\label{fig:contibs7dim}  
\end{figure}

\paragraph{Impact of Field Content} 
Naively one would expect that 9-dimensional operators generate dominantly a short-range contribution at tree level. This is, however, not necessarily the case. For example, the operators $\mathcal{O}_{5,6,7}$ featuring three Higgs doublets contribute dominantly at long-range with
\begin{align}
	\frac{G_F \epsilon_7^{5,6}}{\sqrt{2}} 
		= \frac{v}{16 \pi^2\Lambda^3} + \frac{v^3}{\Lambda^5}, 
	\quad \mathrm{and} \quad
	\frac{G_F \epsilon_7^{7}}{\sqrt{2}} = \frac{v^3}{\Lambda^5},
\label{eq:567}
\end{align}
respectively. The different scaling arises from the different $SU(2)$ structure; for $\mathcal{O}_5 = L^i L^j Q^k d^c H^l H^m \overline{H}_i \epsilon_{jl}\epsilon_{km}$ and $\mathcal{O}_6 = L^i L^j \overline{Q}_k\bar{u^c}H^l H^k \overline{H}_i\epsilon_{jl}$, the Higgs doublets can be additionally closed to a loop. The limit on their effective coupling is already so stringent such that $\Lambda > 4\pi v$ and the contribution $v/(16 \pi^2 \Lambda^3)$ dominates. This is in contrast to $\mathcal{O}_7 = L^iQ^j \bar{e^c}\overline{Q}_kH^k H^l H^m\epsilon_{il} \epsilon_{jm}$, whose structure does not allow for loop closing and thus scales with $1/\Lambda^5$, cf. Fig.~\ref{fig:contibs7dim}. Generically, this leads to a suppressed washout for $\mathcal{O}_{5,6,7}$ in comparison to other 9-dimensional operators and thus to a higher limit on the scale above which baryogenesis can be excluded, cf. Fig.~\ref{fig:washout-all-1stgen}.

\subsection{Competition between Long- and Short-Range Contribution}
\label{sec8:interplay}
For the above operators it was straightforward to decipher which contribution (long- vs. short-range) they generate dominantly. For other higher-dimensional operators this is not necessarily the case due to a non-trivial interplay of different aspects outlined below.

\begin{figure}[t!]
	\centering
	\includegraphics[clip,width=0.32\linewidth]{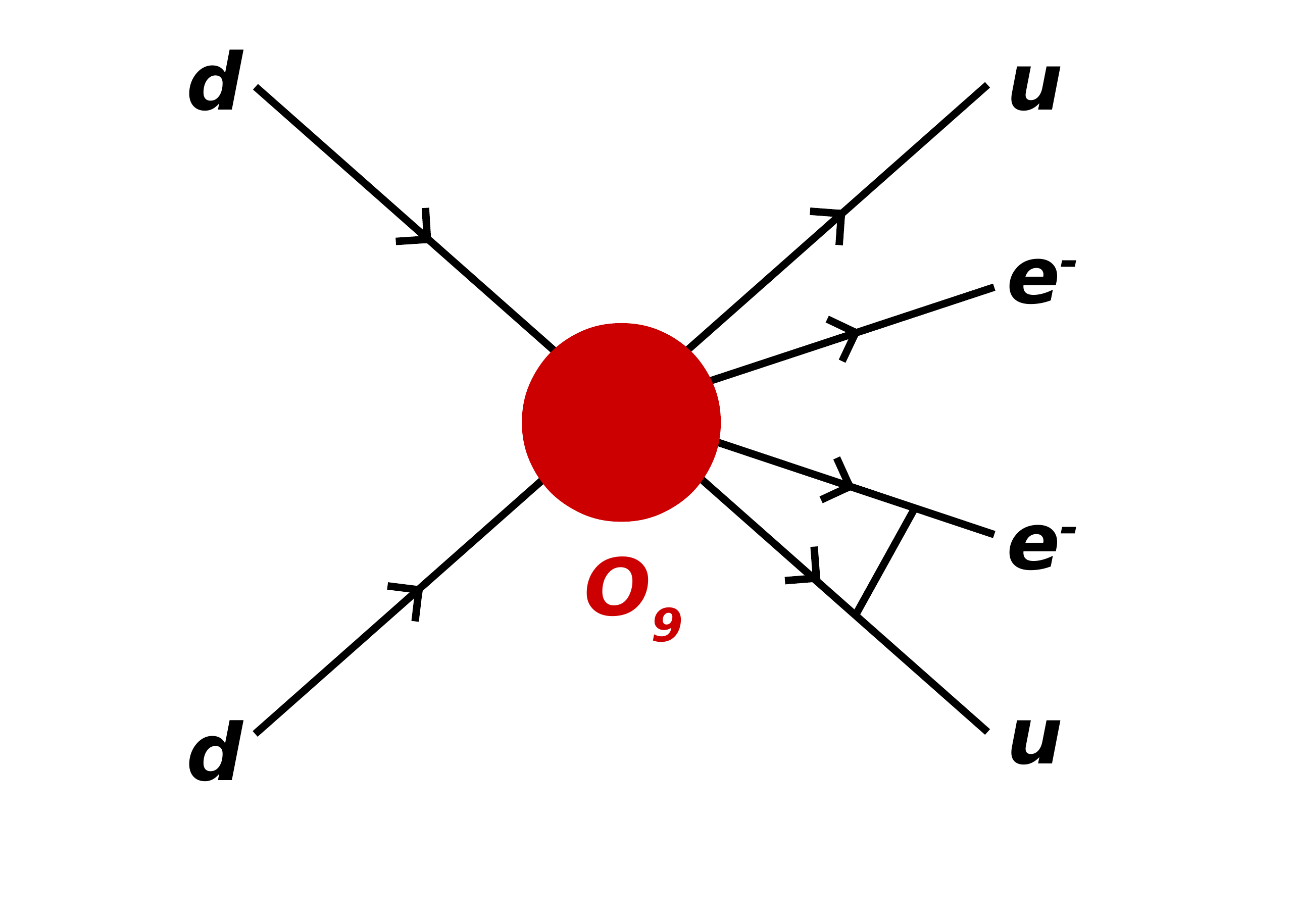}
	\includegraphics[clip,width=0.32\linewidth]{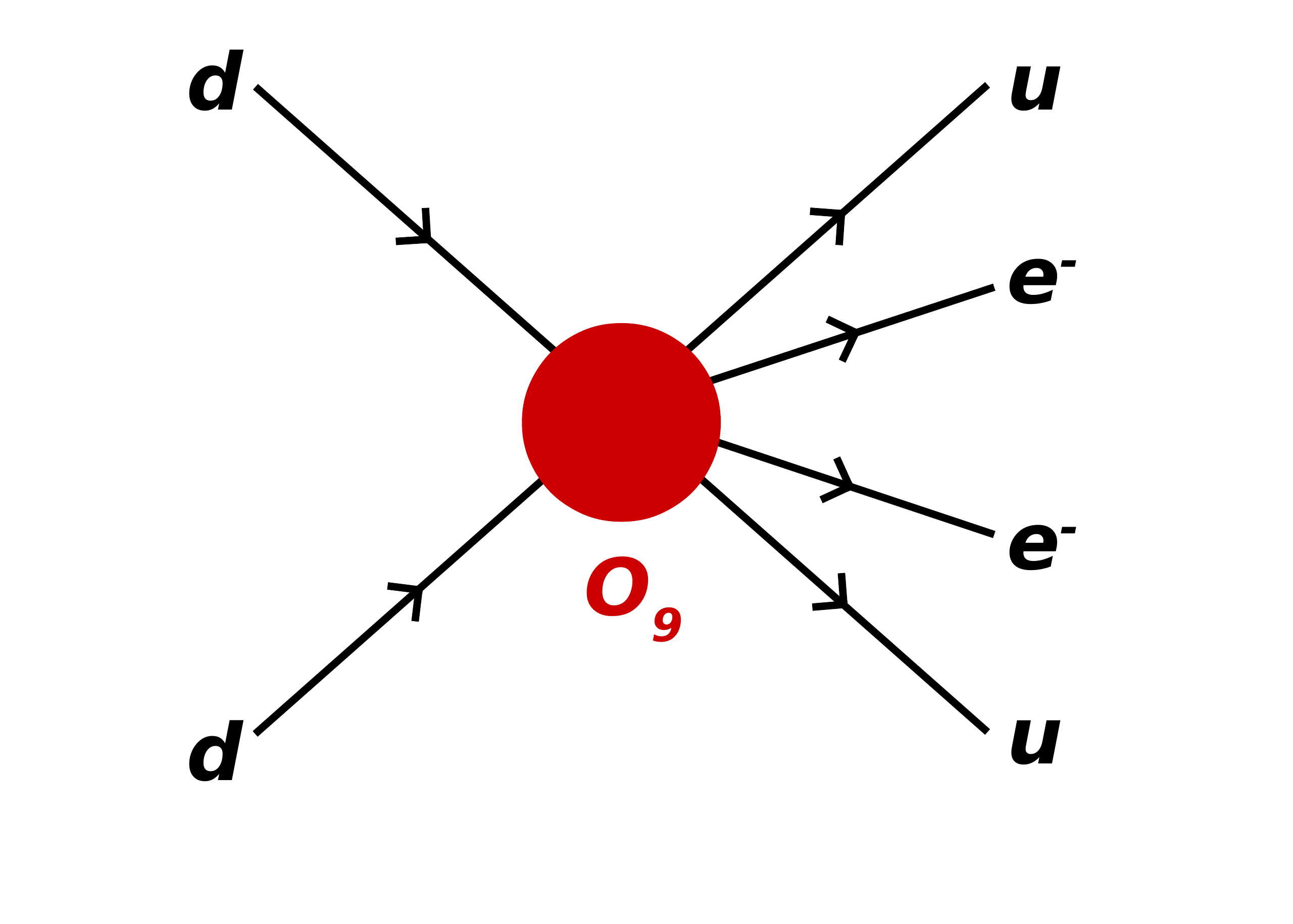}
	\includegraphics[clip,width=0.32\linewidth]{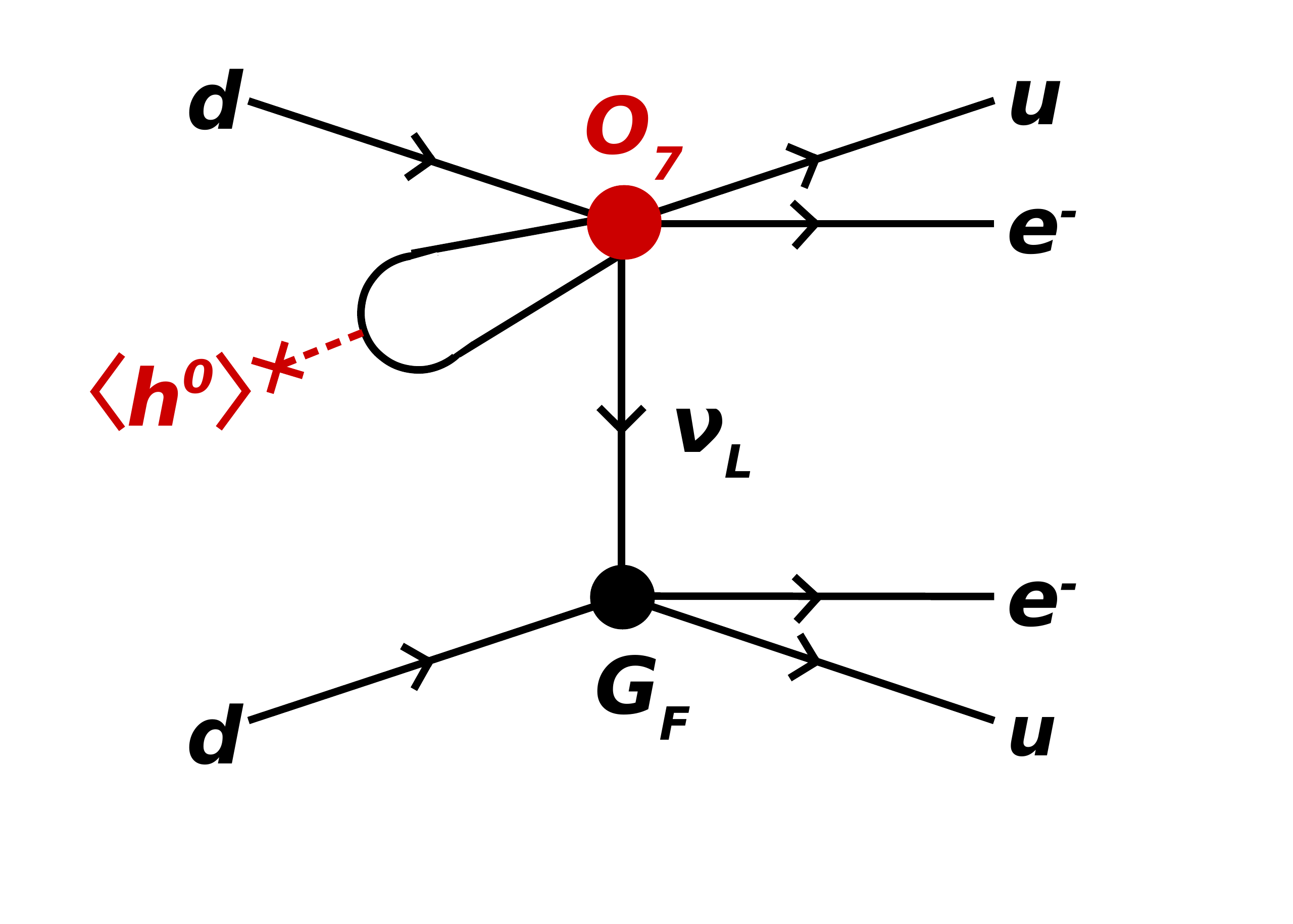}
	\caption{While $\mathcal{O}_{11a}$ contributes only at one loop at short range (left), $\mathcal{O}_{11b}$ contributes directly (centre). Both contribute with the same loop-suppression at long range (right).}
	\label{fig:op11}  
\end{figure}
\paragraph{Impact of $\boldsymbol{SU(2)}$ Structure on Dominant Contribution}
In order to demonstrate the non-trivial interplay, we compare the behaviour of $\mathcal{O}_{11a}$ and $\mathcal{O}_{11b}$ that differ only in their $SU(2)$ structure. While $\mathcal{O}_{11b}$ contributes at tree level to the short-range contribution, $\mathcal{O}_{11a}$ contributes only one-loop suppressed, cf. Fig.~\ref{fig:op11},
\begin{align}
	\frac{G_F^2 \epsilon_9^{11a}}{2 m_p} = \frac{g^2}{16 \pi^2 \Lambda^5}, \quad
	\frac{G_F^2 \epsilon_9^{11b}}{2 m_p} = \frac{1}{\Lambda^5},
\end{align}
with $\epsilon_9^{11a,b} = \epsilon_1$. The corresponding experimental limits are given in Tab.~\ref{tab:limits}. Besides their different short-range contributions, they contribute both identically to the long-range contribution
\begin{align}
\label{eq:11ablongrange}
	\frac{G_F \epsilon_7^{11a,b}}{\sqrt{2}} = \frac{y_d v}{16 \pi^2\Lambda^3}, 
\end{align}
however, with different effective couplings $\epsilon_7^{11a} = \epsilon_{T_R}^{T_R}$ and $\epsilon_7^{11b} = \epsilon_{S+P}^{S+P}$. As depicted in Fig.~\ref{fig:op11}, $\mathcal{O}_{11a}$ and $\mathcal{O}_{11b}$ can be reduced to $\mathcal{O}_{3a}$ and $\mathcal{O}_{3b}$, respectively. 

\begin{figure}[t!]
	\centering
	\includegraphics[clip,width=0.49\linewidth]{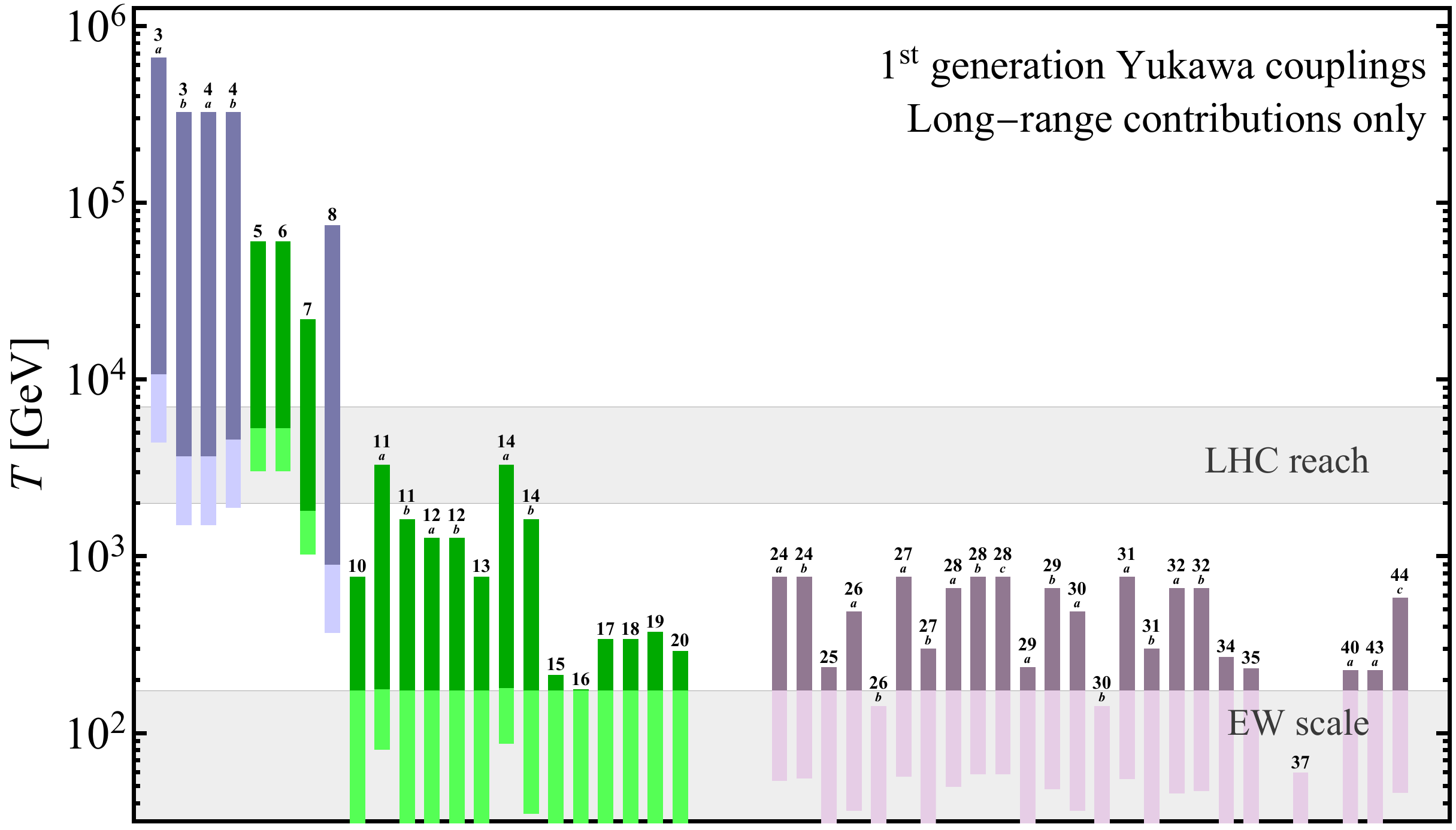}
	\includegraphics[clip,width=0.49\linewidth]{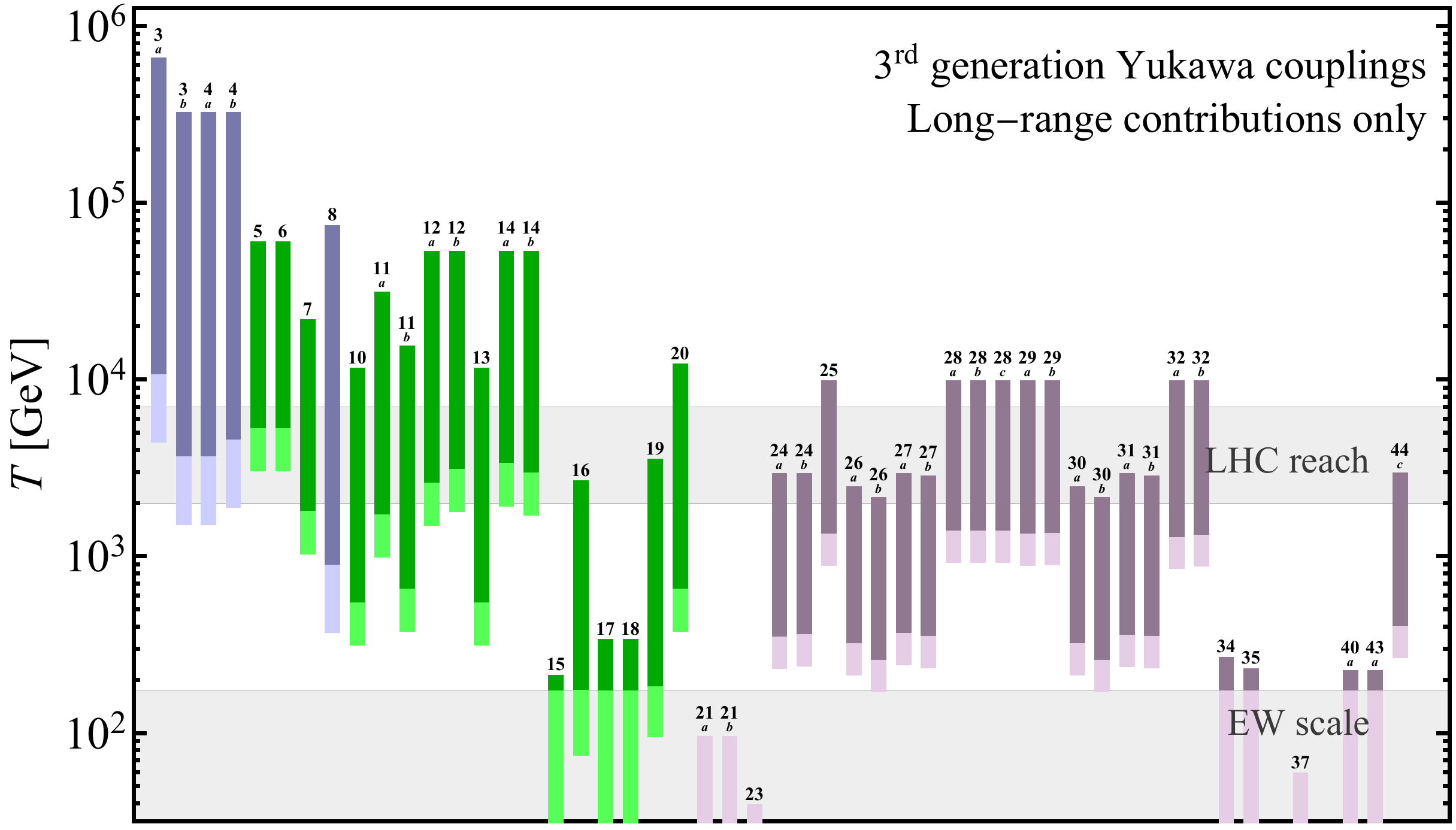}\\
	\includegraphics[clip,width=0.49\linewidth]{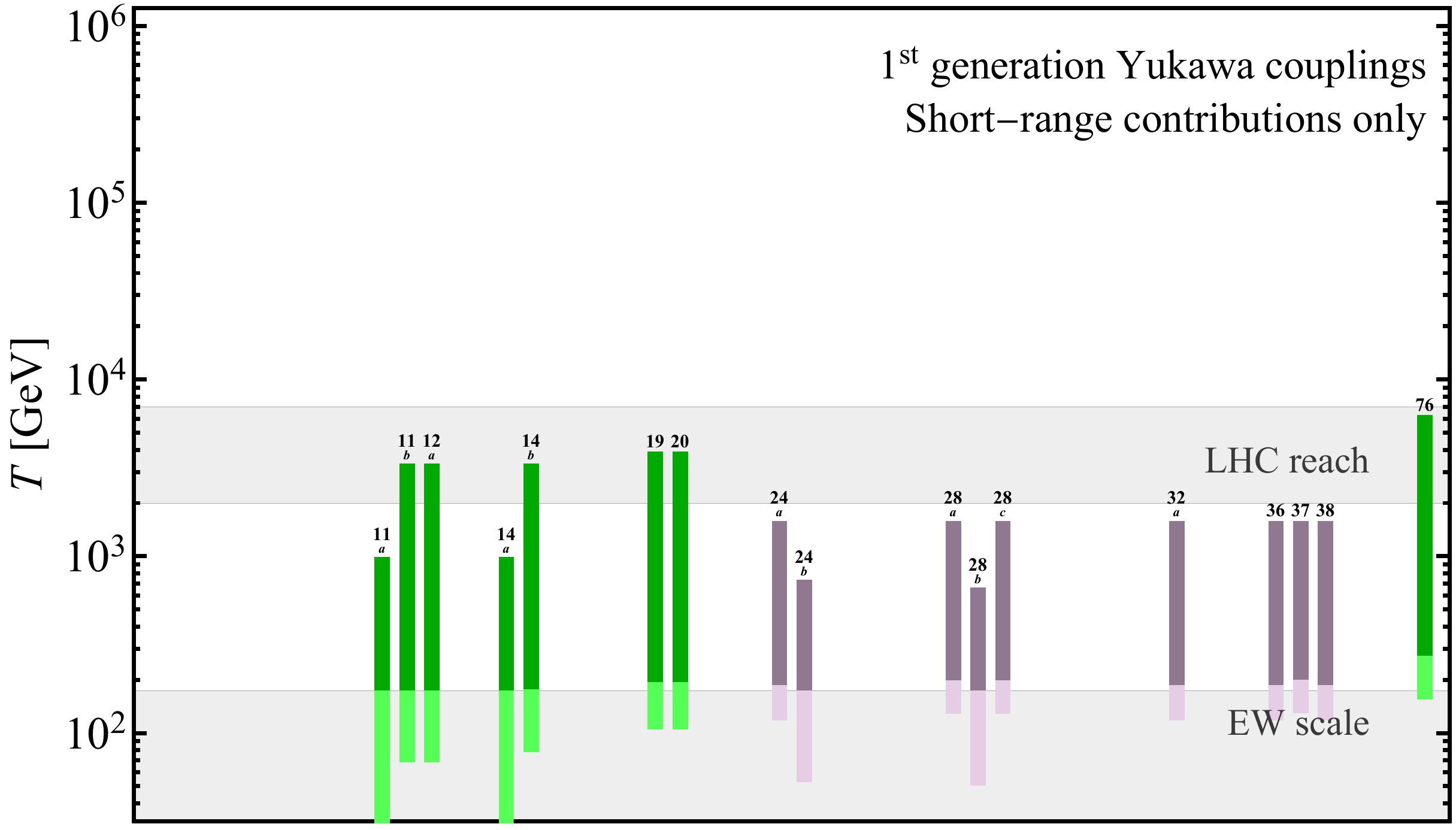}
	\includegraphics[clip,width=0.49\linewidth]{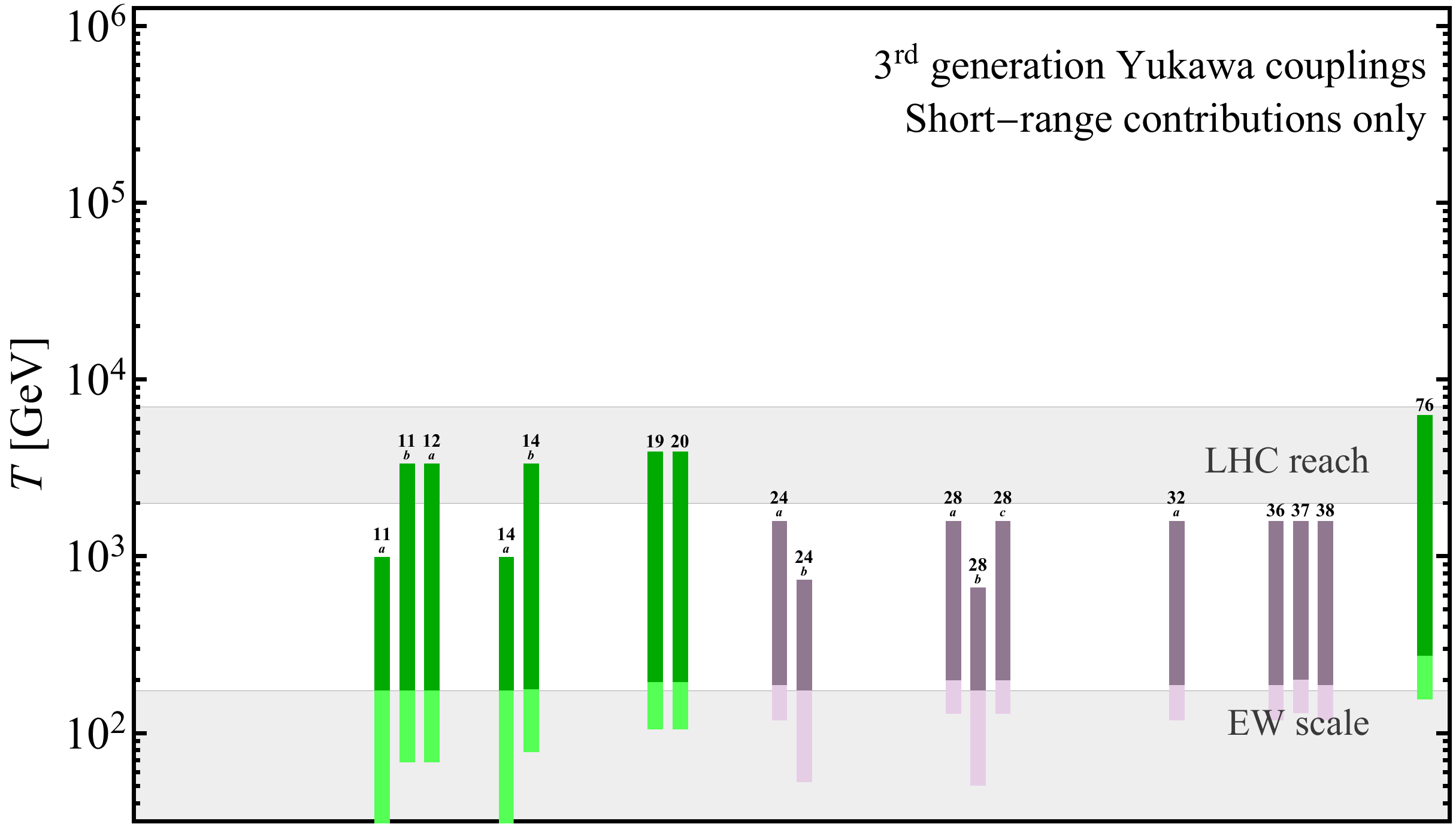}
	\caption{As Fig.~\ref{fig:washout-all-1stgen}, but constraining separately the operators' long-range (top row) and short range contribution (bottom row), both for first generation Yukawa couplings only (left column) as well as for considering internal third generation Yukawa couplings (right column).}
	\label{fig:washout-lr-sr}  
\end{figure}

\begin{figure}[t!]
\centering
\includegraphics[clip,width=0.99\linewidth]{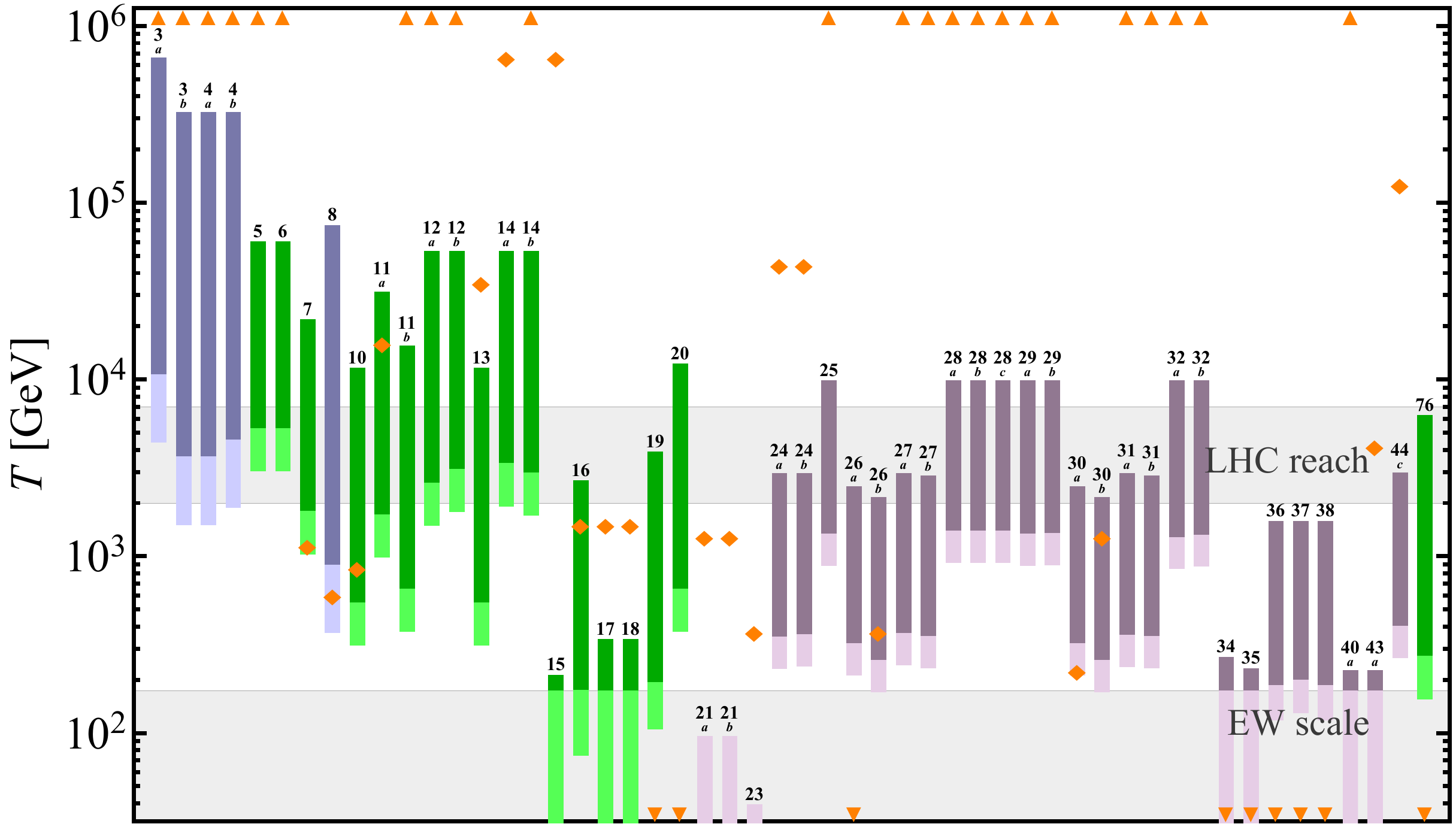}
\caption{As Fig.~\ref{fig:washout-all-1stgen}, but Yukawa couplings at vertices not attached to outer legs in are chosen at their third generation values, while external ones are kept at their first generation values.}
\label{fig:washout-all-3rdgen}  
\end{figure}

We study the competition of long- and short-range contributions by comparing with Fig.~\ref{fig:washout-lr-sr} showing the corresponding washout range when constraining the operators' contribution to long- (upper left) or short-range (lower left) separately. As expected, $\mathcal{O}_{11b}$ reproduces for the short-range contribution the limits of $\mathcal{O}_{12a}$ \cite{Deppisch:2015yqa}. The short-range contribution of $\mathcal{O}_{11a}$ however is loop suppressed, leading to a lower operator scale $\Lambda_9^{11a}$ and thus a stronger washout than for $\mathcal{O}_{11b}$. While both operators contribute similarly at long-range (cp. \eqref{eq:11ablongrange}), their specific SU(2) structure leads to different effective couplings $\epsilon_7^{11a}, \epsilon_7^{11b}$. This results in an operator scale of $\Lambda_7^{11a}=3.3~\mathrm{TeV} > \Lambda_9^{11a}$ and $\Lambda_7^{11b}=1.6~\mathrm{TeV} < \Lambda_9^{11b}$. This is summarized in Tab.~\ref{tab:scales11ab} revealing an interesting effect. The naively expected behaviour is reproduced by $\mathcal{O}_{11b}$: The short-range contribution dominates for the 9-dimensional operator\footnote{As described in Section~\ref{sec:identifydominant}, we can always identify the higher scale with the more dominant contribution. This can be understood as follows: we independently identify the scales $\Lambda_\text{long}$ and $\Lambda_\text{short}$ by constraining the long- and short-range contribution separately assuming observation at $T_{1/2}^\text{Xe} = 10^{27}$~y. Picking the lower scale of ($\Lambda_{\mathrm{long}}$, $\Lambda_{\mathrm{short}}$) would imply exceeding the experimental limit for the contribution with the higher scale, such that taking the higher scale ensures choosing the dominant contribution that is not yet excluded by experiment. By comparing the upper row with the lower row of Fig.~\ref{fig:washout-lr-sr}, we can identify the dominant contribution. The corresponding scale is then taken to compile the final comparison of washout regimes for different operators in Figs.~\ref{fig:washout-all-1stgen} and \ref{fig:washout-all-3rdgen}.}, while the long-range contribution would only dominate for scales $\Lambda \gtrsim 9900$~GeV. This is different for $\mathcal{O}_{11a}$: While it features a similar long-range contribution, the short-range contribution is loop suppressed such that the long-range contribution dominates already above scales $\Lambda > 163$~GeV. In other words, the loop suppression of the short-range contribution leads to the dominance of the long-range contribution. This example demonstrates clearly that the $SU(2)$ contraction for the same operator can lead to significant changes in the identification of the dominant contribution.

\begin{table}[t!]
\centering
\begin{tabular}{|cc|cc|cc|}
	\hline
	{} & {} & \multicolumn{2}{c|}{$\mathcal{O}_{11a}$}& \multicolumn{2}{c|}{$\mathcal{O}_{11b}$}\\
	\hline
	{} & {} & 1st gen & 3rd gen  & 1st gen & 3rd gen \\
	\hline
	\multirow{2}{*}{long-range} & $T^{-1}_{1/2}~\mathrm{GeV}^{-6}$& $1.3\times 10^{-6}\Lambda^{-6}$ & $\Lambda^{-6}$ & $1.8\times 10^{-8}\Lambda^{-6}$ & $0.01\Lambda^{-6}$\\
	{}& $\Lambda$ &  3299 & 31504 & 1623 & 15501 \\
	\hline
	\multirow{2}{*}{short-range}& $T^{-1}_{1/2}~\mathrm{GeV}^{-10}$ & \multicolumn{2}{c|}{$911\Lambda^{-10}$} & \multicolumn{2}{c|}{$1.8\times 10^8\Lambda^{-10}$} \\
	{} & $\Lambda$  &   \multicolumn{2}{c|}{991} & \multicolumn{2}{c|}{3345}\\
	\hline
	\multicolumn{2}{|c|}{dominant $\bigcap$ non-excluded} & long & long & short & long\\
	\hline
\end{tabular}
\caption{For both operators $\mathcal{O}_{11b,b}$ we show $T_{1/2}$~[y] as a function of $\Lambda$~[GeV] considering 1st generation Yukawa couplings or internal 3rd generation Yukawa couplings. The corresponding operator scale is given assuming observation at $T_{1/2} = 10^{27}y$. We can identify the higher scale with the dominant contribution that is not yet excluded by experiment, demonstrating that the short-range contribution is dominant only for $\mathcal{O}_{11b}$ with first generation Yukawa couplings.}
\label{tab:scales11ab}
\end{table}

\paragraph{Impact of Flavour Structure on Dominant Contribution}
As $\ovbb$ decay involves only first generation quarks and leptons, we have assumed so far only first generation Yukawa couplings in our calculations. However, comparing with Figs.~\ref{fig:contibs7dim} and \ref{fig:op11}, Yukawa couplings in loops are not necessarily fixed by external particles such that the final contribution can be summed over loops including all flavours, e.g. in a democratic flavour structure. For simplicity and a first comparison, we repeat our analysis with third generation Yukawa couplings for vertices not attached to outer legs, while keeping first generation Yukawa couplings at external vertices. We can thus assess the potential range in the contribution. The corresponding results are given in Fig.~\ref{fig:washout-all-3rdgen} considering all contributions, and in Fig.~\ref{fig:washout-lr-sr} (right columns), assuming short- and long-range contributions separately.

We take $\mathcal{O}_{11a}$ and $\mathcal{O}_{11b}$ again as examples. While the short-range contribution remains unchanged, the long-range contribution of $\mathcal{O}_{11b}$ is enhanced by the internal third generation coupling such that the long-range contribution dominates now already from scales $\Lambda > 335$~GeV, cf. Fig.~\ref{fig:washout-lr-sr}. This results in a now dominant long-range contribution for $\mathcal{O}_{11b}$ in contrast to a dominant short-range contribution when considering only first generation Yukawa couplings. Due to the loop suppression of the short-range contribution, $\mathcal{O}_{11a}$ features already a comparably low scale above which the long-range contribution dominates that is further lowered when taking third generation Yukawa couplings into account (dominant long-range contribution for $\Lambda > 6$~GeV).

While already a change from an internal first to third generation down quark Yukawa coupling leads to a swap in the dominant contribution, an even stronger effect is expected for an internal up quark Yukawa coupling. This can be observed e.g. for $\mathcal{O}_{20}$ and its corresponding contributions,
\begin{align}
	\frac{G_F^2 \epsilon_9^{20}}{2 m_p} = \frac{1}{\Lambda^5}, \quad 
	\frac{G_F \epsilon_7^{20}}{\sqrt{2}} = \frac{y_u v}{16 \pi^2\Lambda^3},
\end{align}
with $\epsilon_9^{20} = 2\epsilon_5$ and $\epsilon_7^{20} = 2\epsilon_{V+A}^{V+A}$.

The above examples clearly demonstrate that the long-range contribution can dominate for 9-dim operators although one might naively expect that the short-range contribution will be the most dominant one for these higher dimensional operators. The non-trivial interplay of scales is visualized for all operators in Fig.~\ref{fig:washout-lr-sr}, and the final results given in Fig.~\ref{fig:washout-all-1stgen} and Fig.~\ref{fig:washout-all-3rdgen}.

\paragraph{Consequences for Washout and the Observation of LNV at Colliders}
Under the strong assumption that the new physics responsible for the LNV effective operators couples to the first generation fermions only, the observation of the above operators $\mathcal{O}_{11a}$, $\mathcal{O}_{11b}$ and $\mathcal{O}_{20}$ implies a strong washout rate down to the electroweak scale and would thus falsify high-scale baryogenesis models. If internal third generation Yukawa couplings enter the calculation of $\ovbb$ contributions, this picture changes slightly. The corresponding operator scale shifts to higher values and, while still erasing a lepton asymmetry at higher scales, there can emerge a window between the EW scale and $\hat{\lambda}$ where a lepton asymmetry is not washed out efficiently.

The accessibility of new particles at the LHC depends on the assumption of the corresponding internal Yukawa but as well on the specific dominant contribution, as discussed before and visible in Fig.~\ref{fig:washout-all-1stgen} and Fig.~\ref{fig:washout-all-3rdgen}. While for first generation Yukawa couplings $\mathcal{O}_{11a}$ and $\mathcal{O}_{11b}$ might accessible at the LHC, for third generation Yukawa couplings the scale will be probably too high to be detected as resonant particle.

\subsection{Comparison with Additional s-Channel Contributions}
\label{sec8:schannel}
Let us now look at a specific set of 11-dimensional operators in order to describe what we call `$s$-channel' contributions. We compare operators $\mathcal{O}_{33}, \mathcal{O}_{34}, \mathcal{O}_{36}, \mathcal{O}_{37}$ and $\mathcal{O}_{38}$, which, despite having similar structure, differ slightly by their field content and that can lead to different $\ovbb$ decay contributions.

\begin{figure}[t!]
\centering
\begin{minipage}{.45\textwidth}
	\centering
	\includegraphics[scale=0.35]{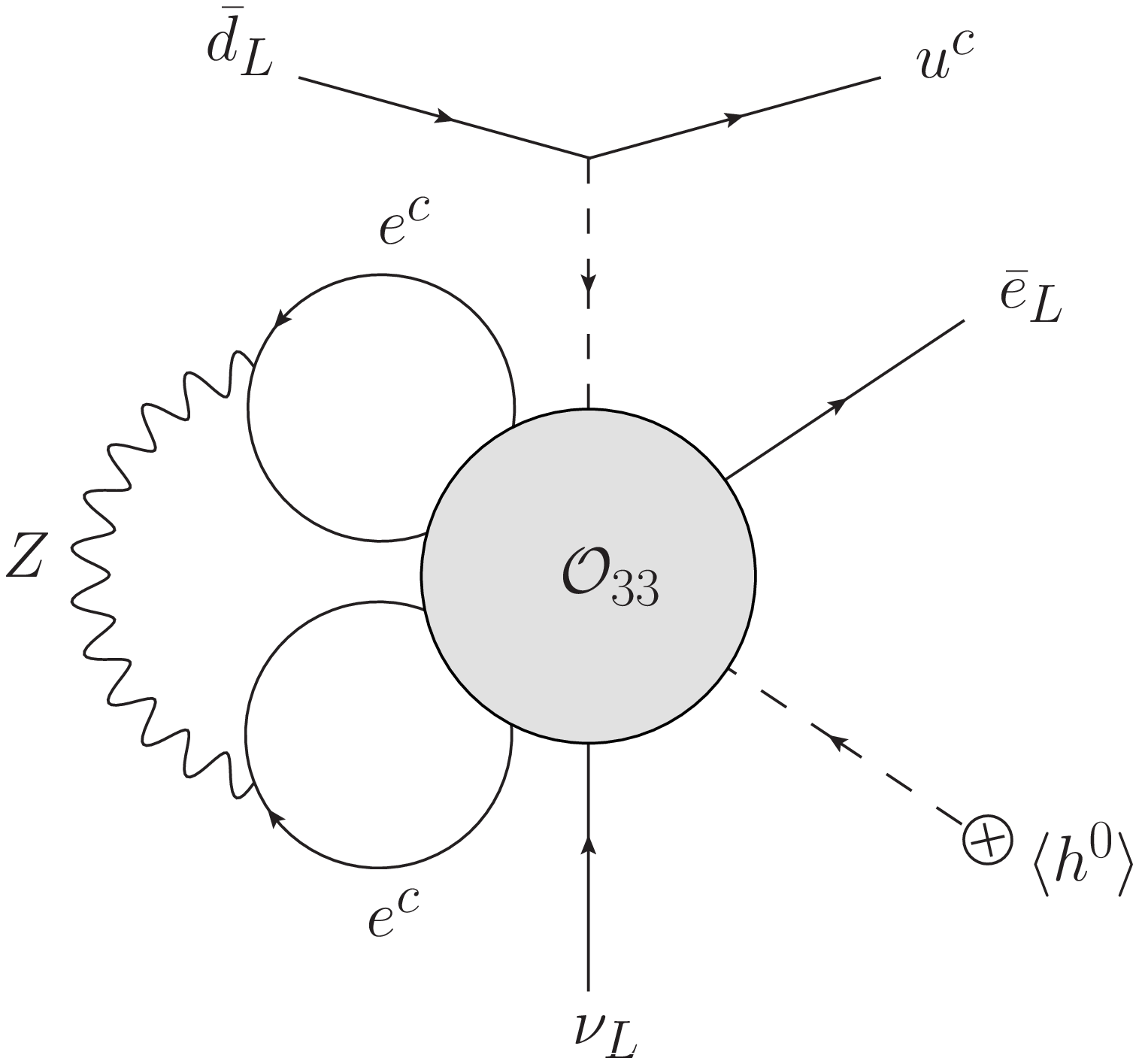}
\end{minipage}
\hspace{1.cm}
\begin{minipage}{0.45\textwidth}
	\centering
	\includegraphics[scale=0.35]{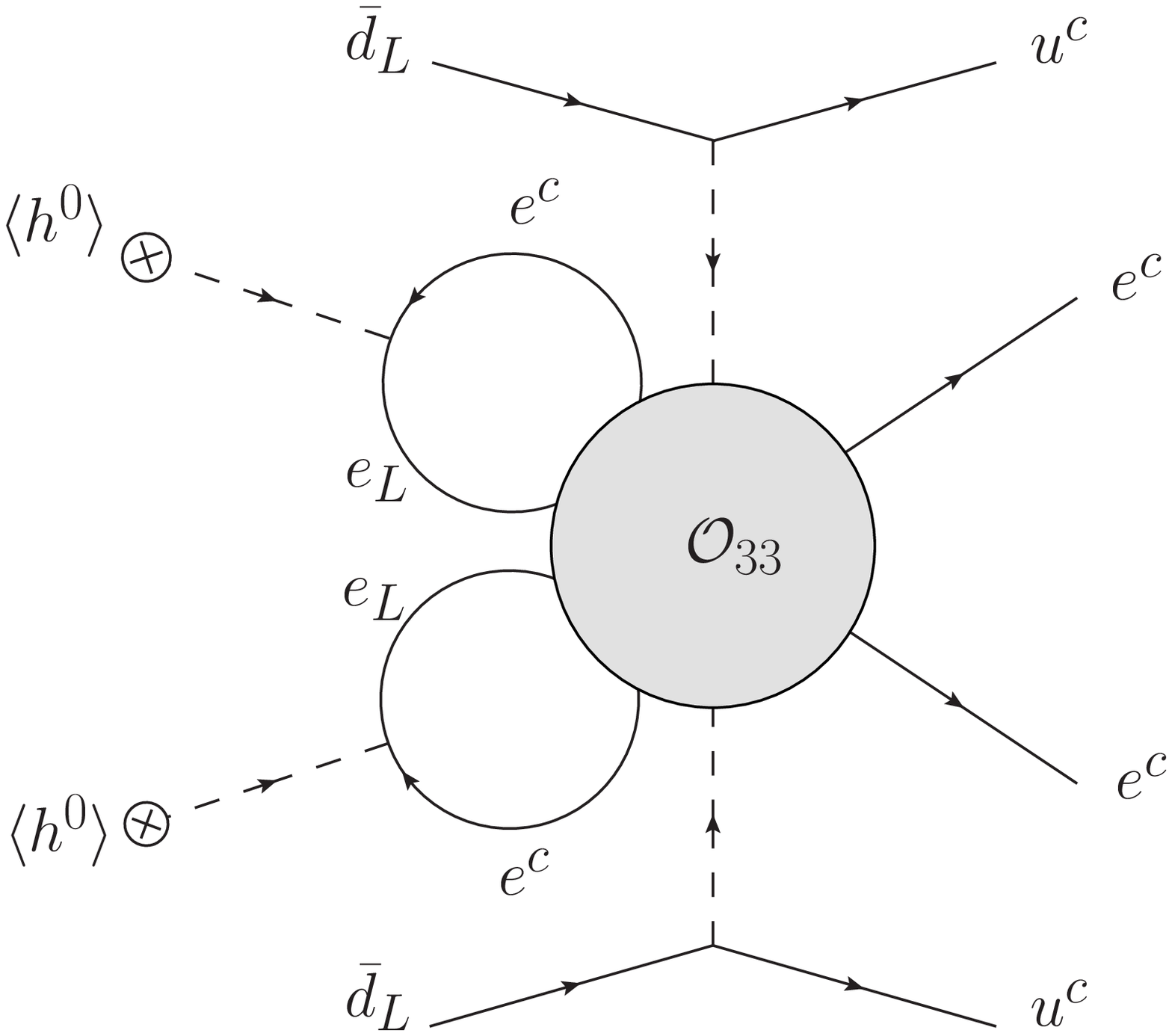}
\end{minipage}
\caption{Long-range (left) and short-range (right) contribution of $\mathcal{O}_{33}$ to $\ovbb$ decay.}
\label{fig:Op33}
\end{figure}
First, focusing on operators $\mathcal{O}_{33}, \mathcal{O}_{34}$ and $\mathcal{O}_{36}$ we see they contain 6, 4 and 2 leptons, respectively. This implies that operator $\mathcal{O}_{33}$ does not trigger $\ovbb$ decay, unless we trade at least two of its leptons for quarks. However, such an exchange requires the propagation of a heavy boson, Fig.~\ref{fig:Op33} (left), and thus suppresses the overall contribution by the square of its mass and we get
\begin{align}
	\mathcal{O}_{33}^{LR} \propto 
	\frac{y_d^\text{ex}g^2 v}{(16\pi^2)^3\Lambda} \frac{1}{m_H^2}.
\end{align}
As apparent from the Feynman diagram shown above, the original 11-dimensional operator is first reduced to the Weinberg operator and then a Yukawa interaction is attached to it. This contribution to $\ovbb$ decay will therefore be sub-leading with respect to the standard mass mechanism. Consequently, we do not include this type of operators (e.g. $\mathcal{O}_{2},\mathcal{O}_{9},\mathcal{O}_{22}$ or $\mathcal{O}_{39}$) in our analysis.

While two quarks are enough for the long-range contribution, one needs four of them for the short-range one. Therefore, two $s$-channel-like transitions must occur in this case leading to an even stronger suppression, see Fig.~\ref{fig:Op33}~(right).

As operator $\mathcal{O}_{34}$ contains 4 leptons and 2 quarks, it leads to a long-range $\ovbb$ decay mechanism at two-loop order without propagation of a heavy boson. On the other hand, for the short-range contribution we need two more quarks and therefore an $s$-channel transition. Interestingly, in both cases one can reduce the operator to different long-range mechanisms corresponding to distinct chiralities of the external particles, but the obtained contribution factors have the same form and differ just in the type of the external Yukawa coupling. Similarly, one can find more short-range contributions for each of these operators, but the long-range contributions always dominate. In contrast, the operator $\mathcal{O}_{36}$ has the right field content, and thus, contributes to a short-range $\ovbb$ decay mechanism at tree level, $\mathcal{O}_{36}^{SR} \propto v^2/\Lambda^7$, while at least two loops are needed in the long-range case,
\begin{align}
	\mathcal{O}_{36}^{LR} \propto 
	\frac{y_e^{ex2}y_dv}{(16\pi^2)^2\Lambda^3} \nba{\frac{1}{16\pi^2} 
	+ \frac{v^2}{\Lambda^2}}.
\end{align}
Moreover, the long-range contribution is further suppressed by external Yukawa couplings; therefore, the leading contribution comes from the short-range mechanism. Since no $s$-channel transition occurs, the contributions triggered by this operator are in general larger than contributions of the operators $\mathcal{O}_{33,34}$. The operator $\mathcal{O}_{38}$ gives a very similar contribution as operator $\mathcal{O}_{36}$. The only difference is that the internal Yukawa coupling appearing in these three contributions is $y_u$ instead of $y_d$.

\begin{figure}[t!] 
	\centering
	\includegraphics[scale=0.36]{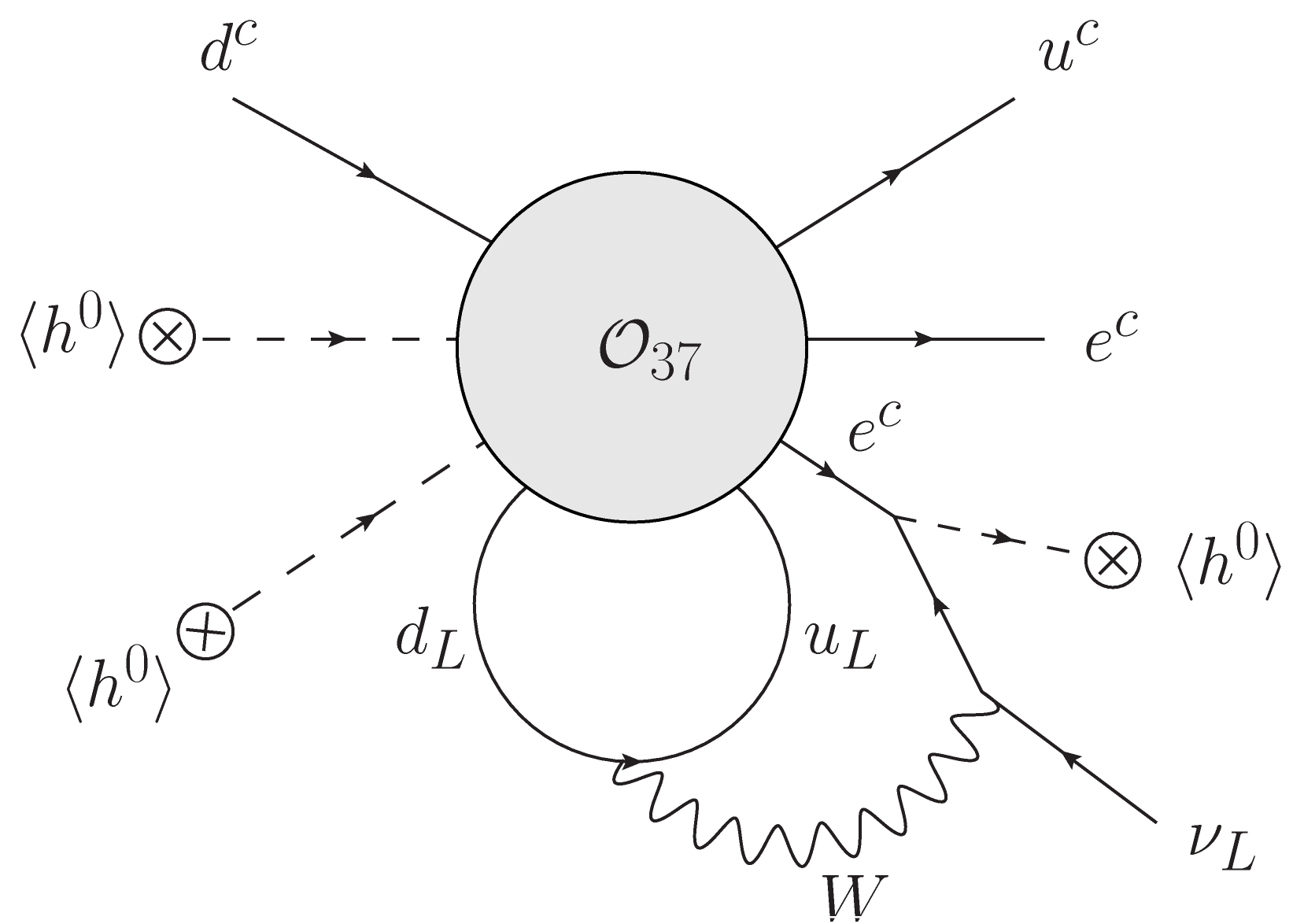}
	\caption{Dominant long-range contribution of $\mathcal{O}_{37}$ to $\ovbb$ decay.}
	\label{fig:Op37}
\end{figure}
An interesting behaviour can be observed for the operator $\mathcal{O}_{37}$. Analogous to operators $\mathcal{O}_{36}$ and $\mathcal{O}_{38}$, it gives long-range contributions proportional to the second power of the external Yukawa couplings, and thus, one would expect them to be similarly suppressed. However, for $\mathcal{O}_{37}$ these contributions are not the dominant ones; there is another long-range contribution containing a vector boson in the loop instead of the Higgs (see Fig.~\ref{fig:Op37}), which makes it proportional to $g^2$ times a single external Yukawa coupling,
\begin{align}
	\mathcal{O}_{37}^{LR} \propto
	\frac{y_e^{\mathrm{ex}}g^2v}{(16\pi^2)^2\Lambda^3}
	\nba{\frac{1}{16\pi^2}+\frac{v^2}{\Lambda^2}}.
\end{align}

\subsection{Comparison with Standard Mass Mechanism}
\label{sec8:Weinberg}
As discussed in Section~\ref{sec:0vbb}, we assumed in our analysis so far that $\ovbb$ decay does not directly originate from the Majorana neutrino mass mechanism but from another LNV new physics contribution. Under this assumption, the $\ovbb$ decay half life gives no direct information about the neutrino mass although any LNV contribution will additionally contribute to the Majorana mass. The corresponding washout interval was derived under this assumption, neglecting the mass contribution for fixing the operator scale.

However, it is still interesting to consider the possibility that also the mass mechanism is triggered only by loop contributions of the higher-dimensional operators in question. We have thus derived as well the corresponding operator scale assuming the observation of $\ovbb$ decay is generated by an underlying loop induced Weinberg-operator. We indicate the corresponding scale with an orange diamond in Fig.~\ref{fig:washout-all-1stgen} and Fig.~\ref{fig:washout-all-3rdgen}. The orange arrows pointing up or down merely indicate a Weinberg operator scale outside the range of the plot. A few comments are in order:

\paragraph{First Generation Yukawa Couplings} Assuming only first generation Yukawa couplings, exotic long- or short-range contributions occur to be mainly dominant, cf. Fig.~\ref{fig:washout-all-1stgen}. Only for around a fifth of the operators listed, the mass mechanism is dominant.
 
\begin{figure}[t!] 
\centering
\includegraphics[clip,width=0.28\textwidth]{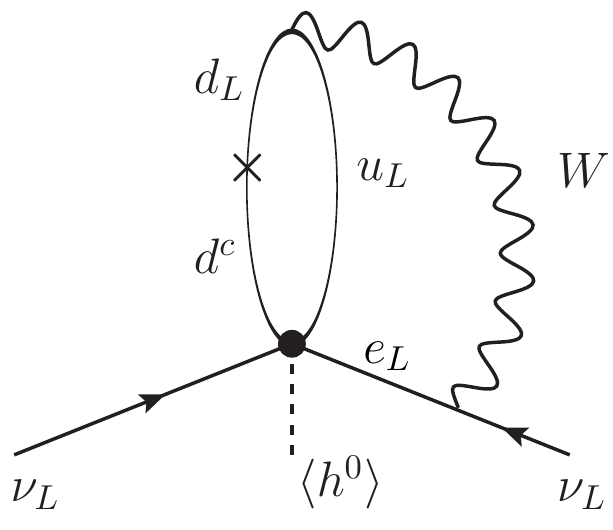}
\includegraphics[clip,width=0.28\textwidth]{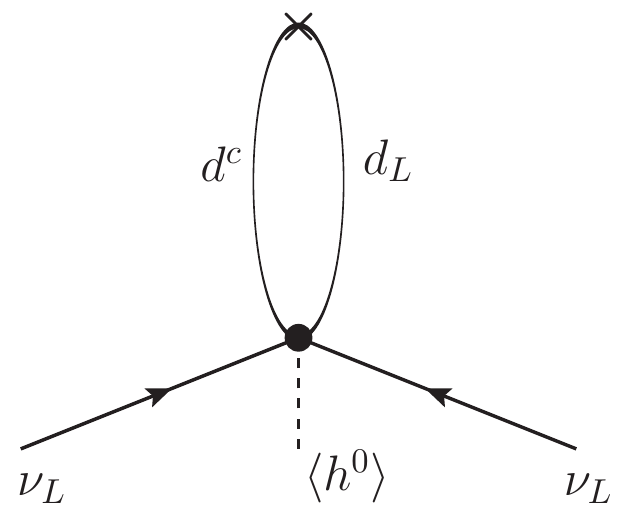}
\caption{Dominant mass contribution of $\mathcal{O}_{3a}$ (left) and $\mathcal{O}_{3b}$ (right) to $\ovbb$ decay.}
\label{fig:op3ab}
\end{figure}

Naively one would expect that the mass mechanism dominates for the 7-dim operators and 9-dim operators with three Higgs doublets (i.e. operators up to $\mathcal{O}_8$) due to reduced loop suppression than for higher operators. This is, however, not necessarily the case. When an operator includes the $SU(2)$ contraction $L_i L_j \epsilon_{ij}$, one lepton leg has to be flipped, as discussed in Section~\ref{sec:loopclosing} and additional loop factors suppress the scale. The complementary structure $L_i L_j \epsilon_{ik} \epsilon_{jl}$ (or similar), however, is less suppressed. To confront both cases, we refer to $\mathcal{O}_{3a}$ and $\mathcal{O}_{3b}$ as examples, cf. Fig.~\ref{fig:op3ab} and the corresponding entries in Tab.~\ref{tab:op7}. Whereas the long-range contribution dominates over the mass mechanism for $\mathcal{O}_{3a}$, it is the opposite for $\mathcal{O}_{3b}$.

In contrast, higher-dimensional operators (9-dimensional and above), however, are generally expected to be less constrained by the neutrino mass due to a higher order loop suppression and small Yukawa couplings. As can be seen in Fig.~\ref{fig:washout-all-1stgen}, the mass mechanism would mostly point to a far too small scale, demonstrating the need of an additional mechanism to generate light neutrino masses. This would in turn imply that a corresponding signal from $\ovbb$ would hint towards a dominant long- or short-range contribution. However, those operators that are proportional to gauge couplings only, as e.g. the previously discussed $\mathcal{O}_{44c}$ in Fig.~\ref{fig:rules2}, can still contribute dominantly to the mass mechanism. This is similarly true for $\mathcal{O}_{27a}$, $\mathcal{O}_{27b}$, $\mathcal{O}_{29a}$, $\mathcal{O}_{29b}$, $\mathcal{O}_{40a}$ based on the same reasoning.
 
\paragraph{Third Generation Yukawa Couplings} For third generation couplings however, the interplay between the mass-mechanism and the long-/short-range contributions is less obvious and more complicated. This results from the interplay between loop suppression on the one hand but large Yukawa couplings on the other hand. As internal and external Yukawa couplings have to be distinguished in each contribution, the situation becomes even less trivial. Comparing with Fig.~\ref{fig:washout-all-3rdgen}, we see that for a third of the studied operators the mass mechanism is not the dominant contribution.
  
\begin{figure}[t!]
\centering
\includegraphics[clip,width=0.28\textwidth]{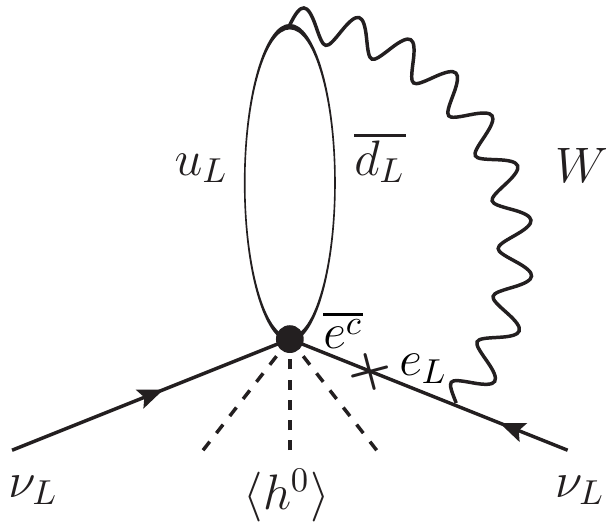}
\includegraphics[clip,width=0.28\textwidth]{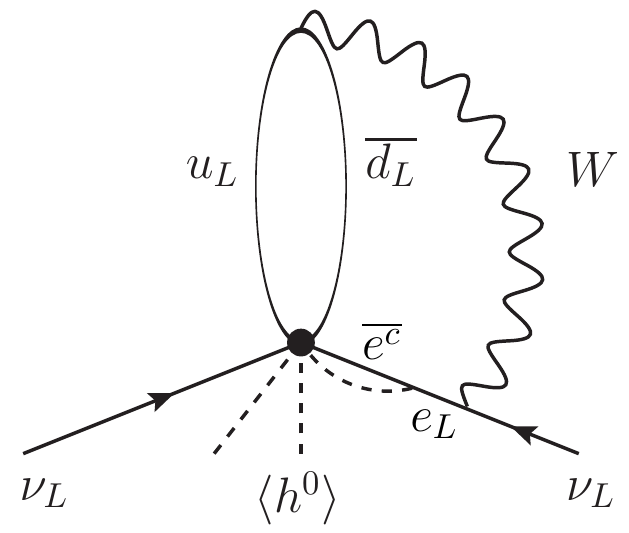}
\includegraphics[clip,width=0.28\textwidth]{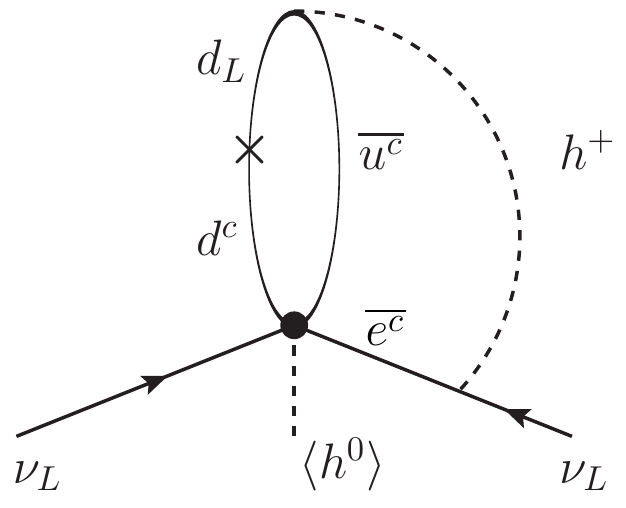} 
\caption{Dominant mass contributions to $\ovbb$ decay generated from $\mathcal{O}_7$ $\propto v^2/\Lambda$ (left), $\propto v^4/\Lambda^3$ (centre) and the corresponding diagram for $\mathcal{O}_8$ (right).}
\label{fig:op78}
\end{figure}

Generally one would now naively expect that the mass mechanism dominates fully for the 7-dim operators and 9-dim operators with three Higgs doublets (i.e. operators up to $\mathcal{O}_8$). This is true for all but $\mathcal{O}_7$ and $\mathcal{O}_8$, for which still the exotic long-/short-range contribution dominates. The reason in case of the 7-dimensional $\mathcal{O}_8$ is that it is the only 7-dim operator where a lepton leg can be flipped by a Higgs boson only with an corresponding external (small) Yukawa coupling, due to the right-handed current structure. For similar operators the left-handed current allows for a flip via gauge bosons featuring a higher operator scale. As example we compare $\mathcal{O}_8$ with $\mathcal{O}_{3a,3b,4a}$, see Fig.~\ref{fig:op3ab} and Fig.~\ref{fig:op78}. Comparing the 9-dimensional operators with three Higgs doublets with each other, one observes a similar reasoning. In order to obtain the mass contribution with $\mathcal{O}_7$, a more complicated loop structure including one external small Yukawa coupling is needed in contrast to $\mathcal{O}_{5,6}$ (cp. Fig.~\ref{fig:op78} as well the corresponding Tabs.~\ref{tab:op7} and \ref{tab:op9}).

\begin{figure}[t!]
\centering
\includegraphics[clip,width=0.28\textwidth]{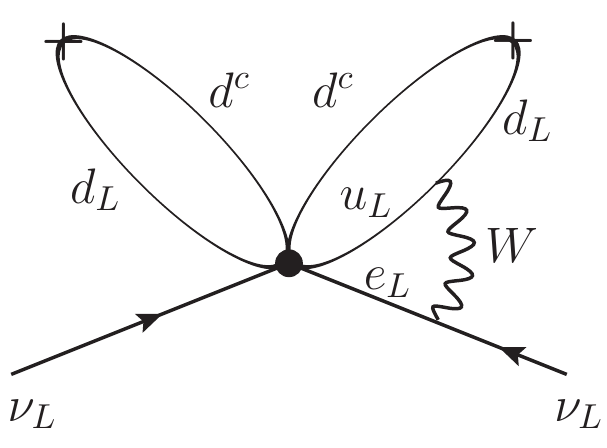}
\includegraphics[clip,width=0.28\textwidth]{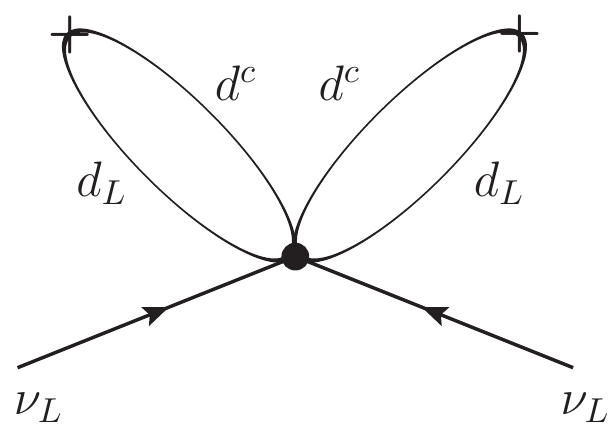}
\caption{Dominant mass contribution of $\mathcal{O}_{11a}$ (left) and $\mathcal{O}_{11b}$ (right) to $\ovbb$ decay.}
\label{fig:op11abmass}
\end{figure}

For higher dimensional operators the interplay is not easy to decipher any more, as the operator scale now highly depends on how much external Yukawa couplings are involved in the mass mechanism and to what kind of exotic long-/short-range contribution it is to be compared. For operators $\mathcal{O}_{10}$, $\mathcal{O}_{11a}$ and $\mathcal{O}_{11b}$, which all feature a similarly strong long-range contribution, the mass mechanism dominates only for $\mathcal{O}_{11b}$. This is due to the fact that its mass contribution is the only one that is not further suppressed by either a small externally fixed Yukawa coupling ($\mathcal{O}_{10}$) or an additional loop suppression ($\mathcal{O}_{11a}$), cf. Fig.~\ref{fig:op11abmass}.

\begin{figure}[t!]
\centering
\includegraphics[clip,width=0.28\textwidth]{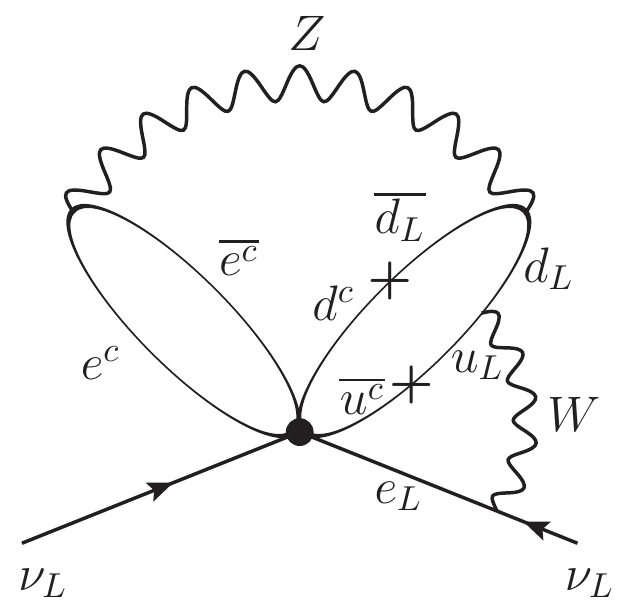}
\includegraphics[clip,width=0.28\textwidth]{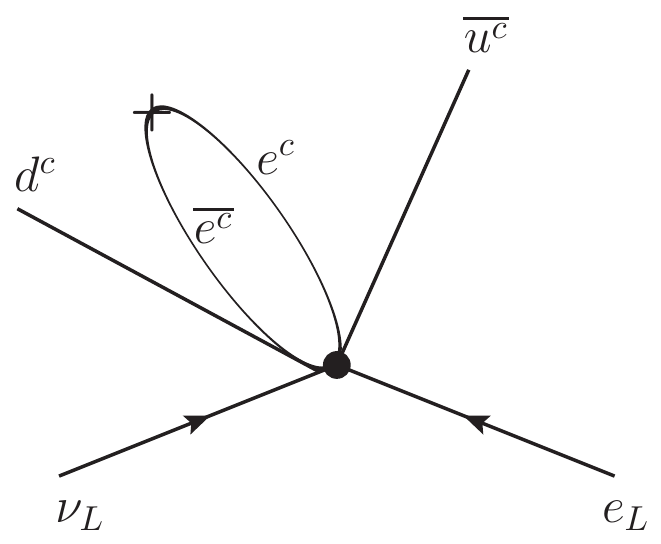}
\includegraphics[clip,width=0.28\textwidth]{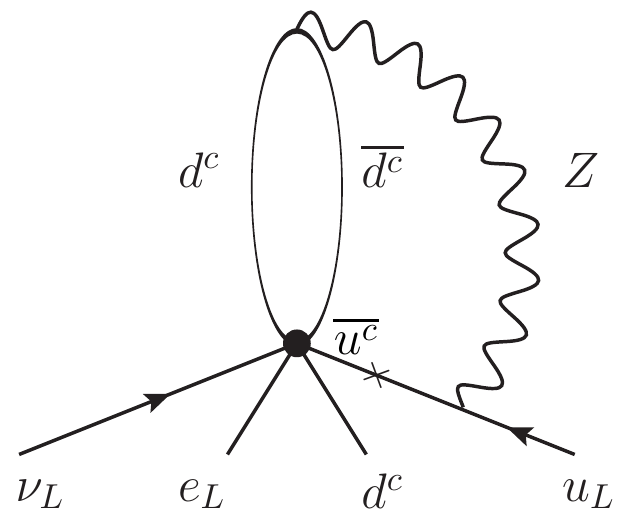}
\caption{Dominant mass contribution of $\mathcal{O}_{16}$ (left) (also similar for $\mathcal{O}_{17}$ and $\mathcal{O}_{18}$) and the dominant long-range contributions of $\mathcal{O}_{16}$ (centre) and $\mathcal{O}_{17}$ (right) to $\ovbb$ decay.}
\label{fig:op161718}
\end{figure}

The operators $\mathcal{O}_{16}$, $\mathcal{O}_{17}$ and $\mathcal{O}_{18}$ demonstrate the opposite picture. While all feature the same mass contribution, their dominant long-range contributions differ, cf. Fig.~\ref{fig:op161718}. While $\mathcal{O}_{17,18}$ feature an additional loop suppression and an external Yukawa coupling, only $\mathcal{O}_{16}$ contributes directly. Thus, the non-standard contribution is still dominant for $\mathcal{O}_{16}$.

The 9-dim operators $\mathcal{O}_{19}$, $\mathcal{O}_{20}$, $\mathcal{O}_{76}$ and the 11-dim operators $\mathcal{O}_{26a,b}$, $\mathcal{O}_{30a,b}$ and $\mathcal{O}_{34-38}$ feature a dominant non-standard contribution. This is mainly triggered by fixed external first generation Yukawa couplings in the Weinberg contribution, suppressing the scale.

While in Tabs.~\ref{tab:op7}, \ref{tab:op9} and \ref{tab:op11} the dominant contribution is shown under the assumption of first generation Yukawa couplings only, we want to stress again that we have taken into account all possibilities to generate the mass, long- and short-range contributions in our full analysis shown in Figs.~\ref{fig:washout-all-1stgen}-\ref{fig:washout-all-3rdgen}. Its importance can be demonstrated by taking $\mathcal{O}_{29a}$ as an example, discussed in Section~\ref{sec:loopclosing} in more detail. While Eq.~\eqref{eq:29a1} dominates only for first generation Yukawa couplings, Eq.~\ref{eq:29a3} is the dominant one when taking into account third generation internal Yukawa couplings, cf. Fig~\ref{fig:op29}. This is especially interesting for the interplay with similar operators as e.g. $\mathcal{O}_{29b}$ that can generate only Eq.~\eqref{eq:29a1} due to its structure. 
%

%% file: conclusions.tex
\section{Discussion and Conclusions}
\label{sec:conclusions}

The observation of $\ovbb$ decay, or LNV in general, would have profound consequences for our understanding of nature. Specifically, it will likely lead us to an understanding of the light neutrino masses and open a portal to new physics beyond the SM. The presence of LNV, or more generally $B-L$ violating interactions, would also impact potential mechanisms generating the observed baryon asymmetry of the universe. In this work, we perform a detailed study on how effective SM invariant $\Delta L = 2$ operators containing SM fermions and Higgs, contribute to $\ovbb$ decay both at tree and loop level. In other words, we derive the effective scale $\Lambda$ associated with the operator from a hypothetical observation of $\ovbb$ decay by considering the contributions an operator generates. The $\Delta L = 2$ operators of dimension-7, 9 and 11 are collected in Tabs.~\ref{tab:op7} - \ref{tab:op11}, adapted from Ref.~\cite{deGouvea:2007xp} and supplemented by the Hilbert Series method \cite{Henning:2015alf}. We mainly focus on operators of dimension-7 and 9 but we also study a representative range of dimension-11 operators, focusing on those that exhibit qualitatively different features from dim-9. Given an operator, the tree-level contribution to $\ovbb$ decay exists as long as the operator contains the required particle content to realize the short-range or long-range interactions. To be more precise, an operator that consists of two up- and two down-quarks as well as two electrons, will have a short-range contribution. It is, however, highly non-trivial to manually exhaust all possibilities of loop diagrams, stemming from the given operator, that trigger $\ovbb$ decay. For example, the operator  $\mathcal{O}_{16}$ includes the term $\nu_L e_L d^c e^c \bar{e}^c \bar{u}^c$ after $SU(2)_L$ decomposition and thus has no short-range contribution at tree level. By connecting $e^c$ and $\bar{e}^c$ with an electron-mass insertion, the resulting lower-dimensional operator $\nu_L e_L d^c \bar{u}^c$ triggers $\ovbb$ via a long-range contribution.  

To explore the large number of possibilities, and to discuss other observables in the future, we have developed a tool which generates the radiative terms by closing loops, if necessary with mass insertions and Higgs/gauge boson emission. There is a rich phenomenology involved in comparing the effect of different operators. Our results are summarized in Tabs.~\ref{tab:op7} -\ref{tab:op11}, where the dominant long-range and short-range $\ovbb$ contribution is given for each operator, as well as with the associated neutrino mass scale generated radiatively in the same fashion. The operator scale $\Lambda$ is then shown in Figs.~\ref{fig:washout-all-1stgen} and \ref{fig:washout-all-3rdgen}, corresponding to a hypothetical observation of $\ovbb$ decay at $T_{1/2}^\text{Xe} = 10^{27}$~y and assuming the Yukawa couplings involved are of first and third generation, respectively.

With the scale of a given $\Delta L = 2$ operator determined from $\ovbb$ decay, we compute the corresponding washout rate on the lepton asymmetry in the early universe.
We solve the Boltzmann equation for the evolution of the lepton asymmetry in the presence of an effective $\Delta L = 2$ operator and the usual SM interactions, including sphaleron transitions. All possible permutations of the particles, contained in the $\Delta L = 2$ operator, corresponding to physically distinctive washout processes, are taken into account. As a consequence, we find a temperature range within which the washout is efficient and also infer a lower bound on the temperature above which any pre-existing lepton or baryon asymmetry will be erased by the operator.

\begin{figure}[t!]
\centering
\includegraphics[clip,width=0.44\linewidth]{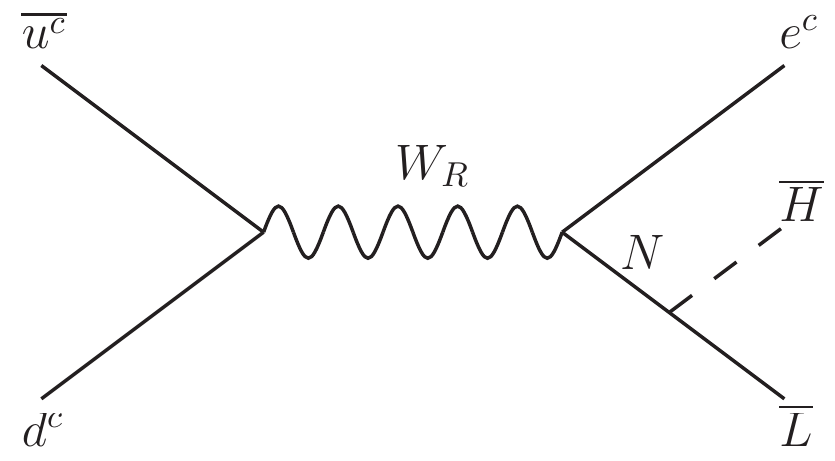}
\includegraphics[clip,width=0.44\linewidth]{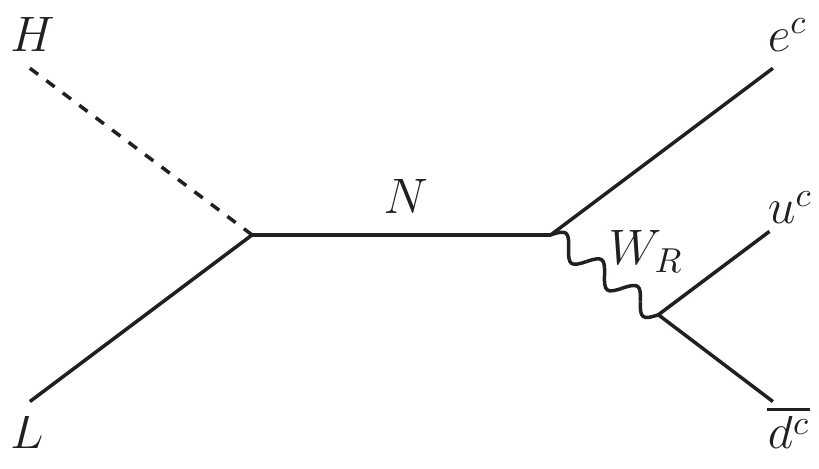}
\caption{Example diagrams in a left-right symmetric model framework that give rise to the effective operator $\mathcal{O}_8 = L^i \bar{e^c} \bar{u^c} d^c H^j \epsilon_{ij}$.}
\label{fig:EFT_UV}
\end{figure}
The operator scale $\Lambda$ and the associated temperature range of strong washout, given an observation of $\ovbb$ at $T_{1/2} = 10^{27}$~y, can be as high as $[\lambda, \Lambda]  \approx [2\times 10^3, 3\times 10^5]$~GeV for dim-7 operators. These dim-7 operators, however, often have strong constraints from the requirement to keep the neutrinos light. On the other hand, dim-9 and dim-11 operators typically washout the lepton number in the range from 100~GeV to 1~TeV if the underlying operator couples to first generation fermions only, in which case the constraint from reproducing light neutrino masses can be evaded in many cases. Surprisingly, many of the dim-9 and dim-11 operators in fact induce sizeable long-range contributions.

To better understand the washout from effective operators, we would like to comment on the difference in terms of lepton number washout rates between the effective operator approach and an underlying UV theory. For illustration, we choose the operator $\mathcal{O}_8 = L^i \bar{e^c} \bar{u^c} d^c H^j \epsilon_{ij}$ and consider a left-right symmetric model (LRSM)~\cite{Mohapatra:1974hk, Mohapatra:1974gc, Senjanovic:1975rk, Mohapatra:1980yp}, which gives rise to this operator after integrating out the right-handed gauge boson $W_R$ and the right-handed neutrino $N$. Moreover, instead of taking into account all permutations of the initial and final state particles, we confine ourselves only to two of them to underscore the impact of the resonant enhancement from on-shell $W_R$ or $N$. The two washout processes of interest are $\bar{u^c} d^c \leftrightarrow \bar{L} e^c \bar{H}$ and $L H \leftrightarrow e^c u^c \bar{d^c}$, respectively. The corresponding Feynman diagrams in the LRSM are shown in Fig.~\ref{fig:EFT_UV}. The computation of the scattering amplitudes is straightforward and the thermal rate can be obtained based on Eq.~\eqref{eq:thermal_rate}. We also calculate the washout rate for these two processes combined according to the effective operator approach. The decay width of $W_R$ and $N$ are estimated to be $\Gamma_{W_R} = m_{W_R} g^2_R/(8\pi)$ and $\Gamma_N = m_N (y^2_\nu + \frac{g_R^4}{24\pi^2} m_N^4/m_{W_R}^4)/(8\pi)$, respectively, where $g_R$ is the right-handed gauge coupling and $y_\nu$ is the neutrino Yukawa coupling. In the limit of the momentum transferred being much smaller than $m_{W_R}$ and $m_N$, the underlying UV theory and the effective operator should produce the same result which can be ensured by requiring $g^2_R y_\nu/(m^2_{W_R} m_N) = 1/\Lambda^3$.

\begin{figure}[t!]
\centering
\includegraphics[clip,width=0.49\linewidth]{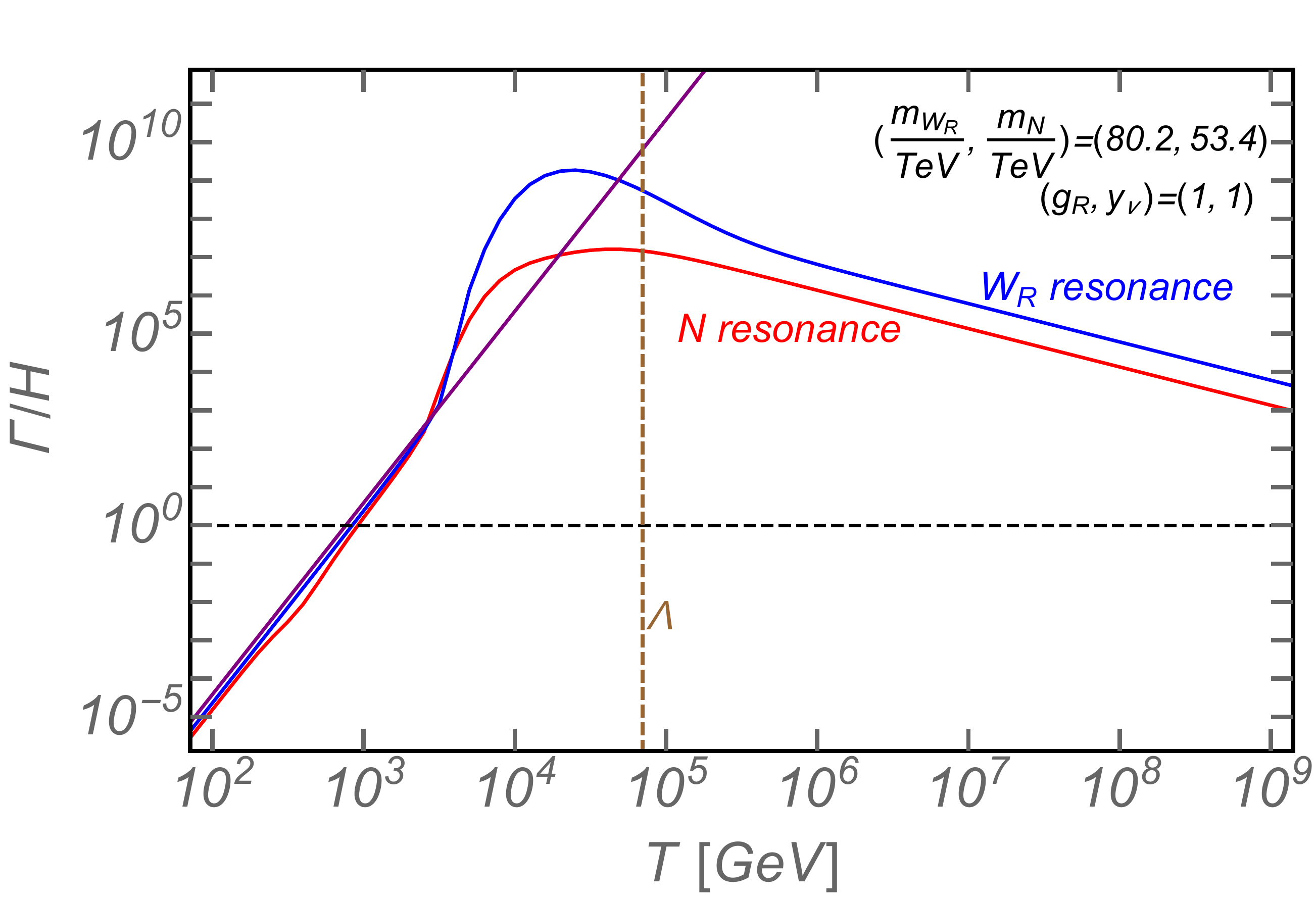}
\includegraphics[clip,width=0.49\linewidth]{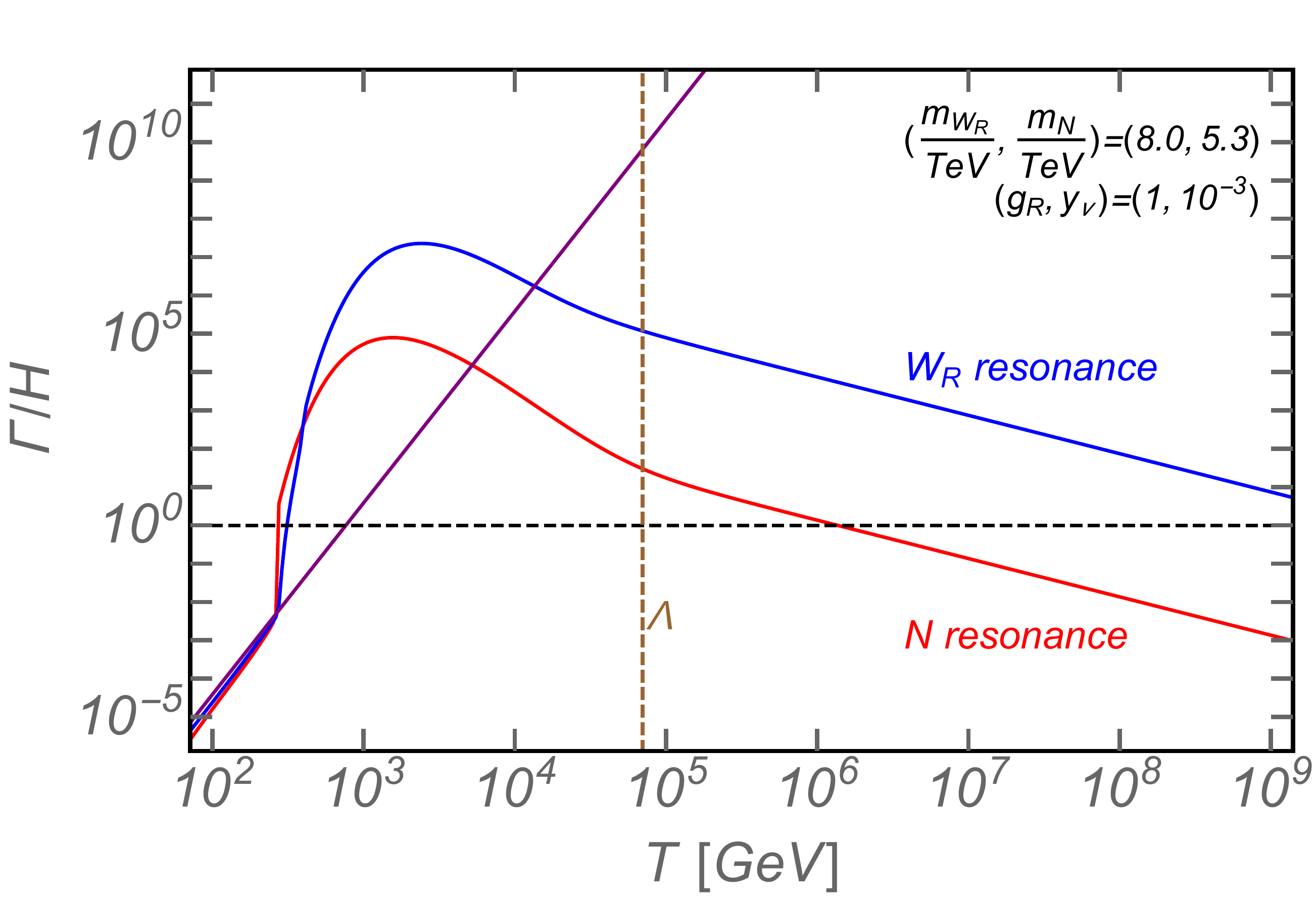}
\caption{Lepton number washout rate as function of temperature from the UV processes $\bar{u^c} d^c \leftrightarrow \bar{L} e^c \bar{H}$ (blue), $L H \leftrightarrow e^c u^c \bar{d^c}$ (\emph{red}), and the sum of the two processes~(\emph{purple}) computed in the effective operator approach with the operator scale set to $\Lambda = 7 \times 10^{4}$~GeV. We show results for two different sets of values for the right-handed gauge coupling $g_R$ and the neutrino Yukawa coupling $y_\nu$. The corresponding heavy particle masses are also shown.}
\label{fig:ga_EFT_UV}
\end{figure}
In Fig.~\ref{fig:ga_EFT_UV}, we show the normalized interaction rates with respect to the Hubble expansion rate $\Gamma_W/H$ as a function of the temperature. The straight purple line indicates the washout rate of the effective operator matched to the sum of the above two UV processes, while the blue and red lines correspond to the rate of the UV process $\bar{u^c} d^c \leftrightarrow \bar{L} e^c \bar{H}$ and $L H \leftrightarrow e^c u^c \bar{d^c}$ in Fig.~\ref{fig:EFT_UV}, respectively. As mentioned above, the couplings and masses in the LRSM are correlated with the effective scale $\Lambda$, for which we use $\Lambda = 7\times 10^4$~GeV taken from Fig.~\ref{fig:washout-all-1stgen}. For definiteness, we accordingly set $m_{W_R} = 1.5\,m_N$ with two different sets of couplings: $(g_R, y_\nu) = (1, 1)$~(left panel) and $(g_R, y_\nu) = (1, 10^{-3})$~(right panel). It is clear that with larger couplings, the masses $m_{W_R}$ and $m_N$ are also larger given the fixed operator scale $\Lambda$ and so the resonance enhancement occurs at higher temperatures. For small temperatures, the UV theory and the effective operator yield consistent results as expected. For temperatures much larger than the masses $m_{W_R}$ and $m_N$, the effective operator approach becomes un-physical while the rate of the UV processes is proportional to $T$. Consequently, the washout process becomes inefficient for much higher temperatures since the Hubble expansion rate is proportional to $T^2$; the smaller $g_R$ and $y_\nu$, the lower the  temperatures above which the $L$ washout is out of equilibrium.

All in all, the lepton number washout remains, in principle, effective above the cut-off scale $\Lambda$, and it can be even stronger than what is predicted by the effective operator approach due to the resonance enhancement from new particles in an underlying UV theory. The details of this are of course model-dependent. Our conclusion drawn based on the effective operator approach that a pre-existing asymmetry above the scale $\hat{\lambda}_D$ will be erased, however, is expected to hold since at low energies both the UV theory and the effective operator approach should yield the same result, unless the coupling constants in the UV theory are so small that the new particles in the UV theory have masses near or below the EW scale. It is also worthwhile to point out that if the LNV arises from spontaneous breaking (as for instance the triplet Higgs VEVs $\langle \Delta_{L,R} \rangle$ breaking $B-L$ in the LRSM), the corresponding symmetry is restored for temperatures above the breaking scale. The $L$ washout processes are then expected to cease to work above the scale of symmetry restoration. 

We conclude the paper by commenting on a few limitations of our approach. First, $\ovbb$ decay involves electrons but not $\mu$ and $\tau$ leptons. To wipe out asymmetries stored in the $\mu$ and $\tau$ flavours, one would also need to establish that those lepton flavour asymmetries are equilibrated as well, e.g. by observing lepton flavour violation effective around the same temperatures~\cite{Deppisch:2015yqa}. Another possibility are processes with LNV directly involving muons or taus such as meson decays and direct searches at the LHC \cite{Deppisch:2013jxa}. Rare lepton flavour violating (LFV) processes are induced to lowest order by 6-dim operators of the form $\mathcal{O}_{\ell\ell\gamma} = \mathcal{C}_{\ell\ell\gamma} \bar L_\ell \sigma^{\mu\nu} \bar{\ell^c} H F_{\mu\nu}$ and $\mathcal{O}_{\ell\ell q q} = \mathcal{C}_{\ell\ell q q} (\bar{\ell} \, \Pi_1 \ell) (\bar{q} \, \Pi_2 q)$ (possible Lorentz structures are represented by $\Pi_i$), with $\ell = e, \mu, \tau$. Each of these operators is associated with a corresponding operator scale that is probed by low energy LFV observables such as the decay branching ratios $\text{Br}_{\mu\to e \gamma} < 5.7\times 10^{-13}$~\cite{Adam:2013mnn}, $\text{Br}_{\tau\to \ell\gamma} \lesssim 4.0\times 10^{-8}$ ($\ell=e,\mu$)~\cite{Agashe:2014kda} and the $\mu-e$ conversion rate $\text{R}^\text{Au}_{\mu\to e} < 7.0\times 10^{-13}$~\cite{Agashe:2014kda} (current limits at 90\% C.L.). The associated operator scales probed by these searches are of the order $\Lambda_{\mu e\gamma}\approx 3\times 10^6$~GeV, $\Lambda_{\tau \ell\gamma}\approx 3\times 10^4$~GeV and $\Lambda_{\mu eqq}\approx 2\times 10^5$~GeV, respectively \cite{Deppisch:2015yqa}. While these operators do not lead to a washout of net lepton number, they will equilibrate the individual flavour number asymmetries within a certain temperature interval $[\lambda_i, \Lambda_i]$ \cite{Deppisch:2015yqa}, in analogy to the treatment of lepton number washout in this work. In case this interval overlaps with the $\Delta L = 2$ washout interval of total electron number (if $0\nu\beta\beta$ is observed), the net number of the either muons or taus will be efficiently washed out as well. In \cite{Deppisch:2015yqa} we have calculated these temperature intervals and we have found that the overlap between LNV washout (of 7,9,11-dim operators) and the LFV operators $\Lambda_{\tau \ell\gamma}$ and $\Lambda_{\mu eqq}$ is indeed large, assuming observation of $\ovbb$ and LFV at near-future experimental sensitivities. On the other hand, the operator scale $\Lambda_{\mu e\gamma}$ already has such a stringent lower limit that the associated flavour equilibration interval does not overlap with the LNV washout interval of most 9-dim and 11-dim operators.

Second, to compute the washout effects in a model-independent way, we simply assume that the baryon asymmetry generation mechanism is not related to the washout process in question. It may be the case that the underlying $L$ violating theory responsible for the washout also creates a lepton number asymmetry in the first place. In other words, our conclusion only applies to those asymmetries generated before the $L$ washout becomes efficient. Finally, as pointed out in \cite{Dimopoulos:1988jw, Sierra:2013kba}, if there exists a decoupled sector which shares the baryon asymmetry with the visible sector, the $L$ washout can not completely erase the $L$ asymmetry in the visible sector. That is because when the decoupled sector communicated to the visible sector at very early times, it shared not only the baryon asymmetry but also a hypercharge $U(1)_Y$ asymmetry. As the $L$ washout processes preserve the $U(1)_Y$ charge, the $L$ asymmetry, which is proportional to $U(1)_Y$ asymmetry in this case, can not be completely destroyed. After the sphalerons cease to work, the asymmetry stored in the decoupled sector can be converted back to the visible sector and thus evade the washout process. The conversion can be realized, for example, if the particle that carries the asymmetry in the decoupled sector is long-lived and decays to SM particles below the electroweak scale. Alternatively, the asymmetry transfer mechanism between the two sectors may only become efficient after the electroweak phase transition due to the scaling of the expansion rate as $T^2/\Lambda_\text{Pl}$.

In any case, the observation of $\ovbb$ will provide a means to test mechanisms of baryogenesis in addition to mechanisms of neutrino mass generation. While the high-scale seesaw mechanism operating at a scale of $\approx 10^{14}$~GeV remains a popular scenario, Majorana neutrino mass mechanisms with an associated breaking of the lepton number close to the EW scale are of strong theoretical interest. Such models generically predict new contributions to $\ovbb$ decay. To apply the reasoning put forward in our paper, it is necessary to differentiate between different mechanisms responsible for $\ovbb$ decay, at least in order to distinguish exotic contributions from the standard neutrino mass mechanism. In the context of $\ovbb$ alone, this can be for example achieved by searching for $\ovbb$ decay in multiple isotopes~\cite{Bilenky:2004um, Deppisch:2006hb, Gehman:2007qg} or by utilizing experiments that are sensitive to the individual electron energies \cite{Doi:1982dn, Doi:1985dx, Tomoda:1986yz, Ali:2006iu, Ali:2007ec}, for instance in the SuperNEMO experiment~\cite{Deppisch:2010zza, Arnold:2010tu}. The presence of non-standard contributions to $\ovbb$ decay could also manifest itself as potential inconsistencies between the results from $\ovbb$ decay and from the determination of the sum of neutrino masses using cosmological considerations \cite{Deppisch:2004kn}. More generally, LNV can also be probed in other observables; for example, searches at high energy colliders for LNV in resonant processes would have the advantage of pinpointing the LNV scale more directly as demonstrated in \cite{Deppisch:2013jxa, Deppisch:2015yqa}.

Our results show that the scale $\Lambda$ of many of the $\Delta L = 2$ operators and the corresponding temperature range of strong washout are $\mathcal{O}(\text{TeV})$ assuming an observation of $\ovbb$ in future or planned experiments with a sensitivity of $T_{1/2} \approx 10^{27}$~y. In this case, there is underlying new physics at work, potentially within the reach of the LHC and future colliders. Together with the $B+L$ violating sphalerons, the presence of LNV can erase a pre-existing baryon and lepton asymmetry generated at high temperatures. As a result, the observation of $\ovbb$ decay will strongly constrain high-scale~($\gtrsim$ TeV) scenarios of baryogenesis and leptogenesis.